\documentclass[twocolumn,numberedappendix,twocolappendix]{openjournal}

\usepackage[utf8]{inputenc}
\usepackage[english]{babel}
\usepackage[usenames,dvipsnames]{xcolor}   
\usepackage{colortbl}                      

\usepackage{hyperref}
\hypersetup{
    unicode,
    colorlinks=true,
    linkcolor=linkcolor,
    citecolor=linkcolor,
    filecolor=linkcolor,
    urlcolor=linkcolor,
}
\definecolor{linkcolor}{RGB}{28, 96, 214}
\DeclareGraphicsExtensions{.bmp,.png,.jpg,.pdf}
\DeclareUnicodeCharacter{2212}{-}           

\usepackage{amsmath,amsfonts,amssymb,float,bm}
\usepackage{physics}    

\usepackage{multirow}
\usepackage{subcaption}
\usepackage{adjustbox}

\usepackage{siunitx}
\DeclareSIUnit \h {\ensuremath{\mathit{h}}}
\DeclareSIUnit \parsec {pc}
\DeclareSIUnit \deg {deg}
\DeclareSIUnit \arcmin {arcmin}
\usepackage{lineno}

\usepackage[capitalize]{cleveref}
\crefformat{footnote}{footnote~#2#1#3}
\usepackage{aas_macros}     

\makeatletter
\usepackage{etoolbox}
\patchcmd\H@refstepcounter{\protected@edef}{\protected@xdef}{}{}
\makeatother

\usepackage{xspace}

\newcommand{\Om}{\Omega_{\rm m}}

\newcommand{\n}{n_{\mathrm{gal}}}
\newcommand{\m}{n_{\mathrm{obj}}}

\graphicspath{ {./figs/} }

\makeatletter
\expandafter\global\expandafter\let\csname longtable*\endcsname\relax
\expandafter\global\expandafter\let\csname endlongtable*\endcsname\relax
\makeatother

\begin{document}
\title{Probabilistic characterization of blending with LSST and application to cluster lensing cosmology}

\author{Manon Ramel}
\email{ramelmn@ucmail.uc.edu}
\affiliation{University of Cincinnati, Cincinnati, OH 45221, USA}
\affiliation{LPSC-IN2P3, Univ. Grenoble Alpes, CNRS, LPSC-IN2P3, 38000 Grenoble, France}

\author{Cyrille Doux}
\affiliation{LPSC-IN2P3, Univ. Grenoble Alpes, CNRS, LPSC-IN2P3, 38000 Grenoble, France}

\author{Marine Kuna}
\affiliation{LPSC-IN2P3, Univ. Grenoble Alpes, CNRS, LPSC-IN2P3, 38000 Grenoble, France}

\author{Michel Aguena}
\affiliation{Istituto Nazionale di Astrofisica, Viale del Parco Mellini No.84 00136 ROMA, Italy}

\author{Céline Combet}
\affiliation{LPSC-IN2P3, Univ. Grenoble Alpes, CNRS, LPSC-IN2P3, 38000 Grenoble, France}

\author{Shuang Liang}
\affiliation{Stanford University, 450 Serra Mall, Stanford, CA 94305, USA}

\author{Constantin Payerne}
\affiliation{Université Paris-Saclay, CEA, IRFU, 91191, Gif-sur-Yvette, France}

\author{Camille Avestruz}
\affiliation{University of Michigan, 500 S State St, Ann Arbor, MI 48109, USA}

\author{Alex I. Malz}
\affiliation{Space Telescope Science Institute, 3700 San Martin Dr., Baltimore, MD 21211, USA}

\author{Marina Ricci}
\affiliation{Université Paris Cité, CNRS/IN2P3, APC, 4 rue Elsa Morante, F-75013 Paris, France}

\author{Nikolina \v{S}ar\v{c}evi\'c}
\affiliation{Duke University, Durham, NC 27708, USA}

\collaboration{LSST Dark Energy Science Collaboration}

\begin{abstract}
Next-generation galaxy surveys, like the Vera C. Rubin Observatory's Legacy Survey of Space and Time (LSST), will deliver unprecedented depth and sky coverage, enabling precise measurements of cosmic probes such as weak lensing and galaxy clustering. However, increased imaging depth leads to significant blending of galaxy images, particularly in dense fields like galaxy clusters. This blending, exacerbated by atmospheric blurring in ground-based observations, contaminates galaxy property measurements and causes source confusion.
To simultaneously capture these effects, we develop a probabilistic framework introducing the \textit{blending entropy}, a metric quantifying the ambiguity in matching detected objects to true galaxies or external reference sources. Using simulated data from the DESC Data Challenge 2 (DC2), we characterize blending in LSST data and quantify its impact on cluster lensing cosmology around cosmoDC2 halos. 
We demonstrate that imposing a blending entropy threshold of $S_b<0.2$ effectively filters out highly blended objects (around 25\%), which are especially prevalent near the survey's magnitude limit and are associated with higher errors in shape measurements and photometric redshifts. Applying this cut substantially reduces blending-induced biases in cluster lensing profiles and mass estimates, thereby mitigating systematic errors in cosmological parameters---most notably reducing tension in $\sigma_8$ estimates. Our method is readily generalizable to other static probes and offers a practical path forward for real data analyses, particularly when leveraging overlapping high-resolution datasets from spaced-based missions such as \textit{Euclid} or the Roman Space Telescope, where these external datasets can act as reference catalogs to improve the identification of blended sources in LSST data.
\end{abstract}



\section{Introduction}

The next generation of galaxy surveys, including the Vera C. Rubin Observatory's Legacy Survey of Space and Time~\citep{Ivezic_LSST}, ESA's \textit{Euclid} satellite~\citep{Laureijs_Euclid}, and NASA's Nancy Grace Roman Space Telescope~\citep{Akeson_Roman}, will open a new window into the fundamental nature of the late-time Universe, improving upon current surveys in depth, wavelength coverage and sky area. In particular, their galaxy catalogs will enable measurements of static cosmic probes---including weak lensing, galaxy clustering, and galaxy cluster abundance, all linked to the underlying late-time 3D dark matter distribution which depends strongly on "growth" parameters such as $\Om$ and $\sigma_8$---to unprecedented precision~\citep{Mandelbaum_LSST_Science, Tutusaus_euclid_xc, Wu_cosmo_gc_wl}. However, as surveys become increasingly deep, images will be crowded with astronomical sources, such that distant multiple galaxies may overlap along the line of sight, an effect called \textit{blending}~\citep[see][for a recent review]{Melchior_blending}. Blending is aggravated by the atmosphere for ground-based surveys like LSST, blurring the galaxy images in addition to the Point Spread Function (PSF), and recent studies estimate that roughly two thirds of LSST galaxies will be blended after $10$ years of observation~\citep{Chang_density_LSST, Dawson_blending, Melchior_blending, Sanchez_blending}.

Blending causes contamination of galaxy property measurements, such as positions, fluxes, photometric redshifts and shapes, but also confusion of sources, whereby certain detected objects may correspond to several real galaxies, which defines \textit{unrecognized} blends.\footnote{Galaxies with complex morphologies may also be shredded into multiple detected objects. Therefore, items in the catalog of extended sources processed by the image pipeline will be designated as (detected) \textit{objects}, whereas the term \textit{galaxies} will refer to the true (real or simulated) astrophysical sources.} In these situations, most \textit{deblender} software tools are unable to separate the contributions from galaxies to the total image flux in each pixel~\citep{Melchior_scarlet, Arcelin_debvader, Biswas_madness, Sampson_scarlet2, Mendoza_btk}. It is estimated that about 10-to-20\% of detected objects will be unrecognized blends in LSST~\citep{Dawson_blending,Troxel_blending}, making it a significant source of uncertainty in cosmology with weak gravitational lensing. Current approaches require extensive image simulation suites to mitigate the effects of blending on photometric redshifts and shear measurements~\citep{MacCrann_bl_shear_z, Li_HSC_Y3_cat, Genc_bl_shear, Zhang_bl_redshift}. In particular, lensing measurements in galaxy cluster fields, used to estimate their masses, will likely be more affected by blending due to their higher density~\citep{Hernandez_blending_cluster,Shenming_thesis, Gaztanaga_blending_cluster}. It is therefore natural to expect that cluster detection, lensing profiles, mass estimates, and subsequently cosmological constraints, will be affected in non-trivial ways by blending.

In this work, we attempt to characterize blending for upcoming LSST data using simulated data from the LSST Dark Energy Science Collaboration (DESC) as part of the DESC Data~Challenge~2~\citep[DC2,][]{Abolfathi_DC2} and quantify its impact on cluster cosmology. 
We identify two major difficulties to do so. First, disentangling blending from other sources of systematic errors is non-trivial as it disrupts every step of the pipeline, including detection, deblending, shape measurements, and photometric redshift estimation. The second challenge is to fully capture the impact of blending, including contaminations of galaxy property measurements and source confusion.
Indeed, most analyses to date are either limited to one-to-one comparison of object and galaxy properties, neglecting unrecognized blends, and/or struggle at defining what constitutes a blend, which depends on whether real or simulated data (where truth is accessible) are considered. For real data, blending metrics include blendedness, as defined in the HSC pipeline~\citep{Bosch_blendedness}, and purity~\citep{Sanchez_blending}. For simulated data, however, blends are often identified solely based on comparison of galaxy and object positions, thereby neglecting the extendedness of galaxies~\citep{Nourbakhsh_blending, Troxel_blending, Levine_blending}. Recent works demonstrated the importance of considering this information. Indeed, \cite{Liang_blending} showed that galaxy magnitudes and sizes were some of the most important training features to consider for the detection of unrecognized blends using Machine Learning methods. We therefore define a new matching scheme using positions, fluxes and shapes of detected objects and true galaxies, allowing us to define an estimate of true blending that is better suited for weak lensing cosmology.

The key insight is that blending introduces some ambiguity when matching objects and galaxies, which is quantified by our new metric that we refer to as \textit{blending entropy}. We demonstrate that it efficiently captures blended objects, allowing us to remove them from the analysis when the true underlying galaxy population is known. Finally, we apply our framework to a simplified, simulated cluster cosmology analysis, roughly following~\cite{Payerne_mass_richness}, and quantify blending-related biases on lensing profiles, mass estimates and cosmological parameters. We immediately note that our approach is generalizable to other static probes, including galaxy lensing and clustering auto- and cross-correlations. We also provide insight into how our methodology can be applicable to real data, matching LSST deep images with space-based data used as reference.
The outline of the paper is as follows. In \cref{sec:Simulated_data}, we describe the simulated catalogs used in this work. In \cref{sec:Identification_blending}, we introduce the matching procedure we developed to identify and characterize blended systems, before applying it to cluster lensing cosmology in \cref{sec:Blending_cluster_lensing}. A summary of this work is provided in \cref{sec:Conclusions}.


\section{Simulated datasets}\label{sec:Simulated_data}

In this section, we detail the different simulated catalogs we used in this work. We first present the cosmoDC2 extragalactic catalog \citep{Korytov_cosmoDC2} and its extended version, SkySim5000 \citep{Abolfathi_DC2} as part of the DESC Data Challenge 2 \citep{Abolfathi_DC2}. We then describe the DC2object catalog, derived from LSST-like images of the cosmoDC2 sky.

\subsection{cosmoDC2 extra-galactic catalog}\label{subsec:cosmoDC2}

To prepare for the upcoming Rubin Observatory Legacy Survey of Space and Time (LSST) data, the LSST Dark Energy Science Collaboration (DESC) has developed a large synthetic galaxy catalog, cosmoDC2~\citep{Korytov_cosmoDC2}, as part of the DESC Data~Challenge~2~\citep[DC2,][]{Abolfathi_DC2}. The cosmoDC2 catalog is derived from the \textit{Outer Rim} dark matter $N$-body simulation~\citep{Heitmann_OuterRim}, from which dark matter halos have been identified using a Friends-of-Friends (FoF) halo finder with a linking length of $b=0.168$ and a minimum of 20 particles per halo~\citep{Korytov_cosmoDC2}. Then, cosmoDC2 halos have been populated by galaxies, using GalSampler~\citep[][]{Hearin_GalSampler} and Halo Occupation Distribution (HOD) models. cosmoDC2 covers \SI{440}{\square\deg}, extending to redshifts $z=3$, with completeness to a \textit{r}-band magnitude depth of 28. It includes detailed representations of galaxy properties, such as \textit{true} positions, redshifts, magnitudes in the 6 \textit{ugrizY} LSST-like filters, intrinsic ellipticities, convergence, and shear components at their locations, obtained via ray-tracing. In addition to galaxy data, cosmoDC2 incorporates information on the host dark matter halos and their underlying properties, such as their redshifts, positions, and central galaxies. Their FoF masses $M_{\rm{FoF}}$ are also detailed, with corresponding $M_{200c}$ masses available in SkySim5000~\citep{Abolfathi_DC2}, an extended version of cosmoDC2 that expands the covered area to \SI{5000}{\square\deg}. 

For this analysis, cosmoDC2 (version 1.1.4) is used as the \textit{truth} or reference catalog. A conservative magnitude cut of 29 is adopted in the \textit{i}-band to retain all relevant galaxies, given that the DC2object catalog is complete up to magnitudes of about 26.5 (see next section). 

\subsection{DC2 object catalog} \label{subsec:DC2object}

A \SI{330}{\square\deg} subset of the cosmoDC2 catalog has been used to generate realistic images of the sky, subdivided in $154$ tracts~\citep{Abolfathi_DC2}, covering six optical bands and simulating 5 years depth out of the 10 years of LSST. DC2 images were processed using the version 23.0 of the Rubin Science Pipeline\footnote{\url{https://pipelines.lsst.io/}} and have been passed through a deblender to detect objects, resulting in the DC2 object catalog, which we will refer to as DC2object hereafter. This catalog is built upon a series of observational effects that affect the reconstruction of galaxy properties, such as the variability of the Point Spread Function (PSF) due to the Earth atmosphere, cosmic rays, dead pixels and other CCD artifacts, asteroids, etc. It is used by the DESC collaboration as a testbed to study their impact on shear calibration, photometric redshift calibration and the effect of blending on weak lensing observables. It contains \textit{measured} data, including positions, redshifts, magnitudes, shapes, and other relevant parameters. The photometric redshift estimates for the DC2 objects were derived from their observed DC2 magnitudes with the BPZ algorithm~\citep{Benitez_BPZ}.~\footnote{Bayesian Photometric Redshift code (BPZ) is a template-fitting method that combines observed photometric data with prior information about galaxy types and redshift distributions. It uses Bayesian framework to compute the posterior probability distribution of a galaxy’s redshift by expressing the likelihood from a set of redshifted Spectral Energy Distribution (SED) models.} Ellipticities were measured from the images using the HSM algorithm~\citep{Hirata_HSM,Mandelbaum_HSM}.~\footnote{based on the second moments of objects' surface brightness.}.

In the weak lensing regime, the ellipticity estimate for a single object is typically decomposed as~\citep{Mandelbaum_HSM_2}:
\begin{equation}
\widehat{\epsilon} \approx \epsilon^{\rm int} + (1+m)\gamma +c,
\end{equation}
where $\epsilon^{\mathrm{int}}$ is the object intrinsic ellipticity and $\gamma$ the weak lensing shear. The additive bias $c$ accounts for PSF uncorrected anisotropies. Other effects such as the uncorrected smoothing of the PSF and/or the impact of (unobserved) galaxy properties are included in the multiplicative bias $m$. HSM shapes need external calibration---usually from simulations \citep{Mandelbaum_HSM_2}---to evaluate these coefficients. 

In this work, DC2object is used as the \textit{observed} catalog, providing observation-like simulated data from the first five years of LSST. For our analysis, we applied a magnitude cut of 26.5 in the \textit{i}-band. The magnitude cut for the cosmoDC2 catalog, fixed at 29 as discussed in \cref{subsec:cosmoDC2}, has been purposely chosen larger for our blending study to remain complete at limited object magnitudes. Additional applied cuts on signal-to-noise ratio (SNR) and HSM shapes, will be further discussed in \cref{subsec:recover_profiles}.


\section{Identification and characterization of blending}\label{sec:Identification_blending}

In order to mitigate the impact of blending, it is essential to properly identify blended objects among the data. This can be achieved by matching reference data that are not impacted by the effect of blending with observations. In this section, we first describe the matching procedure used to identify blended groups by comparing the cosmoDC2 catalog with the DC2object catalog. After having summarized available blending metrics, we define the blending entropy, developed in this work to characterize recovered blends. This section ends with comprehensive studies of this new metric properties.

\subsection{Matching procedure}\label{subsec:matching}

We begin the matching procedure by applying a Friends-of-Friends (FoF) matching algorithm based on position information to associate true galaxies with detected objects. Although this method captures initial groups, it only relies on center coordinates which may not be sufficient for describing complex systems, particularly given the extended nature of galaxies. Therefore, after the use of the FoF algorithm, we refine the initial groups using an ellipse-based formalism that incorporates shape information for more precision.

\subsubsection{Step 1: Friends-of-Friends matching}

\begin{figure*}
\centerline{\includegraphics[scale=0.5]{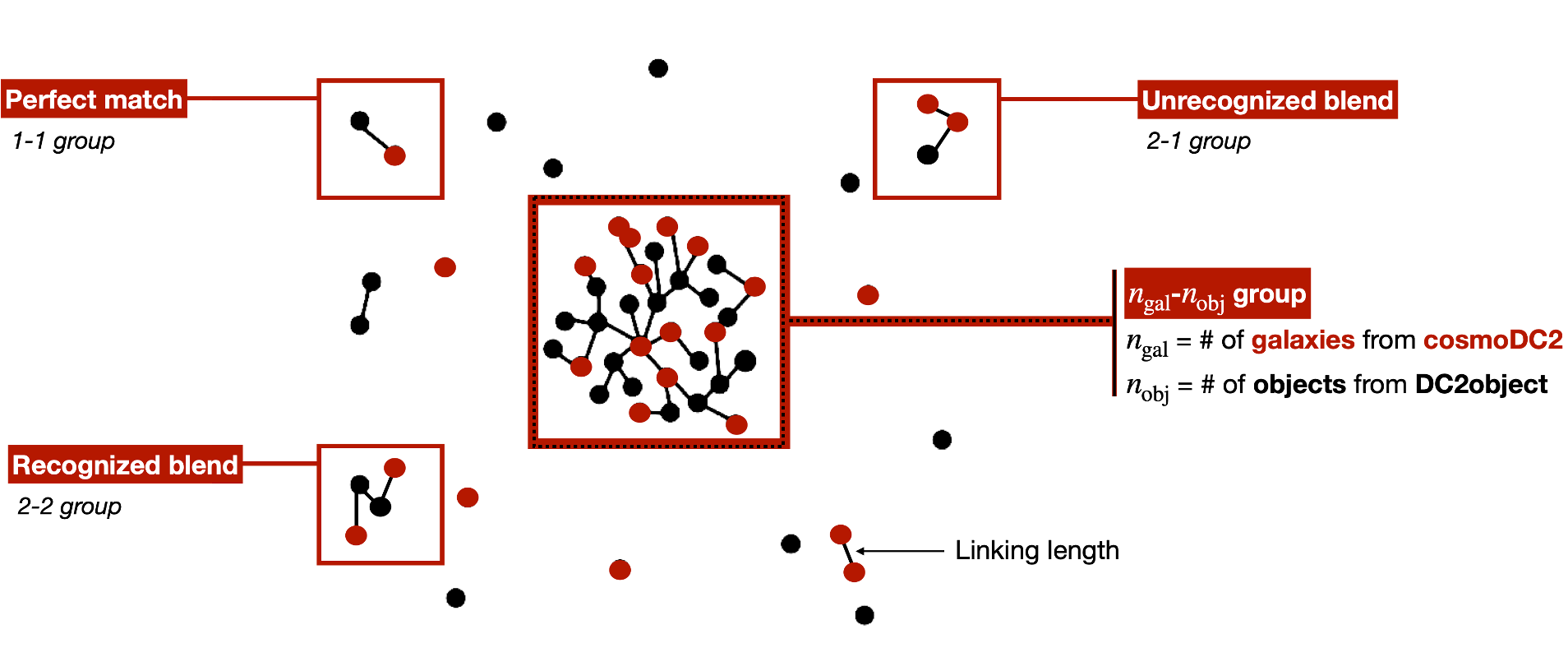}}
\caption{Schematic representation of the Friends-of-Friends matching algorithm between true galaxies from cosmoDC2 (in red) and detected objects from DC2object (in black). Galaxies and objects are represented by dots, linked through edges if they are located within a \SI{2}{\arcsecond} radius. Examples of recognized and unrecognized blends,  and perfect matches are illustrated.}
\label{fig:FoF}
\end{figure*}

\begin{table*}
    \centering
    \subfloat[Percentage of FoF groups]{%
    \hspace{.2cm}
    \begin{tabular}{cc|cccccc}
        & \multicolumn{1}{c}{} & \multicolumn{5}{c}{Number of objects $\m$} \\
        & & 1 & 2 & 3 & 4 & 5 \\
        \cline{2-7}
        \multirow{5}{*}{\rotatebox{90}{Number of galaxies $\n$}}
        &1 & \textbf{23} & 0.026 & 0 & 0 & 0  \\
        &2 & \textbf{23} & \textbf{0.56} & 0.0020 & 0 & 0  \\
        &3 & 19 & \textbf{1.4} & 0.022 & 0 & 0  \\
        &4 & 15 & 2.0 & 0.078 & 0.0010 & 0  \\
        &5 & 12 & 2.5 & 0.17 & 0.0050 & 0  \\
        \cline{2-7}
        & \multicolumn{1}{c|}{} & \multicolumn{5}{c}{Number of groups 
        $N_{\mathrm{groups}}=\num{30522835}$}\\
    \label{tab:FoF_sys_prop}
    \end{tabular}
    \hspace{.2cm}
     }\hspace{.5cm}
    \subfloat[Percentage of \texttt{friendly} groups]{%
    \hspace{.2cm}
    \begin{tabular}{cc|cccccc}
        & \multicolumn{1}{c}{} & \multicolumn{5}{c}{Number of objects $\m$} \\
        & & 1 & 2 & 3 & 4 & 5 \\
        \cline{2-7}
        \multirow{5}{*}{\rotatebox{90}{Number of galaxies $\n$}}
        & 1 & \textbf{44} & 0.020 & 0 & 0 & 0\\
        & 2 & \textbf{29} & \textbf{0.20} & 0.0010 & 0 & 0\\
        & 3 & 15 & \textbf{0.46}& 0.0080 & 0 & 0\\
        & 4 & 6.7 & 0.58 & 0.020 & 0 & 0\\
        & 5 & 3.0 & 0.53 & 0.035 & 0.0010 & 0\\
        \cline{2-7}
        & \multicolumn{1}{c|}{} & \multicolumn{5}{c}{Number of groups 
        $N_{\mathrm{groups}}=\num{71704988}$}\\
        \label{tab:friendly_sys_prop}
    \end{tabular}
    \hspace{.2cm}
    }

\caption{Proportion of $\n - \m$ groups (with $1\leq\n\leq5, \:1\leq\m\leq5$) for a galaxy cut of $i<29$. \textit{Left:} with the Friends-of-Friends matching procedure, using a linking length of $2''$. \textit{Right:} with the \texttt{friendly} matching procedure (i.e. combining the FoF matching with the ellipse overlap test). The percentage of $1-1$ (perfect matches), $2-1$, $3-2$ (unrecognized blends) and $2-2$ (recognized blends) groups are in bold.}
\label{tab:prop_syst}
\end{table*}

We employ a Friends-of-Friends\footnote{\url{https://github.com/yymao/FoFCatalogMatching}} (FoF) matching algorithm, developed in the context of the Satellites Around Galactic Analogs (SAGA) Survey \citep{Mao_fof}. This algorithm identifies groups of nearby truth galaxies and detected objects, based on their celestial coordinates, within a specified linking length, set to \SI{2}{arcseconds} in this work. This relatively large linking length (compared to the LSSTCam pixel size of \SI{0.2}{arcseconds}) was deliberately chosen to capture large groups of nearby galaxies and objects, which can later be refined into smaller and more accurate groups.

Then FoF-recovered blended groups can be identified by comparing the number of detected objects from DC2object with the number of nearby cosmoDC2 true galaxies in each group. In this paper, we categorize these groups using the notation $\n-\m$, where $\n$ represents the number of cosmoDC2 galaxies and $\m$ the number of DC2object detections in the group. Recognized blends are defined as groups where $\n = \m$ and $\n > 1$. Unrecognized blends are characterized by $\n > \m$, where fewer objects compared to galaxies are detected in the group, indicating that some galaxies are not individually resolved. This matching procedure also identifies perfectly matched groups, referred to as $1 - 1$ groups.
The FoF procedure, as well as examples of recovered FoF groups are illustrated schematically in \cref{fig:FoF}. In \cref{tab:FoF_sys_prop}, we indicate the proportion of specific FoF groups for $1 \leq \n \leq 5$ and $1 \leq \m \leq 5$, with highlights on $1-1$, $2-1$, $2-2$ and $3-2$ groups. We do not show the proportion of $0-\m$ groups that correspond to false detections, and of $\n-0$ groups that are numerous due to the 2.5 higher magnitude cut on cosmoDC2. With the FoF matching algorithm, we notice that only $23\%$ of groups are perfect matches ($1-1$), and the same proportion applies to $2-1$ groups (unrecognized blends). The relatively low proportion of perfect matches indicates that the FoF algorithm with a linking length of $2$ arcseconds may not be sufficient in characterizing the different groups and identifying clean detections, motivating the next step.
 
\subsubsection{Step 2: Ellipse overlap test}

\begin{figure}
\centerline{\includegraphics[scale=0.4]{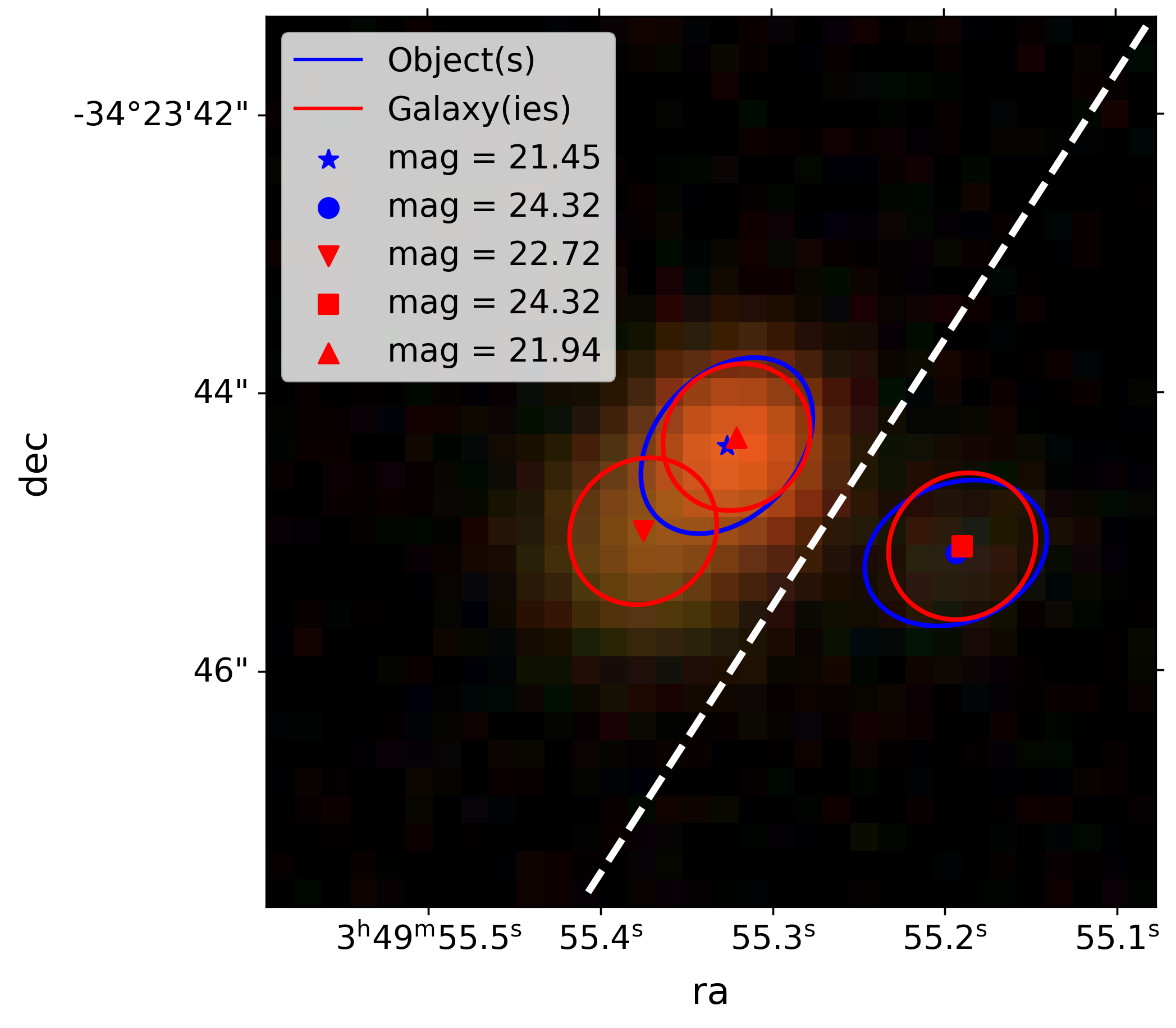}}
\caption{RGB simulated image example of an initial FoF $3-2$ group refined into $2-1$ and $1-1$ groups after the ellipse overlap test, highlighted with the white dashed line. Galaxies are represented with red ellipses. Objects are the blue ellipses.}
\label{fig:EOT}
\end{figure}

The next step of the matching procedure involves refining the initial recovered FoF groups using flux and shape information, in addition to position information used by the FoF algorithm. Since galaxies are extended objects, algorithms only based on the center coordinates are not necessarily sufficient to properly define blended groups of galaxies. To account for it, we model the contours of galaxies and detected objects with an ellipse formalism (using code developed by S. Liang\footnote{\url{https://github.com/LSSTDESC/Cluster\_Blending/}\label{foot:shuang_code}} and equations detailed in \cref{app:ellipse}).

The code works as follows: for galaxies, ellipses are determined using the true semi-major and semi-minor axes, and the orientation angle directly from the cosmoDC2 catalog and converted from degrees to pixels (to match the pixel-level measured quantities from DC2object), based on the \SI{0.2}{\arcsecond} size of the LSSTCam pixels. The center coordinates are converted from (RA, Dec) in degrees to $(x_0, y_0)$ in pixels, relative to the position of one of the detected objects in the group. Finally, the recovered ellipses are convolved with the measured PSF at the location of their FoF groups from the DC2object catalog for comparison with simulated detected objects.
For detected objects, ellipses are determined from second order moments of the flux in the \textit{i}-band using a Gaussian profile, as described in \cref{app:ellipse}.

An example of a refined FoF group using the ellipse overlap test is shown in \cref{fig:EOT}. The initial FoF group is a $3-2$ blended group, composed of three galaxies (respectively two objects) represented by the red (respectively blue) ellipses. After the ellipse overlap test, this group is refined onto two subgroups of overlapping ellipses: a $1-1$ perfect match and a $2-1$ blend. We label the groups---resulting from this refined matching procedure that uses ellipse formalism--- as the \texttt{friendly} groups. In \cref{tab:friendly_sys_prop}, we indicate the proportion of specific \texttt{friendly} groups for $1 \leq \n \leq 5$ and $1 \leq \m \leq 5$, with highlights on $1-1$, $2-1$, $2-2$ and $3-2$ groups. We observe a larger proportion of $1-1$ groups ($44\%$) compared to \cref{tab:FoF_sys_prop}, indicating that some $\n-1$ (with $\n>1$) FoF groups may have been refined after the ellipse overlap test. We notice a larger proportion of $2-1$ groups ($29\%$ against $23\%$) that can be explained by the refinement of some $2-2$ and $3-2$ groups.

\subsection{Available metrics} \label{subsec:metrics}

\subsubsection{Blendedness}

\begin{figure}
    \centering
    \includegraphics[scale=0.35]{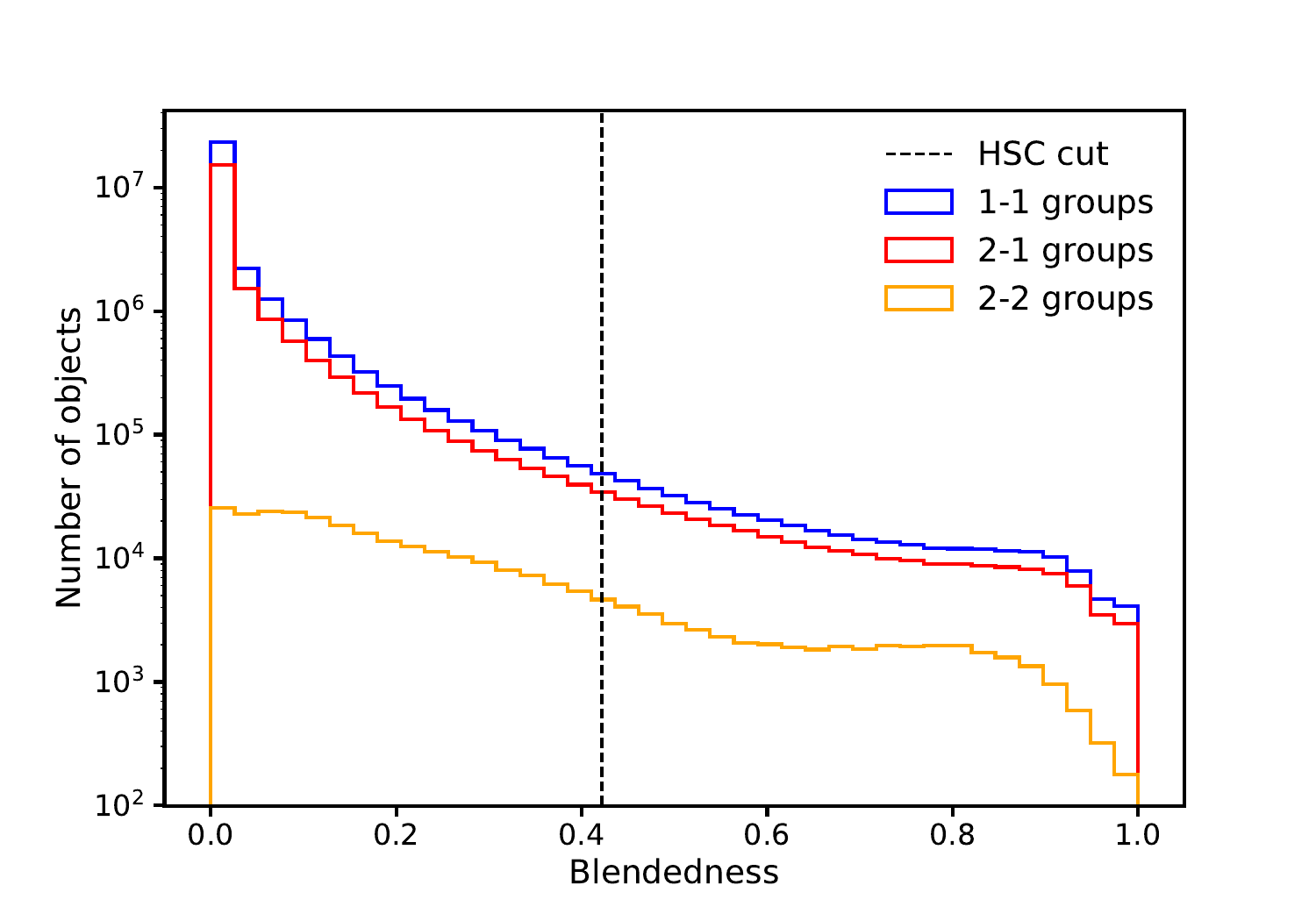}
    \caption{Blendedness distributions for $1-1$ (blue), $2-1$ (red) and $2-2$ (orange) groups. The black dashed line represents the blendedness $b < 10^{-0.375}$ cut defined by HSC \citep[see][]{Mandelbaum_HSC}.}
    \label{fig:blendedness}
\end{figure}

Because deblenders are inherently imperfect, it is essential to quantify the degree to which each detected object is affected by blending. To achieve this, blending can be characterized from data through a metric called the \textit{blendedness}, developed in the context of the HSC pipeline \citep{Bosch_blendedness}. The blendedness is defined as the fraction of the total flux in the neighborhood of a detected object that belongs to its neighbors. Instead of using a fixed boundary, the neighborhood is defined by a Gaussian weight function, whose contour is determined by the source’s own shape. Defining $G$ as the Gaussian weighted total flux, $\Sigma$ the object's moment matrix and $F(\textbf{r})$ the measured flux in pixel \textbf{r}, we can write:
\begin{equation}
    G(F, \Sigma) = \sum_{\textbf{r}}\frac{F(\textbf{r})}{2\pi\sqrt{|\Sigma|}} 
    \exp\left( -\frac{1}{2} \mathbf{r}^\top \Sigma^{-1} \mathbf{r} \right).
\end{equation}
With the \textit{parent} defined as the original blended object and the \textit{child} as the result of the deblending process, the blendedness $b$ is defined using the corresponding flux distributions $F_p$ and $F_c$ as:{\footnote{The same moment matrix $\Sigma$ is used in both measurements, derived from the child image}.}
\begin{equation}
    b = 1- \frac{G(F_c, \Sigma)}{G(F_p, \Sigma)}.
\end{equation}
A well-isolated object has therefore a blendedness of 0 while the maximum value is 1 for highly blended detected objects. With this definition, we notice that the blendedness depends on the actual performance of the deblender, responsible for determining $F_c$ from $F_p$ for each object.

To characterize blended groups, we first use the blendedness, available for all detected objects in DC2object. \Cref{fig:blendedness} presents the distributions of blendedness for $1-1$, $2-1$, and $2-2$ groups. For $1-1$ groups (in blue), where each detected object is matched with one true galaxy, the blendedness distribution is concentrated near zero. This confirms that these objects are well-isolated and experience minimal contamination from neighboring sources. Similarly, the blendedness distribution for $2-1$ groups (in red), where a single detected object is matched to multiple true galaxies, follows a similar trend. This is expected, as the deblender interprets these cases as single and isolated detections. In contrast, objects present in $2-2$ groups (in orange), have a more extended blendedness distribution with no preferred values between 0 and 1. This reflects the fact that these detections are recognized as part of blended groups, capturing a broader range of blending levels. \cite{Mandelbaum_HSC} only includes objects with $b < 10^{-0.375}$ in the shear HSC-SSP weak lensing shear catalog, to avoid spurious detections and highly blended objects with completely unreliable photometry and shape measurements due to contamination by light from nearby bright
objects. This cut is represented with the dashed vertical line in \cref{fig:blendedness} and leads to approximately 3\% of suppressed objects in our framework (see \cref{tab:percentage_suppressed_objects}). While blendedness is a useful metric for quantifying flux contamination from nearby sources, it does not fully differentiate between unrecognized blends and truly isolated objects. Additional metrics are therefore required to characterize completely blended groups and assess the limitations of deblending algorithms. 

\subsubsection{\textit{Top-hat} and \textit{Gaussian} overlap fractions} \label{subsubsec:top_hat_gaussian_overlap}

To go further in the description of blended groups, we first compute the relative overlap surface between two ellipses. This is done by performing a Monte Carlo (MC) sampling (see \cref{app:ellipse}). Knowing the overlap and the surfaces of two ellipses, we can compute the relative overlap fraction. This quantity allows more precise descriptions of the different blended group composition. We label the finite overlap between two ellipses as the \textit{top-hat} overlap.

However, the modeling of galaxies as finite ellipses is a first approximation since galaxies are extended objects with continuous fluxes. In addition to the computation of the \textit{top-hat} overlap, we also implement the overlap surface between two 2D Gaussians, the \textit{gaussian} overlap. Since Gaussians are continuous, we apply a cut of $10^{-3}$ on the overlap in order to suppress links between well isolated objects and galaxies and perform the refinement of the initial FoF groups (see \cref{subsec:matching}). From the Gaussian distributions, we can compute the resulted overlap fraction, given by:
\begin{equation}
    \langle o,g \rangle_{\mathrm{gaussian}} = \frac{\int G_o(x,y)G_g(x,y) \dd{x}\dd{y}}{\sqrt{\qty(\int G_o^2(x,y)\:\dd{x}\dd{y})\times\qty(\int G_g^2(x,y)\:\dd{x}\dd{y})}},
\end{equation}
where $G_o$ and $G_g$ are the 2D Gaussian distributions associated with the object $o$ and the galaxy $g$, based on their shape information.
In the following section, we compare the influence of the overlap type (\textit{top-hat} or \textit{gaussian}) on the identification and characterization of blended groups.

\subsection{Matching probabilities and blending entropy} \label{subsec:blending_entropy}

\begin{figure*}
\centerline{\includegraphics[scale=0.35]{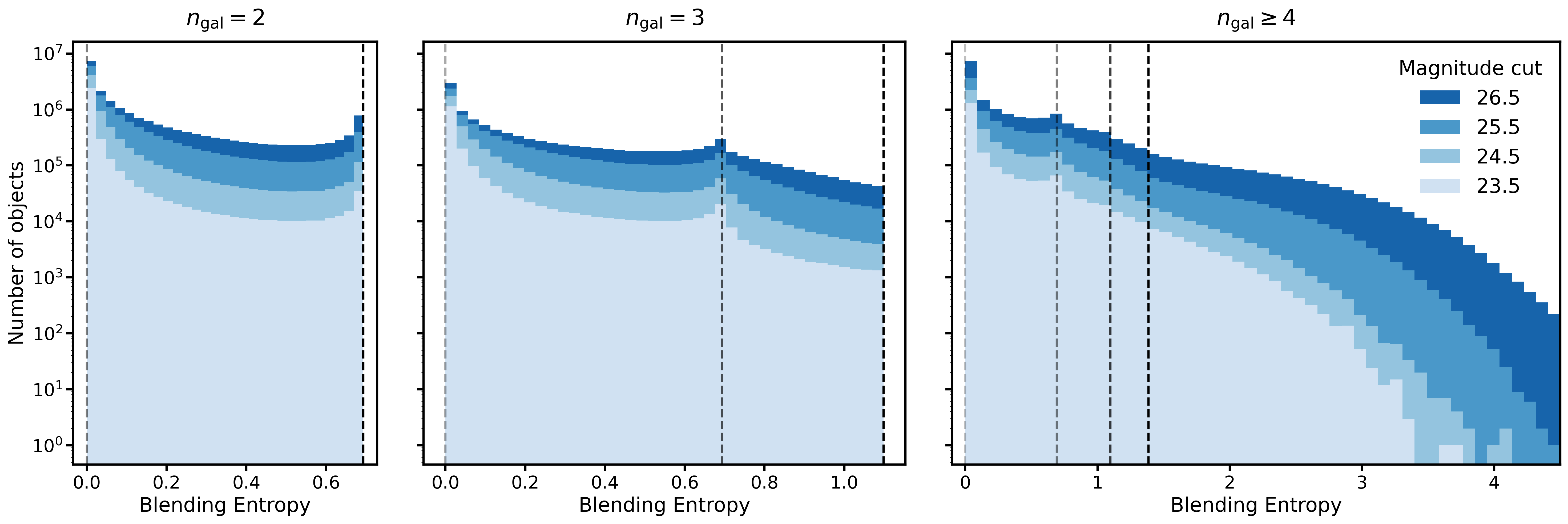}}
\caption{Distributions of the blending entropy for groups composed of $\n$ galaxies where $\n=2$ (first plot), $\n=3$ (second plot) and $\n \geq 4$ (third plot). The different colors represent the cuts on magnitude (in $i$-band) applied to the DC2object catalog. The dashed lines indicate the maximum values of the entropy $S_{b,\rm{max}}=\ln{n}$ for groups composed of $n=1,\dots,\n$ galaxies.} 
\label{fig:Sb_mag_cut}
\end{figure*}

Blending can lead to the misidentification of detected objects with respect to their true underlying galaxies, which in turn biases the measured shapes and redshifts. To address this, we propose quantifying blending as a matching \textit{ambiguity}.


For each detected object $o$, we first assign a relative probability $p_{og}$ of it being matched to a true galaxy $g$ belonging to the same group. In principle, this could be attained through a full statistical model for observed DC2object images. However, we employ a simplified and more tractable approach by designing a catalog-based matching probability that integrates positions and shapes through overlap fractions, and fluxes through magnitude comparison. Specifically, this probability is proportional to the overlap\footnote{\textit{top-hat} or \textit{gaussian}, see \cref{subsec:metrics}} fraction $\langle o, g \rangle$ and downweighted by the (absolute) magnitude difference (in $i$-band) to ensure high matching score between galaxies and objects with similar magnitudes, such that,\footnote{Note that we experimented with a down-weighting coefficient in the form $e^{-\alpha|m_o-m_g|}$ for different values of $\alpha$ but found no improvement over the case with $\alpha=1$.}
\begin{equation}
    p_{og} \propto \langle o, g \rangle e^{-|m_o-m_g|}.
\label{eq:proba_of_matching}
\end{equation}
It is then normalized so ${\sum_g{p_{og}} =1}$. The probability of matching ranges from 0 to 1, with a maximum value for perfectly matched $(1-1)$ groups. For groups composed of detected objects with no corresponding nearby galaxies, we set $p_{og}$ to 0, as these groups are likely artifacts of pixel noise. 

To quantify the level of ambiguity, we then use the (normalized) probabilities of matching to compute a new quantity named the blending entropy{\footnote{We name this quantity entropy because it matches the entropy of a categorical distribution with probabilities $p_{og}$ with ${g=1,\dots,n_{\rm gal}}$. It lives on the $n_{\rm gal}-1$-simplex and measures how concentrated the distribution is towards any or some of the categories (here, galaxies).}} for each detected object $o$, defined as:
\begin{equation}
    S_{b,o} = -\sum_{g \in \rm{gal}}{p_{og} \ln{p_{og}}} \geq 0,
\end{equation}
where $p_{og}$ is the probability for an object $o$ to be matched to the $g$-th galaxy of its group. High blending entropy values characterize objects associated with highly blended groups, while well-matched objects---{where one of the matching probabilities $p_{og}$ is close to 1 with the others close to 0}---have a blending entropy near 0, by definition. Objects with no nearby galaxies in a radius of 2'' are flagged with a blending entropy value of $-1$ and are not considered for this study, as they correspond to detections of noise.

The matching procedure between the two catalogs, the recovery and refinement of the FoF groups, and the computation of the blending entropy are publicly available in the DESC \texttt{friendly} code.\footnote{\url{https://github.com/LSSTDESC/friendly}}\textsuperscript{,}\footnote{To facilitate access to information about blended groups, we use a \texttt{NetworkX}~\citep{Hagberg_networkx} bipartite graph structure (code available at \url{https://networkx.org/documentation/stable/tutorial.html}). Each group is represented with a graph where the nodes correspond to galaxies or objects, which are linked through edges if their associated ellipses overlap.} \newline


{This blending entropy possesses a number of desired properties for a blending metric.
First, it allows us to consider objects and galaxies in groups, bypassing the limitation of one-to-one catalogs comparison. The level of ambiguity in matching objects and galaxies relates to the level of errors in measured properties, as shown in \cref{subsubsec:Sb_top_hat,subsubsec:Sb_gaussian}.
Second, provided that the galaxy magnitude cut is $0.5-1$ higher than that applied to objects---such that all likely matching galaxies are retained---$S_b$ is largely insensitive to the galaxy magnitude cut and to the choice of the FoF linking length.}
{The faintest and isolated galaxies would always have a low probability of matching the detected object of interest and would therefore not impact its blending entropy. This characteristic of the blending entropy is only possible thanks to the defined probabilities of matching that uses both shape and magnitude information (in addition to the commonly used position information).}
{By the same virtue, the blending entropy of objects in groups with a different number of galaxies may be compared on a single scale.}

Since the blending entropy is calculated using the magnitudes of both galaxies and detected objects, it may, however, be influenced by the magnitude cut applied to the DC2object catalog. In this work, we employ a magnitude cut of 26.5 in the \textit{i}-band, corresponding to five years of observations with Rubin-LSST, as described in \cref{subsec:DC2object}. In \cref{fig:Sb_mag_cut}, we present the distributions of the blending entropy across different DC2object magnitude cuts in \textit{i}-band from $23.5$ to $26.5$. We differentiate the distributions according to the number of galaxies $\n$ present in the object's group. 
We find that at low blending entropy values the distributions are sharply peaked, particularly for bright sources, indicating a population of well-detected and reliably matched objects. In contrast, at high blending entropy ($S_b > 3$), the distributions decrease rapidly for bright objects ($i < 24.5$), while the number of high-entropy objects increases significantly as the limiting magnitude is extended to fainter values.
This can be explained by the fact that fainter, thus deeper objects are considered. As they are more likely to be blended with foreground and brighter galaxies, their entropies increase. We indicate with dashed lines the maximum values of the entropy $S_{b,\mathrm{max}}=\ln{n}$ for groups composed of $n=1,\dots,\n$ galaxies. We observe that the blending entropy distribution does not exhibit substantial changes in its overall shape or peak locations when applying different magnitude cuts.

\subsection{Catalog observables vs. blending entropy} \label{subsec:blending_demography}

In this section, we investigate the behavior and properties of the blending entropy $S_b$ by exploring its dependence on both the definition of the overlap used in the matching probabilities (\textit{top-hat} or \textit{gaussian}) and on key observable quantities, available in the cosmoDC2 and DC2object catalogs. This allows us to assess whether $S_b$ behaves consistently with expectations and with known blending-related systematics.

\subsubsection{Blending entropy from \textit{top-hat} overlap}
\label{subsubsec:Sb_top_hat}

\begin{table}
    \centering
    \scalebox{1}{
    \begin{tabular}{c||c|c}
    Group $\n-\m$ & Overlap & $|\Delta \mathrm{mag}|$ \\
    \hline \hline
    \textit{Good} $2-1$  & $\langle g,g \rangle < 0.3$ & (gal,gal) $\geq 2$\\
    \textit{Bad} $2-1$   & $\langle g,g \rangle \geq 0.3$ & (gal,gal) $\leq 2$ \\
    \textit{Good} $2-2$  & $\langle o,g \rangle \geq 0.7$ & (gal,obj) $\leq 2$\\
    \textit{Good} $3-2$  & $\langle o,g \rangle \geq 0.7$ & (gal,obj) $\leq 2$\\
    \textit{Ugly} $>3\; ->3$ & - & -\\
    \end{tabular}
    }
    \caption{Summary table of the defined criteria to differentiate, \textit{good}, \textit{bad} and \textit{ugly} groups. The number of galaxies $\n$ and objects $\m$ within each defined group are indicated. Criteria have been essentially made on the overlap fraction and the absolute difference in \textit{i}-band magnitudes. In case of multiple object-galaxy pairs, the conditions are verified for all the pairs.}
    \label{tab:friendly_groups}
\end{table}

\begin{figure*}
    \centering
    \begin{adjustbox}{max width=\linewidth} 
        \begin{subfigure}{0.3\linewidth}
            \centering
            \includegraphics[width=\linewidth]{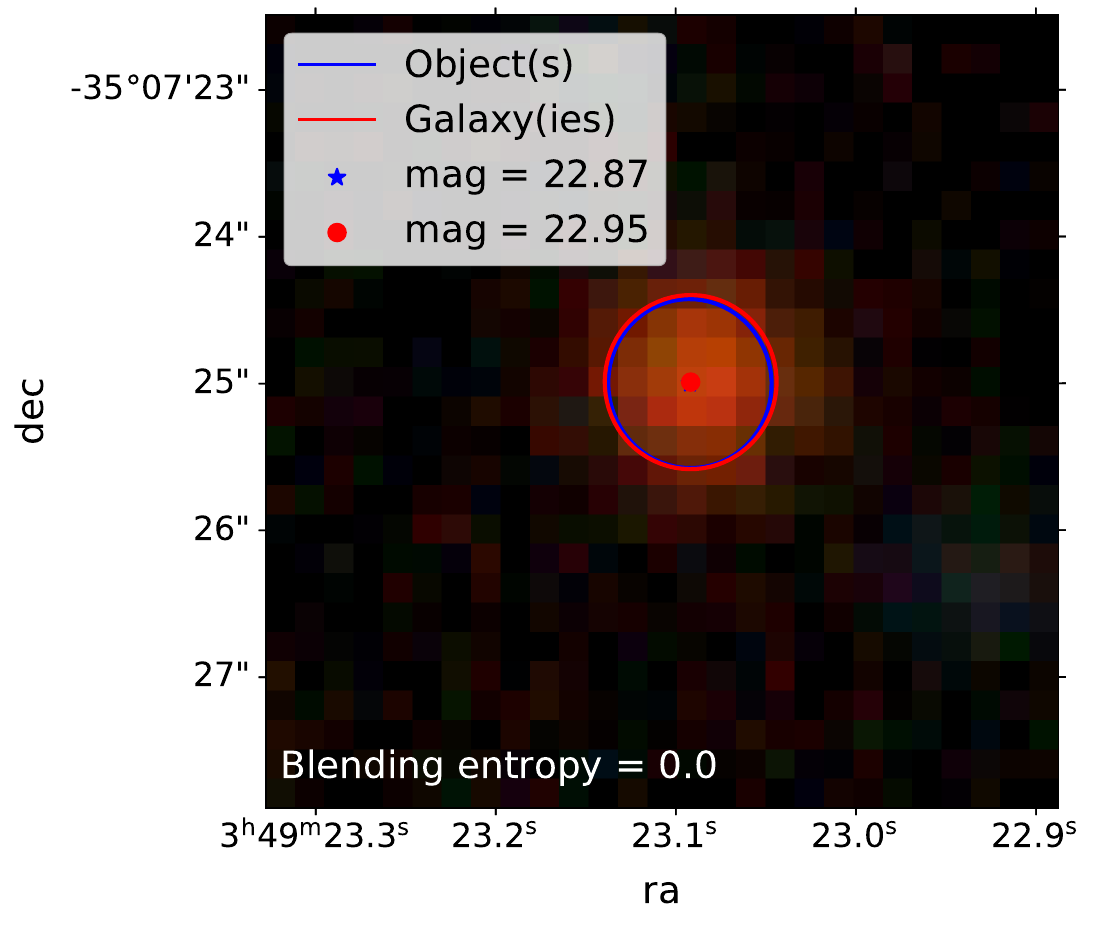}
            \caption{$1-1$ group.}
        \end{subfigure}
        \begin{subfigure}{0.3\linewidth}
            \centering
            \includegraphics[width=\linewidth]{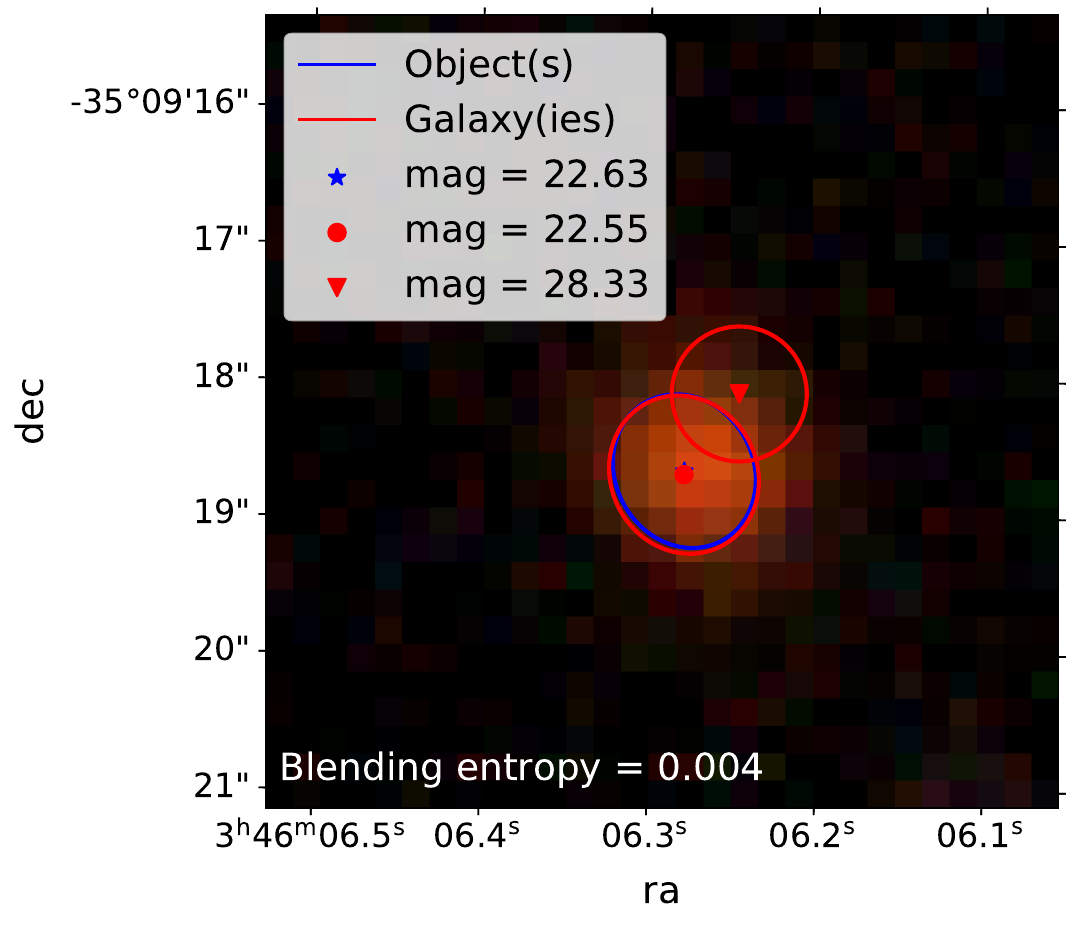}
            \caption{\textit{Good} $2-1$ group.}
        \end{subfigure}
        \begin{subfigure}{0.3\linewidth}
            \centering
            \includegraphics[width=\linewidth]{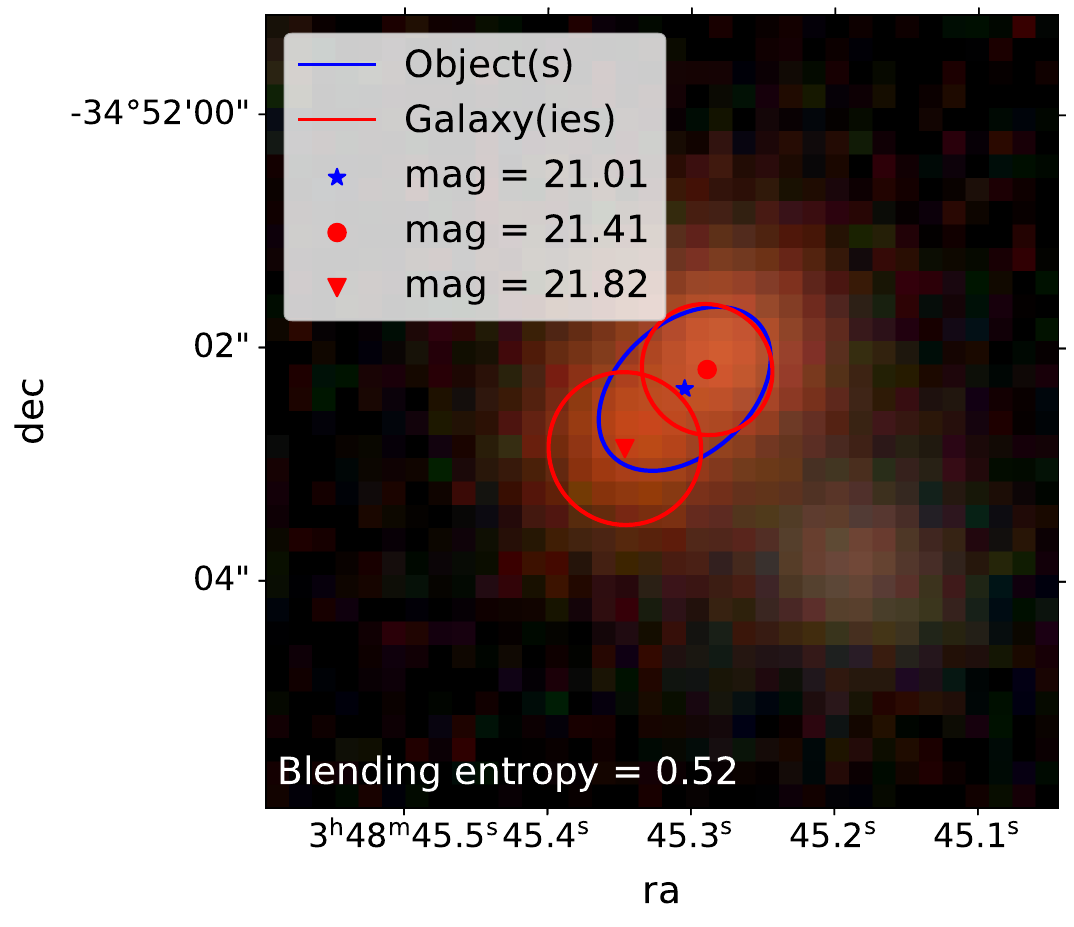}
            \caption{\textit{Bad} $2-1$ group.}
        \end{subfigure}
    \end{adjustbox}

    \vspace{0.4cm} 

    \begin{adjustbox}{max width=\linewidth}
        \begin{subfigure}{0.3\linewidth}
            \centering
            \includegraphics[width=\linewidth]{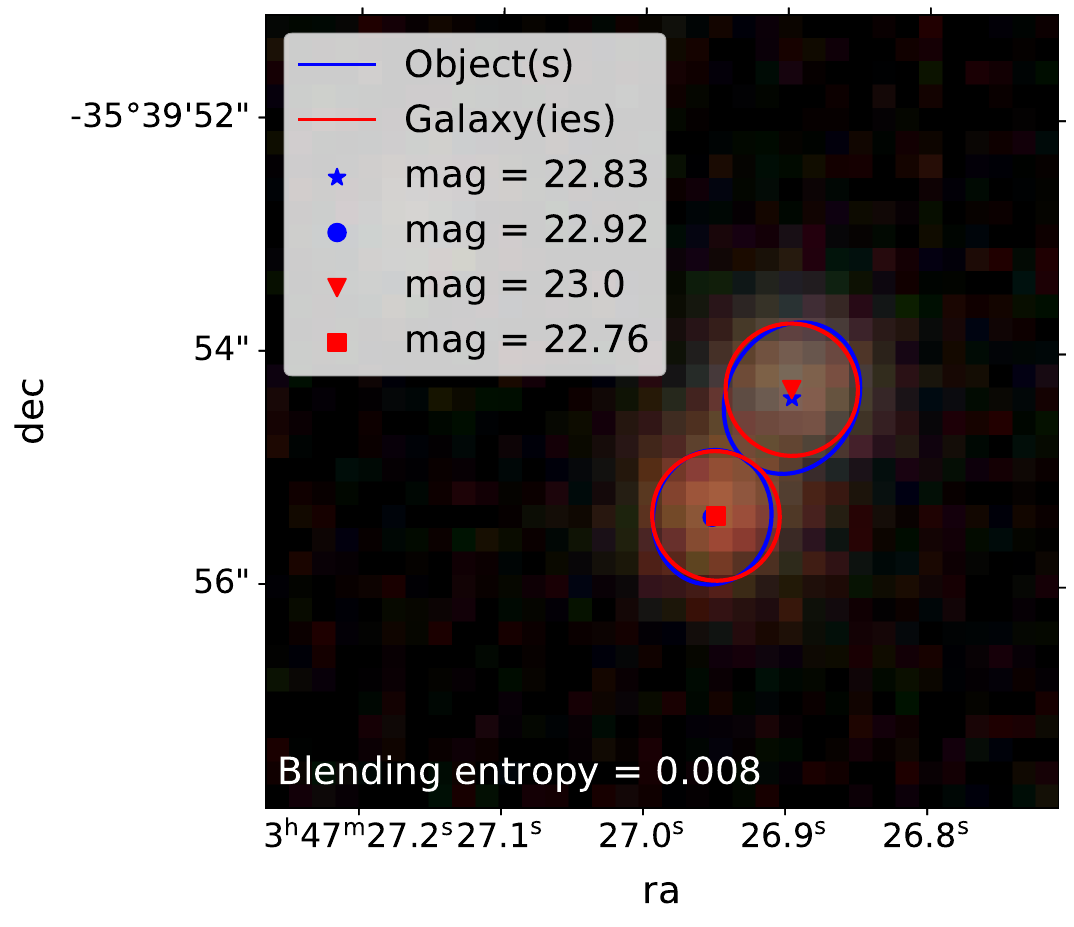}
            \caption{\textit{Good} $2-2$ group.}
        \end{subfigure}
        \begin{subfigure}{0.3\linewidth}
            \centering
            \includegraphics[width=\linewidth]{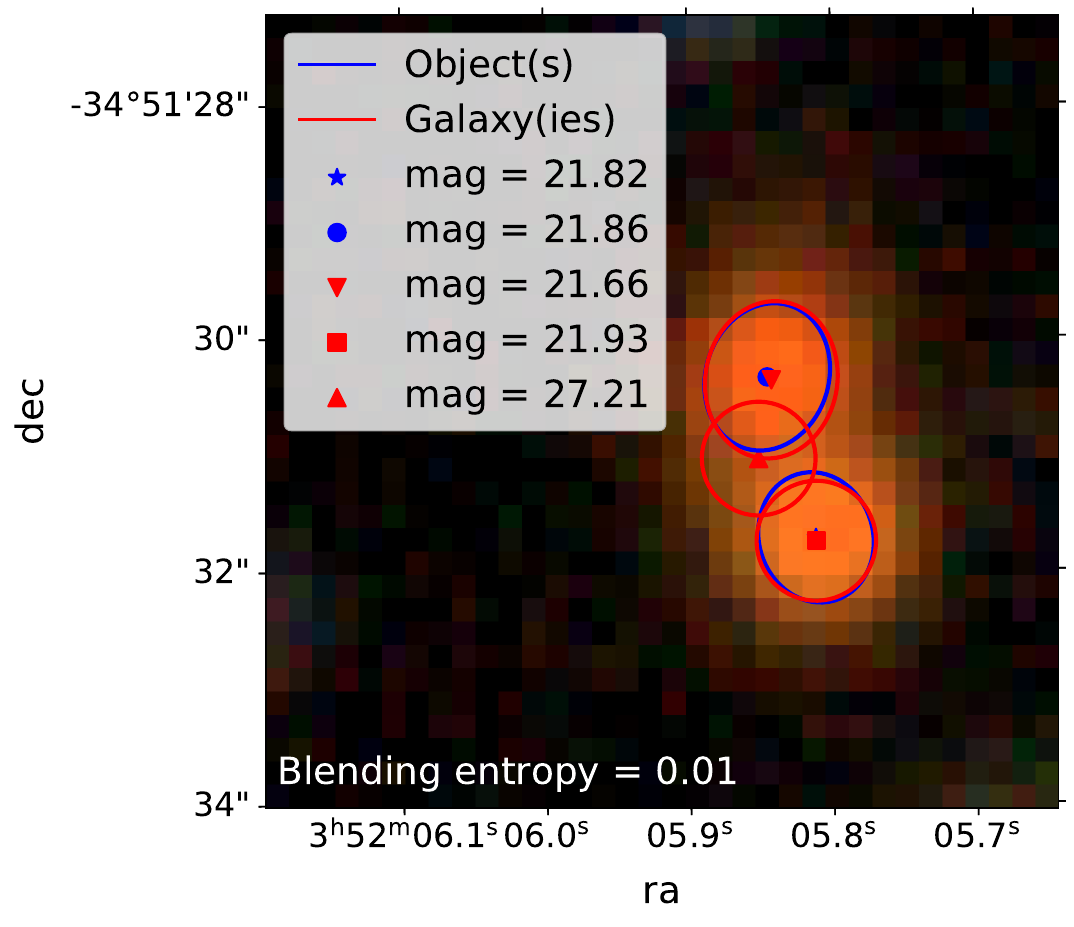}
            \caption{\textit{Good} $3-2$ group.}
        \end{subfigure}
        \begin{subfigure}{0.3\linewidth}
            \centering
            \includegraphics[width=\linewidth]{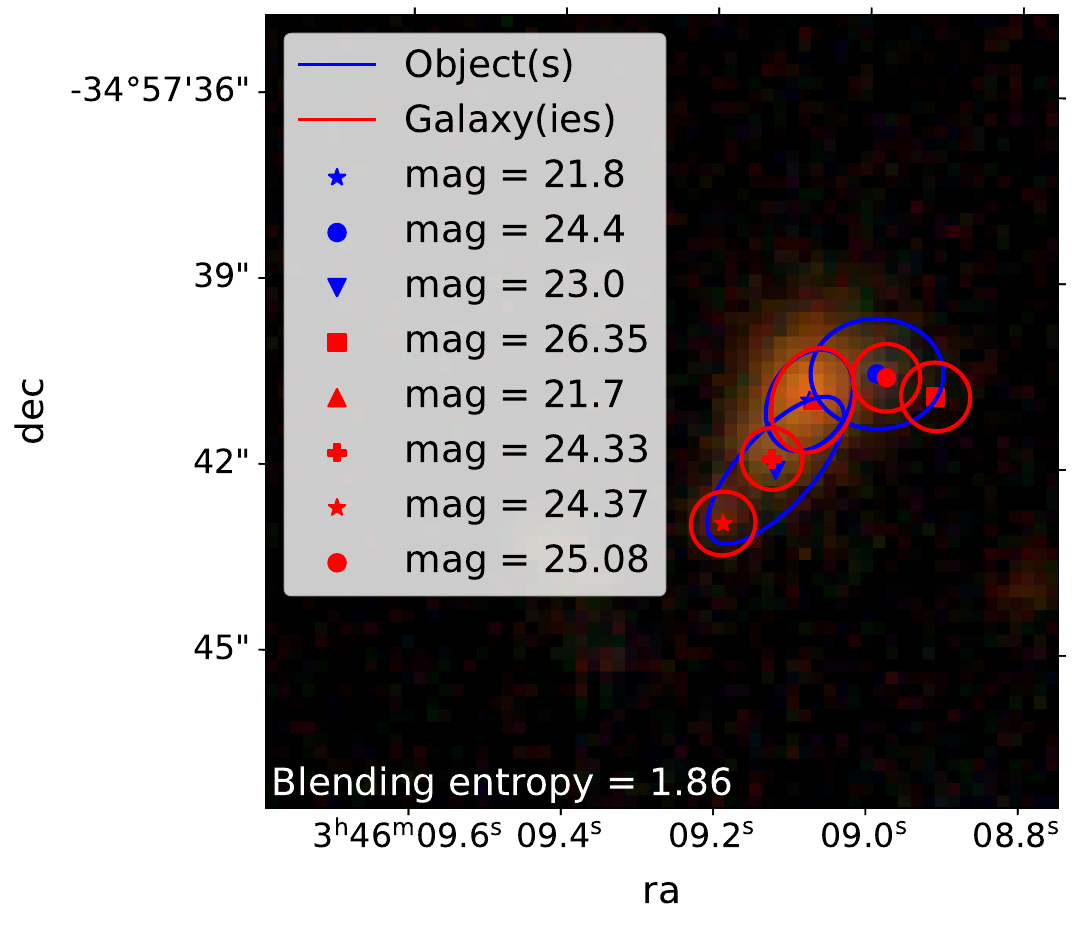}
            \caption{\textit{Ugly} group (illustrative).}
        \end{subfigure}
    \end{adjustbox}

    \caption{DC2 images of \texttt{friendly} groups and their corresponding blending entropies (as the sum of the individual object's entropies). The top images show examples of $1-1$, \textit{good} and \textit{bad} $2-1$ groups, which are the most observed ones among the data. The bottom images illustrate \textit{good} $2-2$, $3-2$ and \textit{ugly} groups. For each image, the red ellipses correspond to galaxies from cosmoDC2. The blue ellipses are the detected objects from DC2object. The position of their centers are also illustrated. The indicated magnitudes are in $i$-band.}
    \label{fig:friendly_groups}
\end{figure*}

\begin{figure}
\centerline{\includegraphics[scale=0.35]{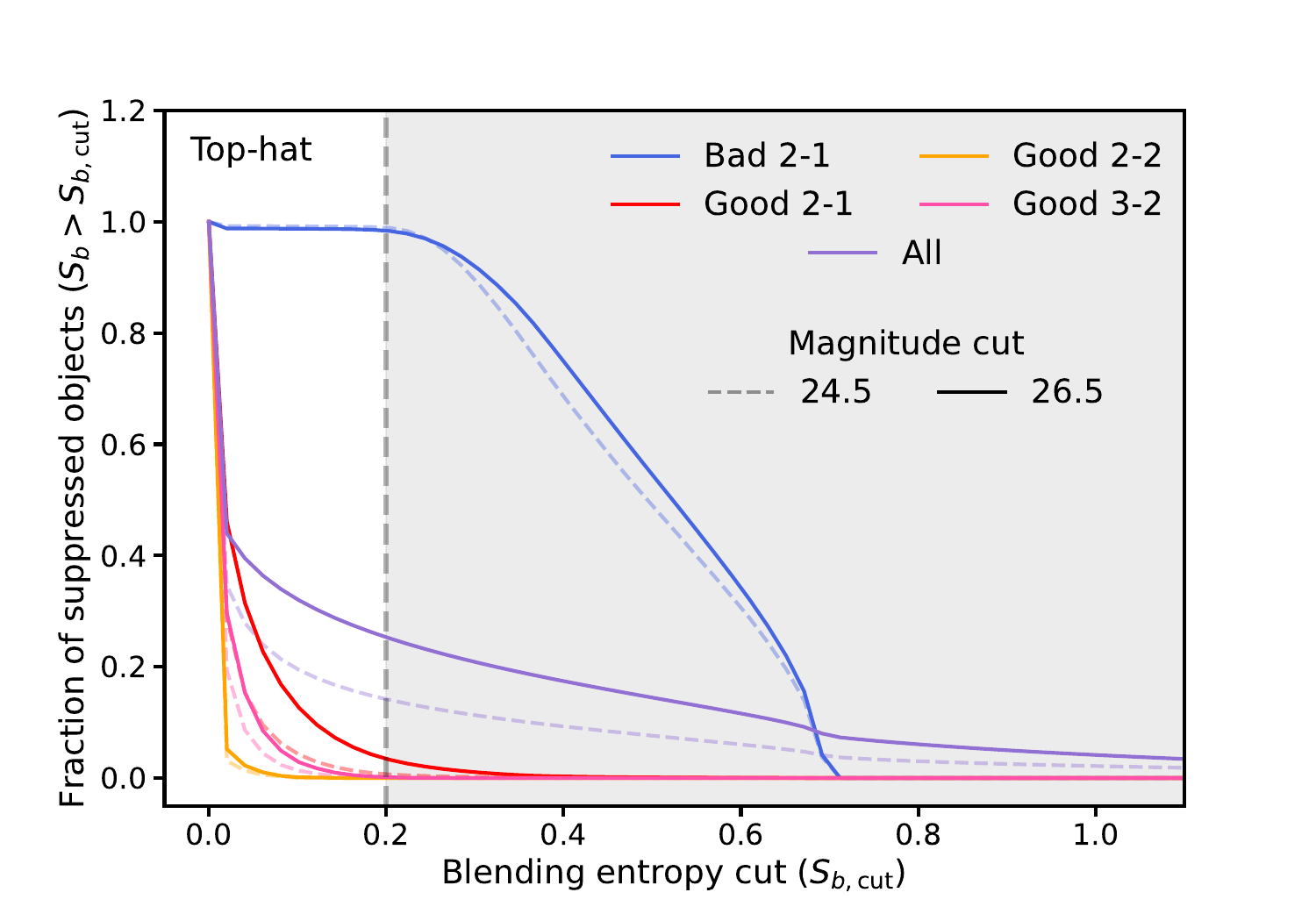}}
\caption{Fractions of suppressed objects belonging to \textit{bad} $2-1$ (blue) and \textit{good} $2-1$ (red), $2-2$ (orange) $3-2$ (pink), and all (purple) groups according to the $S_{b,\ \rm{cut}}$ cut on their blending entropies ($S_b$). The curves have been plotted for two DC2 limiting magnitudes in \textit{i}-band (solid and dashed curves). The dashed vertical line represents the adopted cut of $S_b = 0.2$ used in this work for the separation of \textit{bad} and \textit{good} blended objects. The blending entropy has been computed using the \textit{top-hat} overlap definition.}
\label{fig:Sb_good_bad}
\end{figure}

\begin{figure*}
    \centering
    \begin{adjustbox}{max width=\linewidth} 
        \begin{subfigure}{0.33\linewidth}
            \centering
            \includegraphics[width=1.05\linewidth]{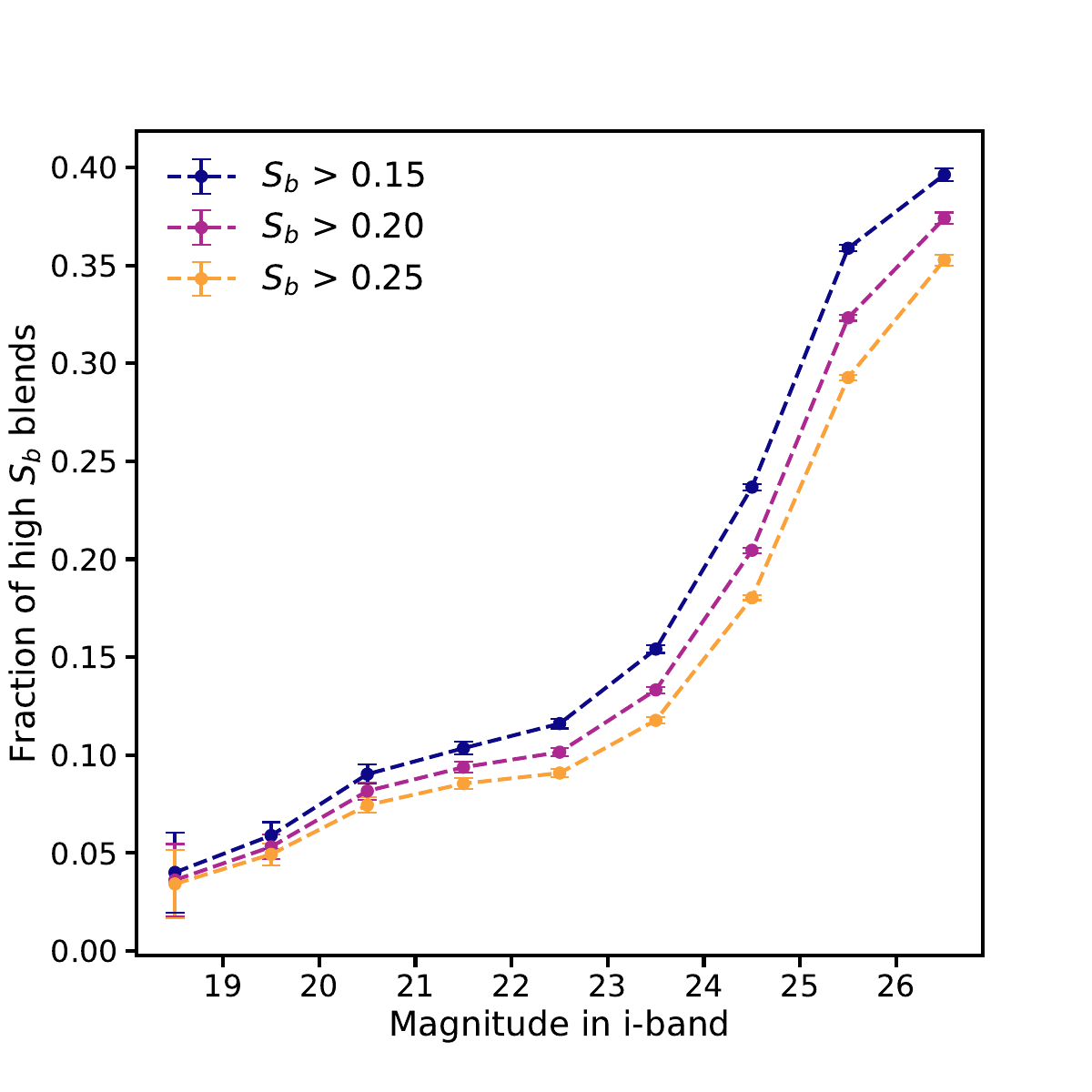}
        \end{subfigure}
        \begin{subfigure}{0.33\linewidth}
            \centering
            \includegraphics[width=1.05\linewidth]{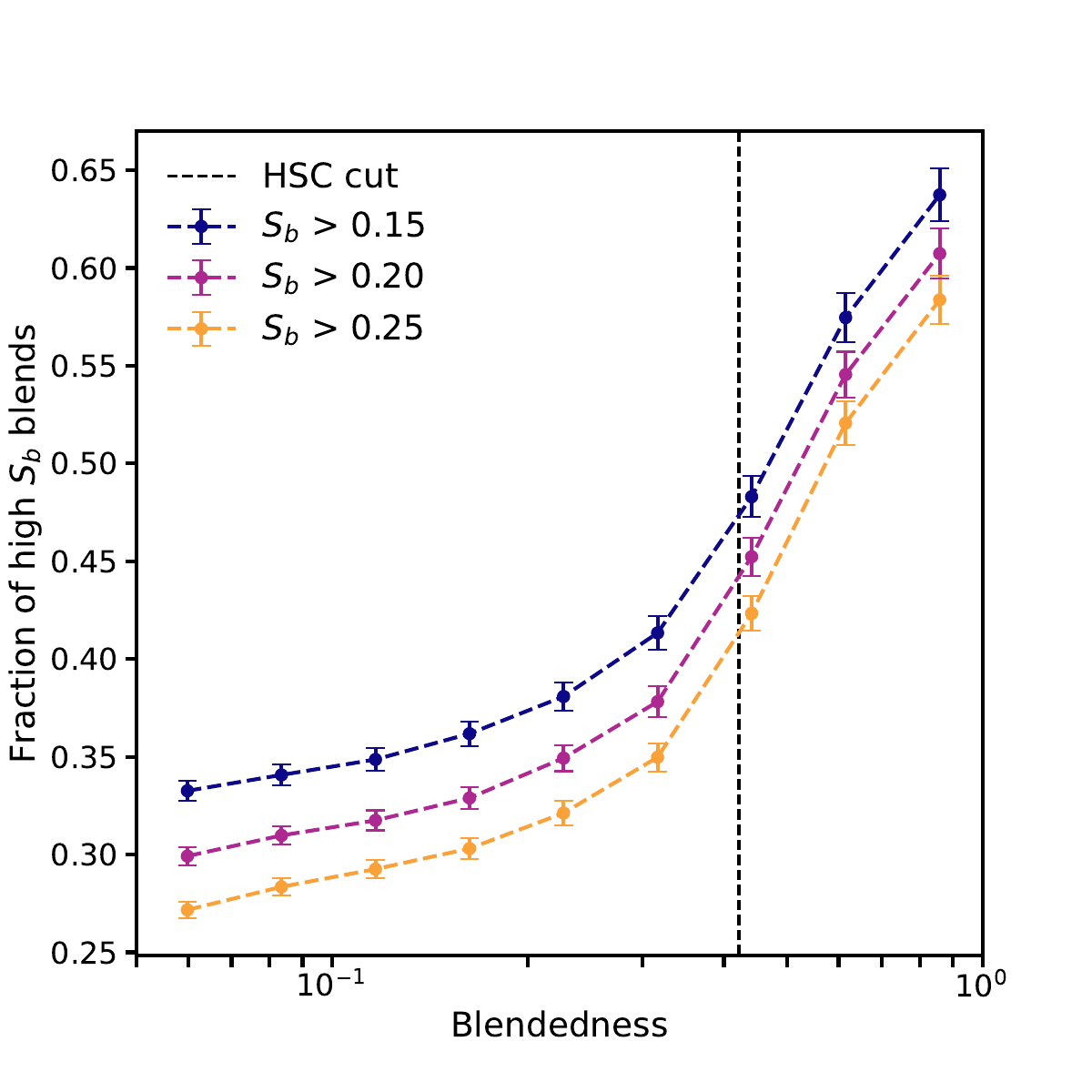}
        \end{subfigure}
        \begin{subfigure}{0.33\linewidth}
            \centering
            \includegraphics[width=1.05\linewidth]{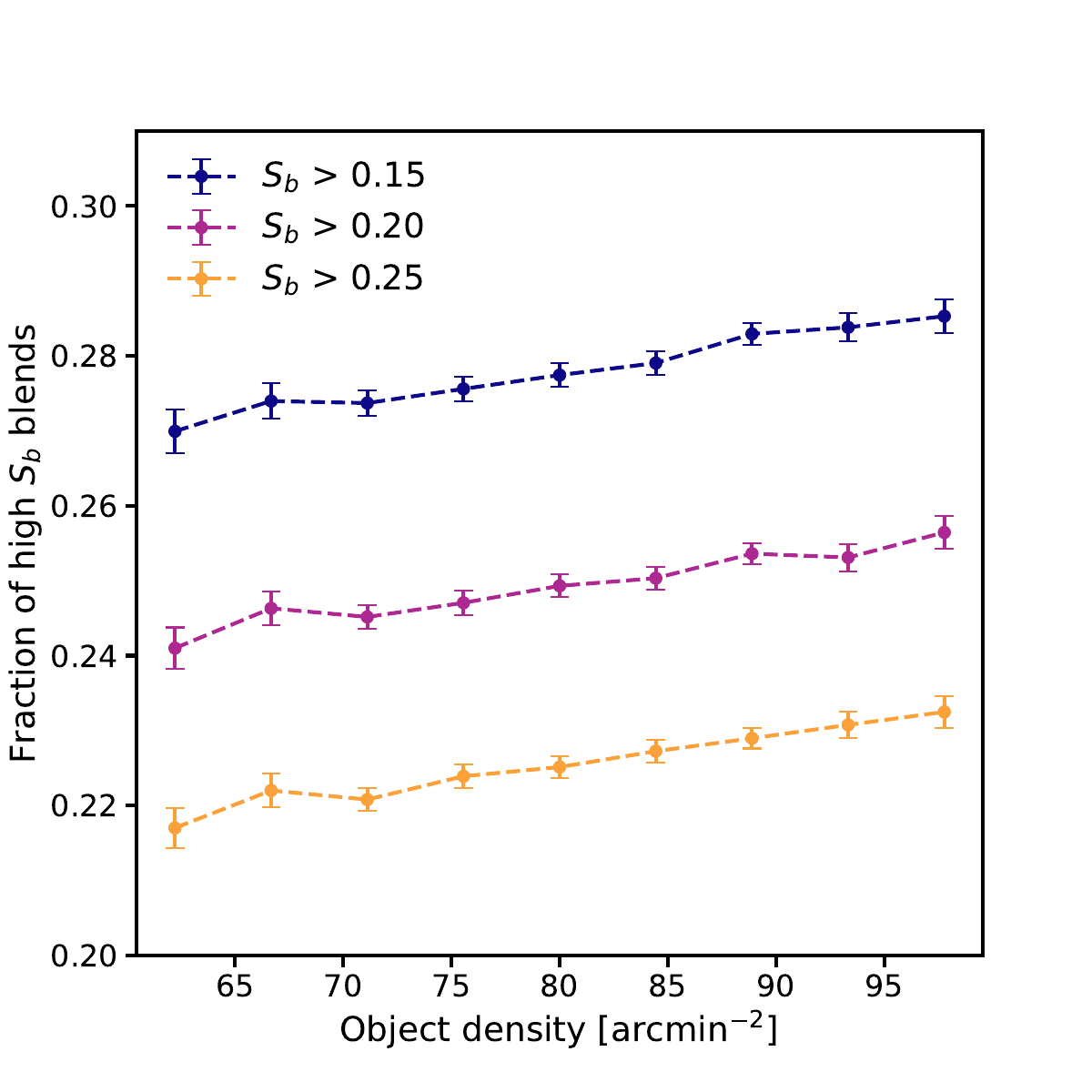}
        \end{subfigure}
    \end{adjustbox}

    \caption{Fraction of high-$S_b$ blended objects (i.e. objects with $S_b > S_{b,\ \rm{cut}}$) binned in DC2object \textit{i}-band magnitude (left), blendedness (center) and density (right) for $S_{b,\ \rm{cut}} = 0.15$, $0.20$ and $0.25$ (different colors). The black dashed line in the center panel represents the blendedness $b < 10^{-0.375}$ cut defined by HSC \citep{Mandelbaum_HSC}.}
    \label{fig:Sb_vs_quantities}
\end{figure*}

\begin{figure}
\centerline{\includegraphics[scale=0.35]{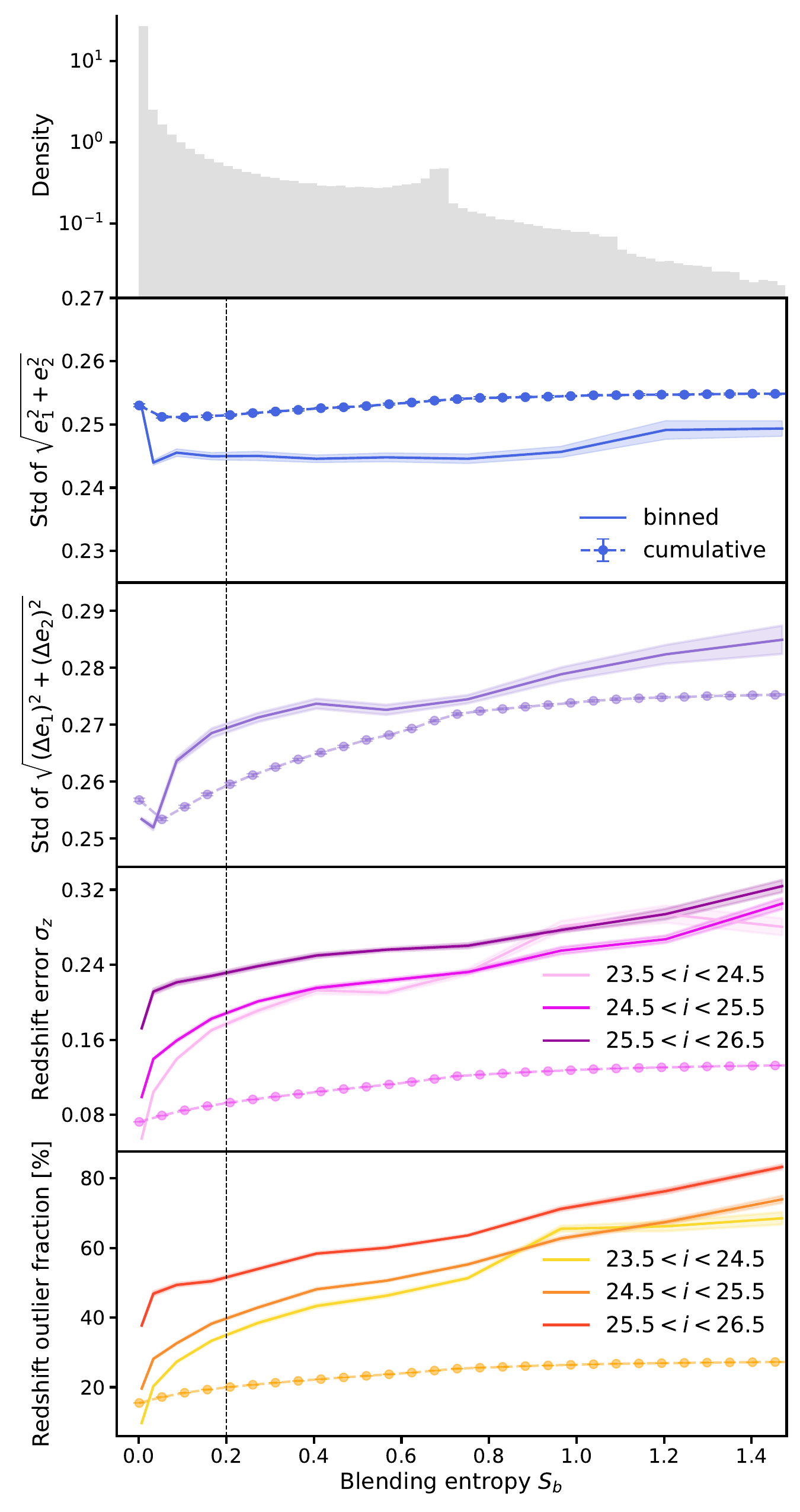}}
\caption{Correlations between the blending entropy $S_b$ and shape and redshift quality metrics. From top to bottom: 
(1) distribution of the blending entropy $S_b$, (2) the standard deviation of DC2 object ellipticities $ \sqrt{e_1^2 + e_2^2} $, 
(3) the standard deviation of ellipticity differences between DC2 objects and matched cosmoDC2 galaxies (i.e. with the largest relative probability of matching) $\sqrt{(\Delta e_1)^2 + (\Delta e_2)^2}$, 
(4) the redshift scatter $\sigma_z$, and 
(5) the redshift outlier fraction (in percent), in bins of DC2 $i$-band magnitudes. 
Each panel displays binned statistics (solid lines with shaded errors) and cumulative statistics (dashed lines with error bars). 
The vertical dashed line indicates the $ S_{b,\mathrm{cut}} = 0.2$ blending entropy cut, used in this work.}
\label{fig:Sb_z_shape}
\end{figure}

In order to gain insight about the blending entropy's ability to distinguish extreme cases of blended systems, we start this section with a set of illustrative and visually obvious examples: the \textit{good}, the \textit{bad}, and the \textit{ugly} blends. These examples are not intended to define a physically unique classification, but rather to provide a controlled framework in which simple, manually defined cuts can be applied to assess whether $S_b$ captures the expected trends. \newline

\textbf{Illustrative example - $\bm{S_b}$ for specific systems:}
in order to illustrate the discriminative power of the blending entropy, we compute its value for objects present in different \texttt{friendly} groups. After a visual inspection, and in addition to the $1-1$ groups denoted as perfect matches, we distinguish different types of blends with $\n>1$ galaxies: the \textit{good}, the \textit{bad}, and the \textit{ugly} groups. We adopt intentionally simple and partly arbitrary definitions of these groups as follows: we define \textit{good} $2-1$ groups as those where the two galaxies have an overlap fraction lower than 0.3 and an absolute difference in \textit{i}-band magnitudes higher than 2, such that the faintest galaxy has little impact on the detection. Conversely, \textit{bad} $2-1$ groups are those where there is more than $30\%$ overlap between the galaxies, as well as similar magnitudes, making it difficult to assign the object to any of the two galaxies. We also define \textit{good} $2-2$ and $3-2$ groups based on one-to-one overlap between pairs of object-galaxy with close magnitudes. We name \textit{ugly} groups, highly populated and therefore more likely blended groups with $\n>3$ or $\m>3$.
These different blended cases are summarized in \cref{tab:friendly_groups}. DC2 simulated images of these groups with their associated entropies---taken to be the sum of the object entropies when $\m>1$---are shown in \cref{fig:friendly_groups}. The galaxies (in red) and the objects (in blue) are represented with ellipse contours on top of the RGB images. The \textit{i}-band magnitudes are also indicated. \newline

In order to analyze the dependencies between these different groups and the blending entropy, we apply several cuts $S_{b,\ \rm{cut}}$ and remove objects with $S_b > S_{b,\ \rm{cut}}$ from the data.
\Cref{fig:Sb_good_bad} shows the fractions of suppressed objects belonging to \textit{bad} 2–1 (blue), and \textit{good} 2–1 (red), 2–2 (orange), and 3–2 (pink) groups, as well as all groups combined (purple), as a function of the cut $S_{b,\ \rm{cut}}$ applied to their blending entropies $S_b$. We observe that nearly all objects in the \textit{bad} 2–1 group ($\sim1$, see blue curve) are suppressed for $S_{b,\ \rm{cut}} \lesssim 0.2$, with the suppression fraction dropping sharply for $0.2 < S_{b,\ \rm{cut}} < 0.7$. This indicates that most \textit{bad} objects have high blending entropies. In contrast, the \textit{good} group curves (red, orange, pink) show a sharp decrease around $S_{b,\ \rm{cut}} \sim 0.1$, suggesting that most \textit{good} objects have low entropy values.

These observations suggest the use of a cut around $S_b \sim 0.2$ (indicated by the vertical dashed line) to effectively separate \textit{bad} from \textit{good} blended objects. Applying this cut allows us to suppress the majority of \textit{bad} objects while retaining most of the \textit{good} ones---only about 5\% of \textit{good} 2–1 objects are removed. The $S_{b,\ \rm{cut}}=0.2$ cut results in a suppression of approximately 25\% of all detected objects, as shown by the purple curve and quantified in \cref{tab:percentage_suppressed_objects}. This cut will be confirmed by our analysis of DC2object observables and galaxy measurement errors in the following paragraphs.

Additionally, the suppression fractions are shown for two magnitude cuts applied to the DC2object catalog: $i < 24.5$ (dashed lines) and $i < 26.5$ (solid lines). As expected, suppression fractions are consistently higher for the fainter cut, since fainter objects are more likely to be blended and thus may have higher blending entropies. When considering all groups, the deeper magnitude cut leads to approximately 10\% more suppression at $S_b = 0.2$, further demonstrating that fainter objects are more affected by blending. Nonetheless, even at deeper cuts, the $S_b$ cut  effectively filters out \textit{bad} blends while preserving most \textit{good} objects, demonstrating the robustness of $S_b$ as a metric for characterizing blending.\newline

\textbf{Blending entropy vs. magnitude:} in the left panel of \cref{fig:Sb_vs_quantities}, we show the fraction of high-$S_b$ blended objects per bins of (object) \textit{i}-band magnitude, using $0.15$, $0.20$, and $0.25$ cuts on the blending entropy. The fraction of high-$S_b$ objects systematically increases as they become fainter. For bright objects ($i \sim 19$), the fraction of high-$S_b$ blends is below 5\%, whereas for faint objects ($i \sim 26$), it reaches approximately 35–40\%, depending on the blending entropy cut. This confirms that blending effects are more prominent at higher depths, where the density of sources increases, making it more likely for objects to be contaminated by neighbors. The value of the $S_b$ cut has a small impact on the resulting blend proportions, especially at high magnitudes ($i \sim 26.5$) where we observe only 5\% more of high-$S_b$ blended objects for $S_b>0.15$ compared to $S_b>0.25$, whereas this difference remains lower at smaller magnitudes. This shows that we have some flexibility in the choice of the blending entropy cut, defined in this work at $S_b = 0.2$. \newline

\textbf{Blending entropy vs. blendedness:} the blending entropy can also be compared to the blendedness (see \cref{subsec:DC2object}). To study correlations between these two blending metrics, the middle panel of \cref{fig:Sb_vs_quantities} shows the fraction of high-$S_b$ blended objects per bins of blendedness, using $0.15$, $0.20$, and $0.25$ cuts on the blending entropy. We observe that this fraction consistently increases with blendedness for the three entropy cuts.
At low blendedness ($b < 0.1$), the fraction of high-$S_b$ blends remains relatively low, around 30–35\%, regardless of the chosen entropy cut.
As the blendedness extends to higher values, the fraction increases, reaching proportions between 55\% and 65\% in the highest blendedness bin. Within all bins, there is 10\% more of high-$S_b$ blends for $S_b>0.15$ than for $S_b>0.25$. The HSC blendedness cut ($b \sim 10^{-0.375} \approx 0.42$) lies at an intermediate point where the fraction of high-$S_b$ blends increases significantly. However, $23$~\% of detected objects with low blendedness values ($b < 10^{-0.375}$) have a blending entropy $S_b>0.2$, therefore characterized as high-$S_b$ blended objects. This suggests that the HSC blendedness cut is reasonable for isolating blended objects but might not fully capture more subtle blending effects, which blending entropy can help refine in this simulated setup (discussion about applications to real data is provided in \cref{sec:Conclusions}). Since the correlations between blendedness and blending entropy are not straightforward, the development of this new metric which accounts for the cosmoDC2 truth information is promising and allows us to distinguish blended cases more precisely with the objective of mitigating their impact on cluster lensing cosmology. \newline

\textbf{Blending entropy vs. density:} 
in the third panel of \cref{fig:Sb_vs_quantities}, we show how the fraction of high-$S_b$ blended objects evolves with the object density in arcmin$^{-2}$, using $0.15$, $0.20$, and $0.25$ cuts on the blending entropy. The local object density was estimated using a HEALPix\footnote{\url{https://healpix.sourceforge.io/}} pixelation with \texttt{nside=4096}. The density assigned to each object corresponds to the number of detected objects within its pixel divided by the pixel area, yielding a surface density in units of arcmin$^{-2}$.
As expected, we observe that the fraction of high-$S_b$ blends increases with increasing object density which reflects that higher densities lead to more overlapping galaxies and hence more blending. 
Even if we reach a higher fraction of high-$S_b$ blends when the cut is lower---as we include more objects---the three curves follow the same trend with relative increases of $\sim 6\%$.
Note that the density reported here is a \textit{raw} object number density $n$, computed by considering all objects with $i<26.5$, with a mean value of $n=\SI{80}{\per\arcmin\squared}$. For the effective number of galaxies $n_{\rm eff}$ used for weak lensing measurements, which accounts for shape measurement uncertainties, we obtain a mean value $n_{\rm eff}=\SI{31}{\per\arcmin\squared}$, consistent with \cite{Chang_density_LSST}.\newline

\textbf{Blending entropy vs. shape:}
\cref{fig:Sb_z_shape} illustrates the impact of the blending entropy $S_b$ on several quality metrics for galaxy shape and redshift estimations. The plots show both binned (solid lines with shaded uncertainty) and cumulative (dashed lines with error bars) statistics as a function of increasing $S_b$. The vertical dashed line indicates the fiducial blending entropy cut $S_{b,\mathrm{cut}} = 0.2$, used throughout this work to identify and suppress high-$S_b$ blended objects. The distribution of the blending entropy is shown in the top panel.
The second top panel of \cref{fig:Sb_z_shape} shows the standard deviation of DC2 object ellipticities $\sqrt{e_1^2 + e_2^2}$ as function of the blending entropy. We observe that it remains relatively constant across different entropy values, with a slight increase from $\sim 0.245$ at low  $S_b$ to $\sim 0.256$ for $ S_b > 1.0$, for the binned curve. This suggests that high-$S_b$ blends do not exhibit more dispersed ellipticity values compared to well-matched and isolated objects ($S_b<0.2$), therefore they are not necessarily rounder or more elongated.
The third panel from the top shows the standard deviation of the ellipticity differences $\Delta e_i$ between DC2 objects and their corresponding matched galaxies (i.e. with the largest relative probability of matching). We observe that the curves increase, showing that blended objects---with high $S_b$ values---tend to have higher shape errors compared to their true cosmoDC2 counterparts. This confirms that higher blending entropy correlates with degraded shape recovery, potentially due to larger magnitudes, and fainter detected objects.\newline

\textbf{Blending entropy vs. redshift:}
the fourth panel of \cref{fig:Sb_z_shape} shows the redshift error $\sigma_z$, computed through the central 68\% interval of $e_z = (z_{\rm{obj}}-z_{\rm{gal}})/(1+z_{\rm{gal}})$. It should be noted that the BPZ photometric redshifts in DC2 are significantly noisier than expected for actual LSST data. However, because our weak lensing analysis presented in \cref{sec:Blending_cluster_lensing} does not rely on the photometric redshift error distribution, we do not expect this to significantly affect our main results.
The redshift scatter shows a strong dependence with the blending entropy. The binned values grow significantly, from $\sim 0.08$ at low $S_b$ to $\sim 0.45$ beyond $S_b = 1.0$. This indicates a severe degradation in redshift estimation for highly blended objects. In contrast, the cumulative curve stays relatively flat, around $\sim 0.11$, indicating that the majority of the DC2 object population retains a stable redshift precision under the chosen cut.
In the bottom panel we show the redshift outlier fraction, defined as $|(z_{\rm{obj}}-z_{\rm{gal}})/(1+z_{\rm{gal}})|>0.15$ for all object-galaxy pairs. We observe that the fraction of catastrophic outliers increases sharply with $S_b$. The binned curve grows from $\sim 15\%$ to over $75\%$, while the cumulative curve levels off around $\sim 27\%$ at high entropy. This again confirms that the redshift estimate fails with objects with high $S_b$ values.
Finally, we observe a clear correlation between photometric redshift errors, outlier fraction, and the blending entropy when binning by DC2object magnitude bins, as shown in the bottom panels of \cref{fig:Sb_z_shape}. As expected, fainter objects exhibit larger redshift uncertainties and are therefore more likely to be classified as outliers. 
In the following sections, we apply additional quality cuts on galaxy shapes and signal-to-noise ratio to select background sources for the cluster weak lensing analysis (see \cref{subsubsec:background_source_sel}). These so-called weak lensing (WL) cuts effectively set a magnitude limit of $i \simeq 25$ on the background sample, which reduces the fraction of redshift outliers and decreases photometric redshift uncertainties, in addition to the application of the blending entropy cut at $S_{b,\mathrm{cut}} = 0.2$.

Altogether, these results demonstrate that the blending entropy effectively traces systematic degradation in shape and redshift measurements. The $S_b < 0.2$ cut succeeds in excluding highly problematic objects, leading to improved data quality while preserving the majority of the DC2 detected object population.

\subsubsection{Blending entropy from \textit{gaussian} overlap} \label{subsubsec:Sb_gaussian}

\begin{table}
    \centering
    \scalebox{0.9}{
    \begin{tabular}{l|c|c|c}
        Metric & Cut & Suppressed objects & Suppressed sources\\
      \hline \hline
      \rowcolor{gray!20}
      \textit{Top-hat} $S_b$  & $< 0.2$ & $25.3\%$ & $22.7\%$\\
      \textit{Top-hat} $S_b$   & $<0.1$ & $32.1\%$ & $29.4\%$ \\ 
      \textit{Top-hat} $S_b$  &$ <0.3$ & $20.9\%$ & $17.2\%$\\
      \textit{Gaussian} $S_b$   & $<0.3$ & $61.3\%$ & $62.4\%$\\
       Blendedness $b$  & $< 10^{-0.375}$ & $2.66\%$ &  $5.41\%$\\
       \hline
       & & $N_{\mathrm{obj}}=\num{85248408}$ & $N_{\mathrm{src}}=\num{29324361}$
    \end{tabular}
    }
    \caption{Blending metrics used in this work with their associated cuts and percentage of suppressed DC2 objects and DC2 background sources. As explained in the body of the text, DC2 objects are all the detections from DC2object catalog, whereas sources are the selected objects used for weak lensing studies (with the so-called WL cuts, see \cref{subsubsec:background_source_sel}). The total number of objects and sources are indicated. The gray-colored row shows the main analysis.}
    \label{tab:percentage_suppressed_objects}
\end{table}

In the previous subsection, the $S_{b,\mathrm{cut}} = 0.2$ cut was defined using a \textit{top-hat} overlap fraction in the computation of the relative matching probabilities (see \cref{eq:proba_of_matching}), which directly enters the definition of the blending entropy.
As mentioned in \cref{subsec:metrics}, we can compare it to the \textit{gaussian} overlap, computed using 2D Gaussian representation of galaxies instead of ellipses.
Using this new definition, we are able to recover the global distribution of the blending entropy, as well as the fractions of suppressed objects per \textit{good} and \textit{bad} groups depending on the blending entropy cut, as studied previously with the \textit{top-hat} overlap fraction. These studies are shown in \cref{app:Sb_gaussian_distrib}. With this overlap definition, we fix $S_b>0.3$ in order to identify and select problematic blends. With this cut, $\sim 61\%$ of the detected objects are removed, as indicated in \cref{tab:percentage_suppressed_objects}. This level of suppression is significant, suggesting that modeling galaxies with Gaussian profiles---and using the resulting \textit{gaussian} overlap fraction---might induce difficulties in defining and separating \textit{good} and \textit{bad} groups, thereby complicating the isolation of blending effects. Nevertheless, this number is on par with previous studies finding that roughly two thirds of LSST detections will be blended to some extent~\citep{Sanchez_blending}. We will show in the next section that both overlap definitions allow us to mitigate the impact of blending on cluster lensing cosmology.


\section{Application to cluster lensing cosmology}\label{sec:Blending_cluster_lensing}

\subsection{Challenges of blending in weak lensing}

\begin{figure}
\centerline{\includegraphics[scale=0.35]{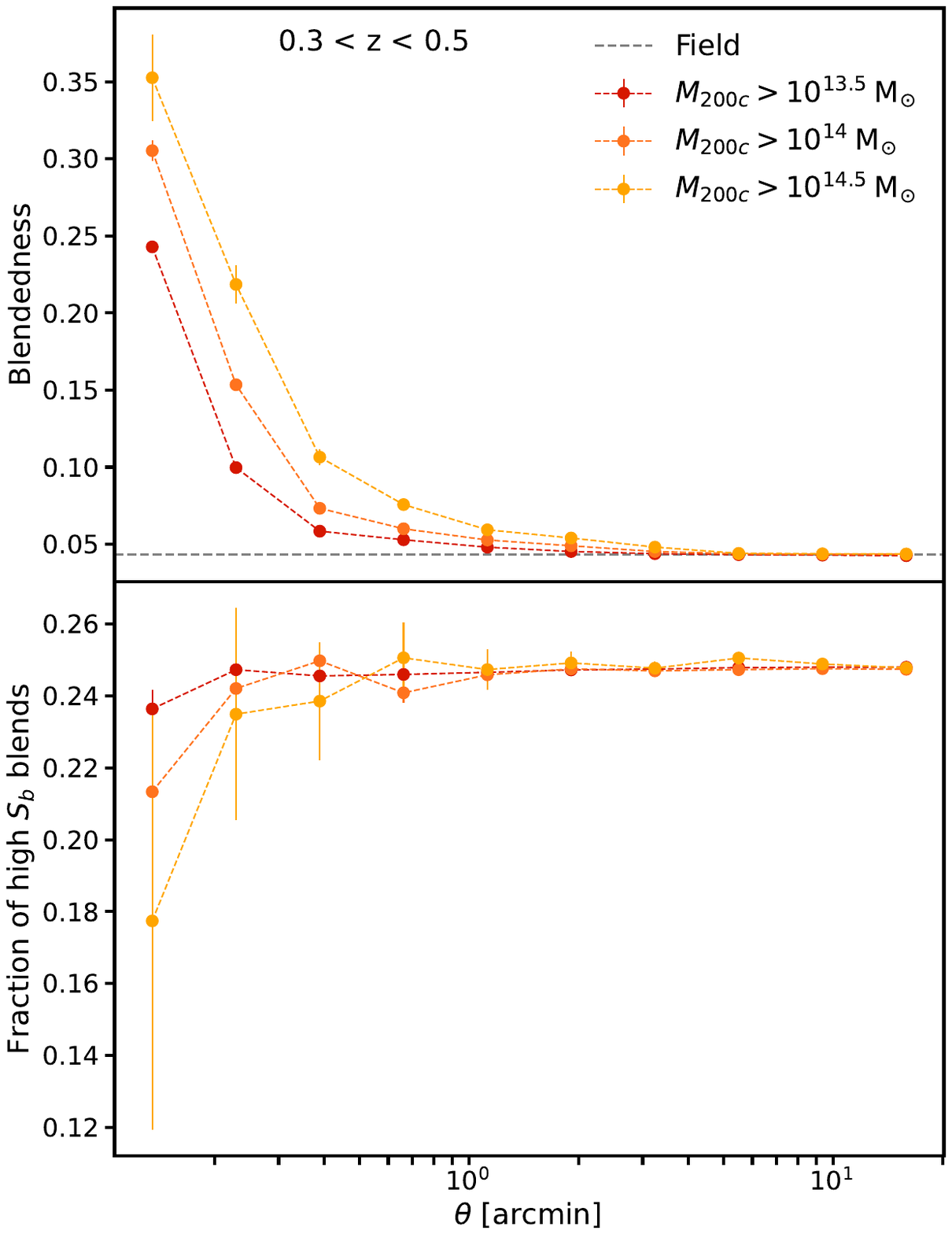}}
\caption{\textit{Top:} binned DC2object blendedness as function of the angular distance to the centers of the nearest cosmoDC2 halos in arcminutes ($\SI{1}{\arcminute} \sim \SI{0.29}{\mega\parsec}$ at $z = 0.4$). The mean blendedness (i.e. without binning) is represented with the horizontal dashed gray line.
\textit{Bottom:} fraction of high-$S_b$ blends (i.e. objects with $S_b>0.2$) as function of the angular distance to the centers of the nearest cosmoDC2 halos in arcminutes.
For both panels, the curves are plotted for 3 halo mass thresholds:
$M > 10^{13.5}~\rm{M_\odot}$ (in red), $M > 10^{14}~\rm{M_\odot}$ (in orange) and $M > 10^{14.5}~\rm{M_\odot}$ (in yellow). The halos are selected to have redshifts $z$ between $0.3$ and $0.5$.}
\label{fig:blending_R}
\end{figure}


Galaxy clusters are the densest regions of the Universe in terms of dark matter particles and baryons which lead to dense galaxy environments and high masses with $M>10^{13.5} \:\rm{M_{\odot}}$. Therefore, we expect more blended galaxies inside the core of clusters compared to the field (i.e. outside galaxy clusters), especially for massive clusters. This could affect the measurement of individual galaxy properties---such as shapes and redshifts---both in the field and more significantly within galaxy cluster regions. Moreover, in cluster fields, the high projected density of cluster member galaxies enhances the superposition and blending of background sources. Therefore, we expect blending to impact the measurement of the cluster weak gravitational lensing signal and bias their mass estimates, as well as $\Om$ and $\sigma_8$ cosmological parameters.

In the top panel of \cref{fig:blending_R}, we show the mean DC2 blendedness value (see \cref{subsec:DC2object}) as a function of the projected angular distance $\theta$ from the centers of cosmoDC2 dark matter halos, expressed in arcminutes. Only halos within the redshift range $0.3 < z < 0.5$ are considered. In order to show how blending tends to increase with cluster mass, we vary the minimum mass of the considered halos from $M > 10^{13.5}~M_\odot$  to  $M > 10^{14.5}~M_\odot$, demonstrating for the presence of blending in denser regions in comparison with the field (dashed gray line). 

A clear increase in the blendedness is observed toward halo centers ($\theta < \SI{1}{\arcminute}$), particularly for the most massive ones. This trend indicates that a significant fraction of detections near cluster cores are affected by flux contamination from neighboring objects. Since galaxies located near the cluster center tend to be brighter, it results in more extensive flux leakage into neighboring detections, and therefore higher blendedness values.

On the bottom panel of \cref{fig:blending_R}, we present the fraction of high-$S_b$ blends (i.e., objects with blending entropy $S_b > 0.2$) as a function of projected angular separation $\theta$ from halo centers, for the same previous halo mass cuts. Contrary to the trend seen for the blendedness, the fraction of high-entropy blends decreases toward the halo centers. For halos with $M > 10^{13.5}~M_\odot$, this fraction remains approximately constant across the full range of separation angles, with values around $25\%$, consistent with the field-level high-$S_b$ fraction observed in \cref{tab:percentage_suppressed_objects} of \cref{subsec:blending_demography}. However, for higher mass halos, the fraction drops sharply near the cluster centers. This can be explained by the fact that objects in halo centers tend to be both brighter and larger. As a result, they are more reliably matched to their true counterparts due to higher signal-to-noise ratios and more well-defined shapes. This precise matching leads to lower entropy values, even in crowded regions.

This comparison highlights the difference and complementarity of the two metrics. The blendedness quantifies the level of flux contamination from neighboring sources in the images. In contrast, the blending entropy measures the ambiguity in object-galaxy associations. Cluster cores may contain blends with substantial flux contamination, but they are often qualified as \textit{good} and correctly matched, reducing the blending entropy score. In both cases, it is necessary to evaluate the impact of blending---using the blendedness or the blending entropy---on cluster lensing profiles and cosmological parameters.

\subsection{Cluster weak lensing mass estimation} \label{subsec:WL_theory}

The observed (complex) ellipticity\footnote{Using the $\epsilon$ ellipticity convention \citep{Schneider_ellipse_def}}, \( e^{\mathrm{obs}} \), of a lensed source galaxy is related to its intrinsic (unlensed) shape, \( e^{\mathrm{int}} \), by the following expression \citep{Seitz_ellipticityWL}:  
\begin{equation}
    e^{\mathrm{obs}}=\frac{e^\mathrm{int}+g}{1+g^*e^{\mathrm{int}}} \quad\mathrm{where}\quad g=\gamma/(1-\kappa)
\end{equation}
is the reduced shear ($g^*$ is the complex conjugate of $g$) with $\gamma$ being the shear and $\kappa$ the convergence. In the weak lensing regime, where $\kappa\ll1$, we can approximate the observed ellipticity using a first-order Taylor expansion in $\gamma$, yielding $e^{\mathrm{obs}} \approx \gamma + e^\mathrm{int} $. Assuming that intrinsic ellipticities are randomly oriented in the sky (i.e. neglecting intrinsic alignments), the mean value of the intrinsic ellipticity is $\expval{e^\mathrm{int}} \approx 0$, leading to $\expval{e^\mathrm{obs}} \approx \gamma$. Thus, in the weak lensing regime, the local shear can be statistically estimated from the observed ellipticities of background galaxies. The two shear components for each galaxy are conveniently expressed in terms of the tangential and cross components $\gamma_+$ and $\gamma_\times$, respectively. The tangential ellipticity $e_+$ is therefore given by:
\begin{equation}
    e_{+} = -[e_1 \cos{(2\phi)} + e_2 \sin{(2\phi)}],
\end{equation}
with $\phi$ the polar angle of the source galaxy relative to the cluster center. Under the assumption
of a spherical matter density distribution, the average tangential ellipticity $e_+$ relates to the
tangential shear $\gamma_+$ by $\langle e_+ \rangle|_R = \gamma_+(R)$ where $R$ is the projected radius to the cluster center.

\subsubsection{Theoretical modeling}

The projected matter density of a cluster along the line of sight $y$ can be expressed as:
\begin{equation}
    \Sigma(R)=\int_{-\infty}^{+\infty}\rho\left(\sqrt{y^2+R^2}\right) \dd{y},
\end{equation}
where $\rho$ is the 3D dark matter density and $R$ the projected radius to the cluster center. From this quantity, we can derive the excess surface mass density $\Delta\Sigma(R)$, expressed as:
\begin{equation}
    \Delta\Sigma(R)=\Sigma(<R)-\Sigma(R) 
\end{equation}
where
\begin{equation}
\Sigma(<R)=\frac{2}{R^2}\int_0^RR'\Sigma(R') \dd{R'}.
\end{equation}

The total predicted excess surface density is given by the sum of two contributions:
\begin{equation}
    \Delta\Sigma(R)=\Delta\Sigma_{\mathrm{1h}}(R)+\Delta\Sigma_{\mathrm{2h}}(R).
\end{equation}
The 1-halo term $\Delta\Sigma_{\mathrm{1h}}$ represents the contribution of the cluster itself to the matter density, while the 2-halo term $\Delta\Sigma_{\mathrm{2h}}(R)$ denotes the contribution of the surrounding halos. In this work, we will neglect the contribution of the 2-halo term, by considering radii $R<\SI{3.5}{\mega\parsec}$ \citep[][]{Payerne_mass_richness}. Under the spherical assumption, $\Delta\Sigma_{\mathrm{1h}}$ can be fitted with a 3D Navarro-Frenk-White~\cite[NFW,][]{Navarro_NFW} dark matter density profile given by:
\begin{equation}
    \rho^{\mathrm{nfw}}(r)=\frac{\rho_s}{\frac{r}{r_s}\left(1+\frac{r}{r_s}\right)^2},
\end{equation}
where $\rho_s$ is the scale density, and $r_s$ the scale radius is expressed through $r_s=r_{200c}/c_{200c}$ with $r_{200c}$ (respectively $c_{200c}$) being the radius (respectively concentration) of a sphere with a matter density 200 times higher than the critical density $\rho_c(z)$. The NFW profile can then be parametrized via the concentration $c_{200c}$ and the overdensity mass $M_{200c}$, allowing for an estimation of the lens mass, defined as:
\begin{equation}
    M_{200c}=\frac{4\pi r_{200c}^3}{3}200\rho_c(z).
\label{eq:M200c}
\end{equation}

\subsubsection{Weak lensing estimator}\label{subsubsec:DeltaSigma_estimates}

In practice, the excess surface density, \(\Delta\Sigma(R)\), is commonly used as a weak lensing shear estimator and is related to the tangential shear, \(\gamma_+(R)\), through \citep[e.g.,][]{Melchior_DeltaSigma,Mandelbaum_DeltaSigma,McClintock_DeltaSigma}:
\begin{equation}
    \Delta\Sigma(R) = \Sigma_{\mathrm{crit}}(z_s,z_l)\,\gamma_+(R),
\end{equation}
where \(\Sigma_{\mathrm{crit}}(z_s,z_l)\) is the critical surface mass density for a source with redshift \(z_s\) and a lens with redshift \(z_l\). It is given by
\begin{equation}
    \Sigma_{\mathrm{crit}}(z_s, z_l) = \frac{c^2}{4\pi G}\,\frac{D_A(z_s)}{D_A(z_l)\,D_A(z_s,z_l)},
\label{eq:sigma_crit}
\end{equation}
with \(D_A(z_l)\), \(D_A(z_s)\), and \(D_A(z_s,z_l)\) denoting the physical angular diameter distance to the lens, to the source, and between the lens and the source, respectively.

In this work, we estimate the excess surface density signal around an ensemble of clusters rather than individual clusters. This stacking strategy increases the signal-to-noise ratio, thereby enabling more precise measurements of the lensing profiles, especially for low-mass clusters. The maximum likelihood estimator of \(\Delta\Sigma\) when considering a stack of \(N_l\) lenses with \(N_{ls}\) background sources is given in a radial bin \([R, R+\Delta R]\) by \citet{Shirasaki_DeltaSigma}:
\begin{equation}
    \widehat{\Delta\Sigma}_+(R) = \frac{1}{\sum_{l=1}^{N_l}\sum_{s=1}^{N_{ls}}w_{ls}}\,\sum_{l=1}^{N_l}\sum_{s=1}^{N_{ls}}w_{ls}\,\Sigma_{\mathrm{crit}}(z_s, z_l)\,e_+^{l,s},
\label{eq:DeltaSigma}
\end{equation}
where \(e_+^{l,s}\) is the tangential ellipticity of a background source $s$ of redshift $z_s$ associated with a lens $l$ of redshift $z_l$. This expression is only valid for point estimates of \(z_s\) (mean of the redshift probability density function, for example). The weights \(w_{ls}\) are applied to maximize the signal-to-noise ratio of the \(\widehat{\Delta\Sigma}\) estimator \citep{Sheldon_WL_weights} and are defined for each lens-source pairs as
\begin{equation}
    w_{ls} = w_{ls}^{\mathrm{geo}}\,w_{ls}^{\mathrm{shape}},
\end{equation}
with
\begin{align}
    w_{ls}^{\mathrm{geo}} & = \Sigma_{\mathrm{crit}}(z_s, z_l)^{-2},\\
    w_{ls}^{\mathrm{shape}} & = \frac{1}{\sigma_{\mathrm{rms}}^2(e_s^+) + \sigma_{\mathrm{meas}}^2(e_s^+)},
\end{align}\label{eq:DeltaSigma_estimate}
where \(\sigma_{\mathrm{rms}}(e_s^+)\) is the standard deviation of the tangential ellipticity and \(\sigma_{\mathrm{meas}}(e_s^+)\) is the measurement error on the shape.

\subsection{Cluster lensing profiles in DC2}\label{subsec:recover_profiles}

\subsubsection{Lens model}\label{subsubsec:lens_model}

\begin{figure}
\centerline{\includegraphics[scale=0.35]{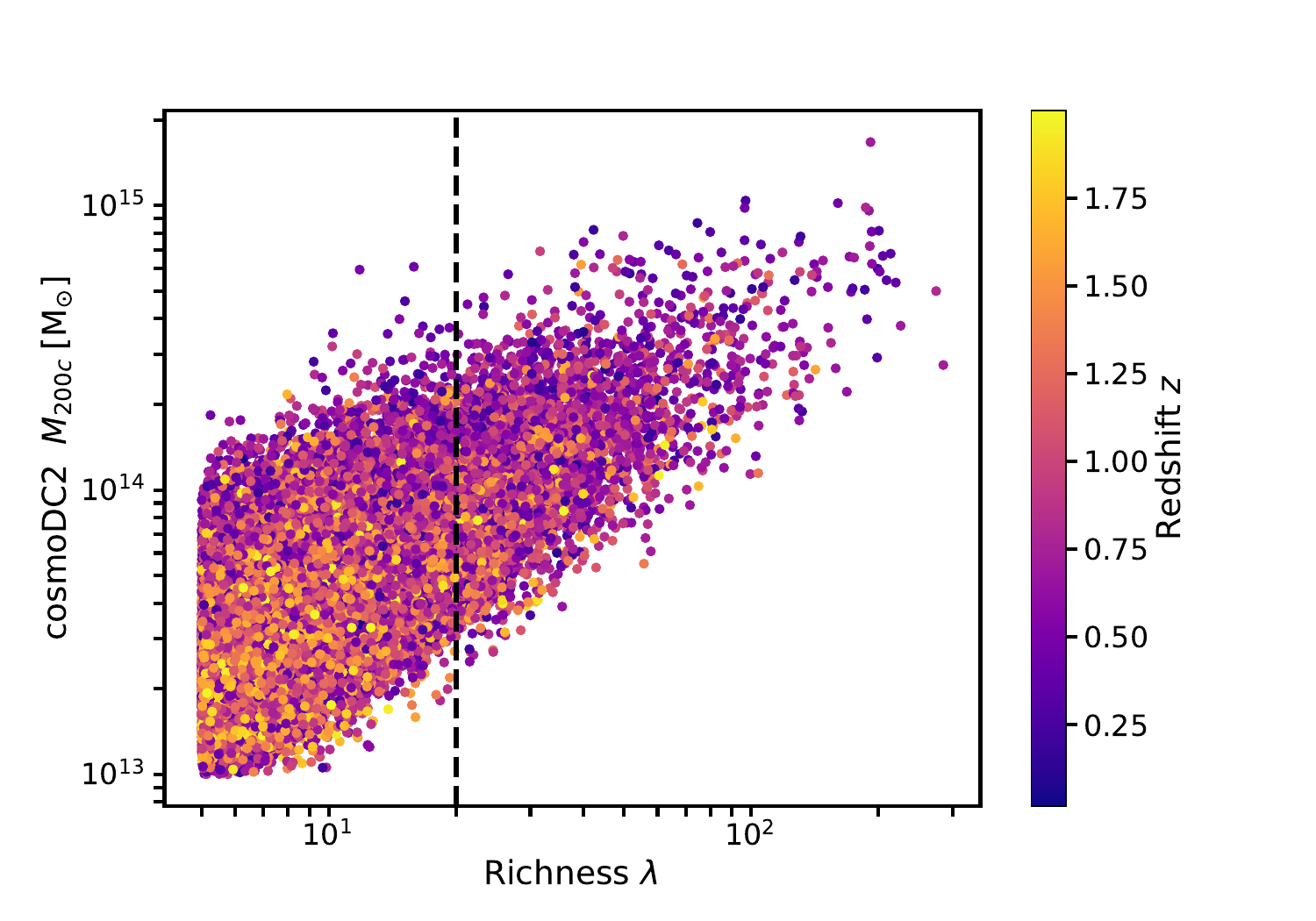}}
\caption{cosmoDC2 halo $M_{200c}$ masses vs. corresponding richnesses $\lambda$. The richnesses are derived from the~\protect\cite{Payerne_mass_richness} mass-richness relation. The points are colored with the cosmoDC2 redshifts of the halo central galaxies. The vertical dashed line represents the low richness cut applied to the halo catalog for weak lensing studies.}
\label{fig:richness}
\end{figure}

For simplification purposes,\footnote{In this study, we do not focus on the impact of blending on cluster detection algorithm.} and to isolate the effect of blending on lensing profiles and mass estimates, we decide to only consider the cosmoDC2 dark matter halos from the \textit{Outer Rim} simulation (see \cref{subsec:cosmoDC2}). The dark matter halos are described by their $M_{200c}$ masses\footnote{We only have access to their FoF masses (i.e. the sum of the particles masses that compose them) from the simulation. In order to access a physical definition of the mass, we match the cosmoDC2 halos to SkySim5000 ones (see \cref{subsec:cosmoDC2}) to recover their $200c$ masses, only available in SkySim5000.} (see \cref{eq:M200c}) and the positions of their central galaxies.

We attribute to each cosmoDC2 halo---with $M_{\rm FoF}>10^{13}\:\rm{M_{\odot}}$ and $z < 2$---a richness $\lambda$ based on its $M_{200c}$ mass and redshift. We consider a log-normal scaling relation~\citep{Murata_mass_richness,Sunayama_mass_richness} with 6 free parameters ($\ln{\lambda_0}, \mu_z, \mu_m, \sigma_{\ln{\lambda_0}}, \sigma_z, \sigma_m$), given by:
\begin{equation}
P(\lambda|m,z) = \frac{1}{\lambda \sqrt{2\pi}\sigma_{\ln{\lambda}|m,z}}\exp{-\frac{[\ln{\lambda}-\langle\ln{\lambda|m,z}\rangle]^2}{2\sigma_{\ln{\lambda|m,z}}^2}},
\end{equation}\label{eq:mass_richness_relation}
where the mean richness has both mass and redshift dependencies, such that:
\begin{equation}
\langle \ln{\lambda|m,z} \rangle = \ln{\lambda_0} + \mu_z \log{\left(\frac{1+z}{1+z_0}\right)}+\mu_m\log_{10}{\left(\frac{m}{m_0}\right)},
\end{equation}
and the dispersion of the log-normal scaling relation is given by:
\begin{equation}
\sigma_{\ln{\lambda|m,z}} = \sigma_{\ln{\lambda_0}} + \sigma_z\log{\left(\frac{1+z}{1+z_0}\right)} + \sigma_m\log_{10}{\left(\frac{m}{m_0}\right)}.
\end{equation}

For the mass-richness parameters, we use the values obtained from the redMaPPer \citep[][]{Rykoff_redMaPPer} mass-richness scaling relation defined in \cite{Payerne_mass_richness}, estimated from the matching between richness of redMaPPer-detected clusters and halo masses from cosmoDC2. Their fiducial (and the $m_0, z_0$ pivot) values are shown in Table 2 of \cite{Payerne_mass_richness} and reported in \cref{tab:priors}. In \cref{fig:richness}, we show the recovered scaling relation using halos with $M_{200c} > 10^{13}\,\mathrm{M_\odot}$ and $z < 2$, and with a minimum richness  $\lambda > 5$. This relatively low mass threshold is chosen to ensure that the halo sample is complete for systems with $\lambda > 20$, which defines the lens selection used for the weak lensing analysis (dashed line). This cut also ensures a reasonable lower limit on $M_{200c}$ masses for galaxy cluster cosmology studies \citep[][]{Lesci_KIDSY3_clusters,DES_Y3_cluster}. We impose the lenses to have redshifts between $0.2$ and $2$,\footnote{With LSST, we will observe galaxy clusters up to $z\sim1.2$. However, in this work we consider halos with redshifts up to $2$ in order to explore the impact of blending on lensing profiles at high redshifts} and their concentrations are derived from the \cite{Duffy_mass_concentration} mass-concentration scaling relation. We obtain a total of $4175$ selected halos.

\subsubsection{Background sources selection}\label{subsubsec:background_source_sel}

We want to compare $\Delta\Sigma$ lensing profiles recovered from both cosmoDC2 background galaxies and DC2object background detections. As explained in \cref{subsec:WL_theory}, the measurements of the source lensed ellipticities and redshifts are necessary to estimate the lensing profiles (see \cref{eq:DeltaSigma}). For cosmoDC2, the source galaxy redshifts are the true redshifts from the simulation. The lensed ellipticities $e_{\rm obs}$ are constructed from the true intrinsic shapes $e_{\rm int}$ (see \cref{subsec:WL_theory}) and the evaluation of the true shear and true convergence at their galaxy's locations, resulting in no associated shape measurement errors. The intrinsic shape noise is taken into account and computed from the two components $e_1, e_2$ of the lensed ellipticities, with a mean value of $0.25$.

For DC2 detected objects, the photometric redshifts (photo-$z$) were estimated from BPZ~\citep{Benitez_BPZ}. Photo-$z$ point estimates as well as a 1D redshift probability density functions (PDF) are both available in the catalog. In this analysis, we define the source redshift as the mean (i.e., expectation value) of its redshift probability distribution. From the redshift estimates, we select background sources with $z_\mathrm{source}>z_\mathrm{lens}+0.2$ in a circular aperture of radius $R=\SI{10}{Mpc}$, for both cosmoDC2 and DC2object catalogs. The object shapes are estimated using HSM ellipticities,\footnote{The ellipticities have been converted from the $\chi$ to the $\epsilon$ ellipse definition to match the cosmoDC2 convention (see \cite{Schneider_ellipse_def})} as provided in the DC2object catalog (see \cref{subsec:DC2object}). Here we perform a simplified calibration based on the matching between cosmoDC2 galaxies and DC2object detections (see \cref{app:HSM_calib}).

In addition to the calibration of HSM ellipticities, cuts can be applied to the DC2object catalog in order to consider higher quality data. In this work, we refer to these cuts as WL (Weak Lensing) cuts. Details about the different source selection cuts are provided in \cref{app:clean_cuts}. In the following sections, WL cuts are always applied to the background objects when considering DC2object lensing profiles.

\subsubsection{Constructing the stacked lensing profiles}\label{subsec:stacked_lensing_profiles}

In order to recover the individual and stacked lensing profiles from the lensed shapes of background galaxies, we use the DESC Cluster Lensing Mass Modeling code~\cite[CLMM,][]{Aguena_CLMM}. It provides various tools for source-lens system modeling, as well as measurements, estimates and fits of cluster lensing profiles. For each halo, we consider the 10 log-spaced radial bins from \SI{0.5}{Mpc} to \SI{10}{Mpc} and estimate the individual $\Delta\Sigma$ profiles. Due to the noise in measurements of the individual profiles, we use the stacking procedure described in \cref{subsubsec:DeltaSigma_estimates}. To estimate the statistical uncertainties of the stacked $\Delta\Sigma$ profiles, we employ a bootstrap resampling method~\citep[see][]{Simet_bootstrap}. We generate 400 bootstrap realizations by randomly resampling the lensing profiles with replacement. For each realization, we compute the new stacked profile. The error bars are then estimated from the square root of the diagonal elements of the covariance matrix.

For $R<\SI{1}{\mega\parsec}$, the stacked lensing profiles are progressively attenuated in the innermost regions due to the limited resolution of the ray tracing that computes the lensing shear and convergence at each galaxy position~\citep[see][]{Abolfathi_DC2,Kovacs_validatecosmodc2}. Therefore, we only use the $R>\SI{1}{\mega\parsec}$ region for the analysis. In order to recover the corresponding stacked halos mass, we fit the stacked profile using a NFW halo model within the radial range $\SI{1}{\mega\parsec}<R<\SI{3.5}{\mega\parsec}$ to avoid the two-halo term~\citep[i.e. signal contribution from surrounding halos, see][]{Payerne_mass_richness}.

\subsection{Impact of blending on cluster lensing profiles and mass estimates}
\label{subsec:blending_lensing_profiles}

In this section, we investigate the impact of blending on cluster lensing profiles and mass estimates. In order to mitigate this impact, our baseline analysis uses a cut on the blending entropy $S_b<0.2$, defined using the \textit{top-hat} overlap (see \cref{subsubsec:Sb_top_hat}). We present the recovered stacked $\Delta\Sigma$ lensing profiles in \cref{subsubsec:lensing_profiles} and the estimated weak lensing masses in \cref{subsubsec:mass_estimates}. Moreover, complementary tests have been made in order to compare the influence of some cuts or modeling on the $\Delta\Sigma$ lensing profiles. In \cref{subsubsec:varying_Sb}, we study the impact of $S_b<0.1$ and $S_b<0.3$ cuts ---in addition to $S_b<0.2$---on lensing profiles. We also investigate the influence of using a cut on the blendedness ($b<10^{-0.375}$) in comparison with the use of the blending entropy.

\subsubsection{Recovery of the DC2 lensing profiles} \label{subsubsec:lensing_profiles}

\begin{figure*}
\centerline{\includegraphics[scale=0.34]{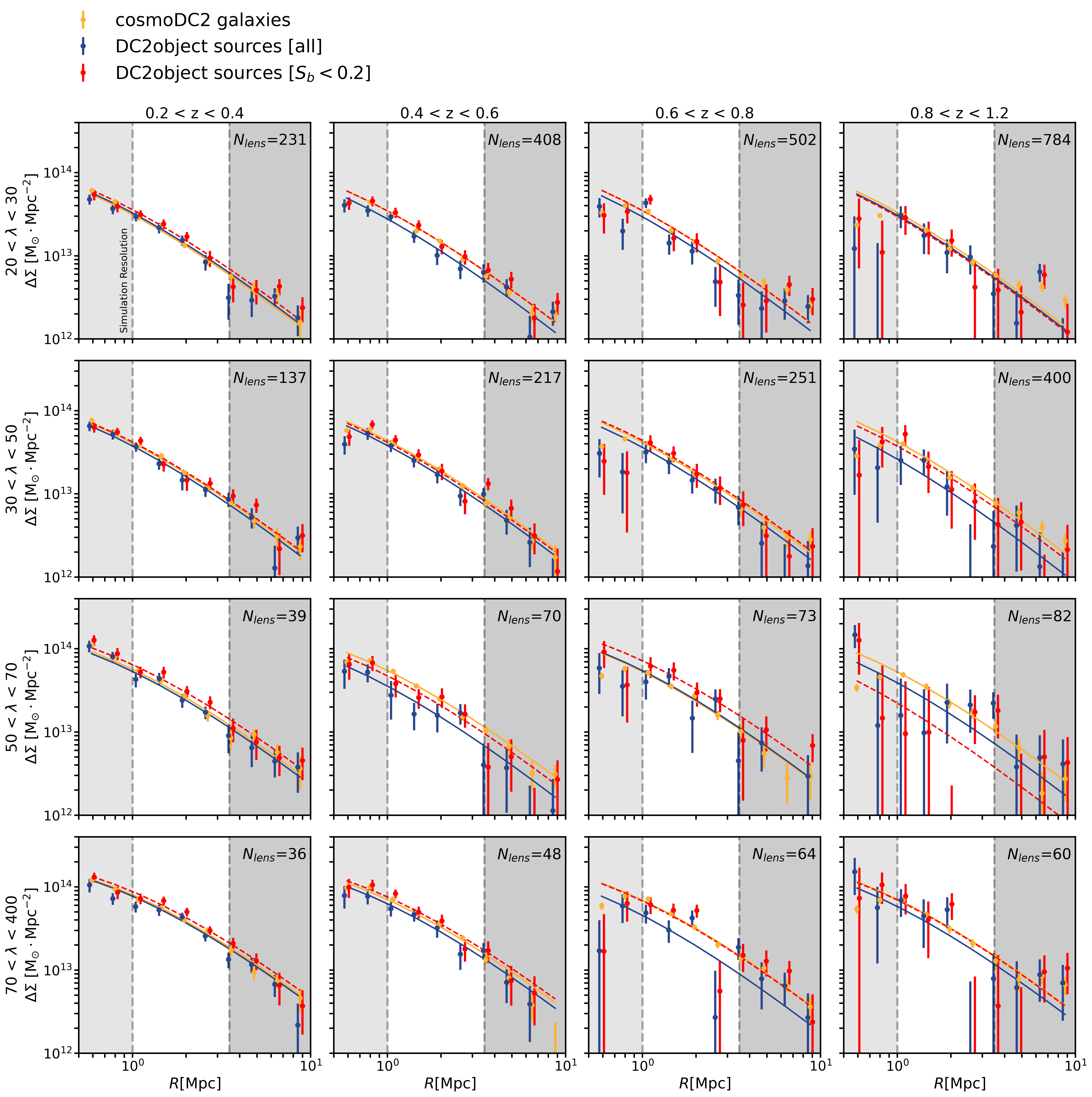}}
\caption{Measured stacked $\Delta\Sigma$ lensing profiles per bins of richness and redshift. The orange profiles are measured from cosmoDC2 background galaxies. The red (respectively blue) profiles are measured with (respectively without) a $0.2$ blending entropy $S_b$ cut on DC2object data. The associated NFW fits are represented with plain lines. The gray left panels indicate the regions ($R<1$ Mpc) where the simulation has not enough resolution to be reliable (see \protect\citealt{Kovacs_validatecosmodc2}). The right ones correspond to the $R>3.5$ Mpc radii, where the two-halo term starts to contribute to the signal. The fits have been performed within $1<R<3.5$ Mpc (i.e. white panel). The number of stacked halos in each bin is indicated for each panel. An offset to the bin points is applied for the readability of the plot.}
\label{fig:binned_profiles}
\end{figure*}

To construct the data vector for the cosmological analysis, we bin the stacked $\Delta\Sigma$ lensing profiles per bins of richness and redshift. In this work, we consider the redshift bin edges $z_i = \{0.2, 0.4, 0.6, 0.8, 1.2\}$
and the richness bin edges $\lambda_i = \{20, 30, 50, 70, 400\}$ for a total of 16 bins. For each bin of \cref{fig:binned_profiles}, we show the stacked lensing profiles either using the cosmoDC2 galaxies (orange) or DC2object detections (blue) as background sources. In addition, we select high-$S_b$ blended objects---using the $0.2$ cut on the blending entropy defined in \cref{subsec:blending_demography}---and remove them from the background sources data (red profile). It allows a direct comparison between DC2object lensing profiles with or without blending. This blending entropy cut induces a $\sim 23\%$ suppression of background DC2object sources (see \cref{tab:percentage_suppressed_objects}). As explained in \cref{{subsec:recover_profiles}}, the DC2object HSM lensed ellipticities have been calibrated beforehand and WL cuts have been applied to improve and boost the measured lensing signal (see \cref{app:HSM_calib} and \cref{app:clean_cuts}).

In the presence of blended sources, the lensing signal is weaker, resulting in a lower amplitude of the DC2object lensing profiles compared to the cosmoDC2 ones. Indeed, blending---by mixing different galaxy ellipticities onto one effective measured ellipticity---tends to smooth out measured shapes and add noise in a direction that is not correlated with the signal, and thereby reducing it. The recovered error bars of the DC2object profiles are also larger compared to the cosmoDC2 ones, due to the lower number of background DC2 detections than input cosmoDC2 galaxies, especially when considering high redshift halos (last column of \cref{fig:binned_profiles}). We are also assuming a \textit{perfect} redshift knowledge as the errors on the BPZ photometric redshift estimates are not taken into account in this work.

Applying a source selection based on the blending entropy (i.e. removing objects with $S_b>0.2$) mitigates the blending-induced bias. The resulting lensing profiles are almost always consistently shifted upwards when removing high-$S_b$ blended background sources, approaching the cosmoDC2 reference profiles. This result suggests that filtering out highly blended sources partially restores the lensing signal, allowing for a more accurate measurement of the stacked profiles and underlying masses. The correction is effective across nearly all richness and redshift bins, demonstrating that blending entropy serves as a valuable metric for identifying and mitigating the effects of blending. 

\subsubsection{Mass estimates} \label{subsubsec:mass_estimates}

\begin{figure*}
\centerline{\includegraphics[scale=0.35]{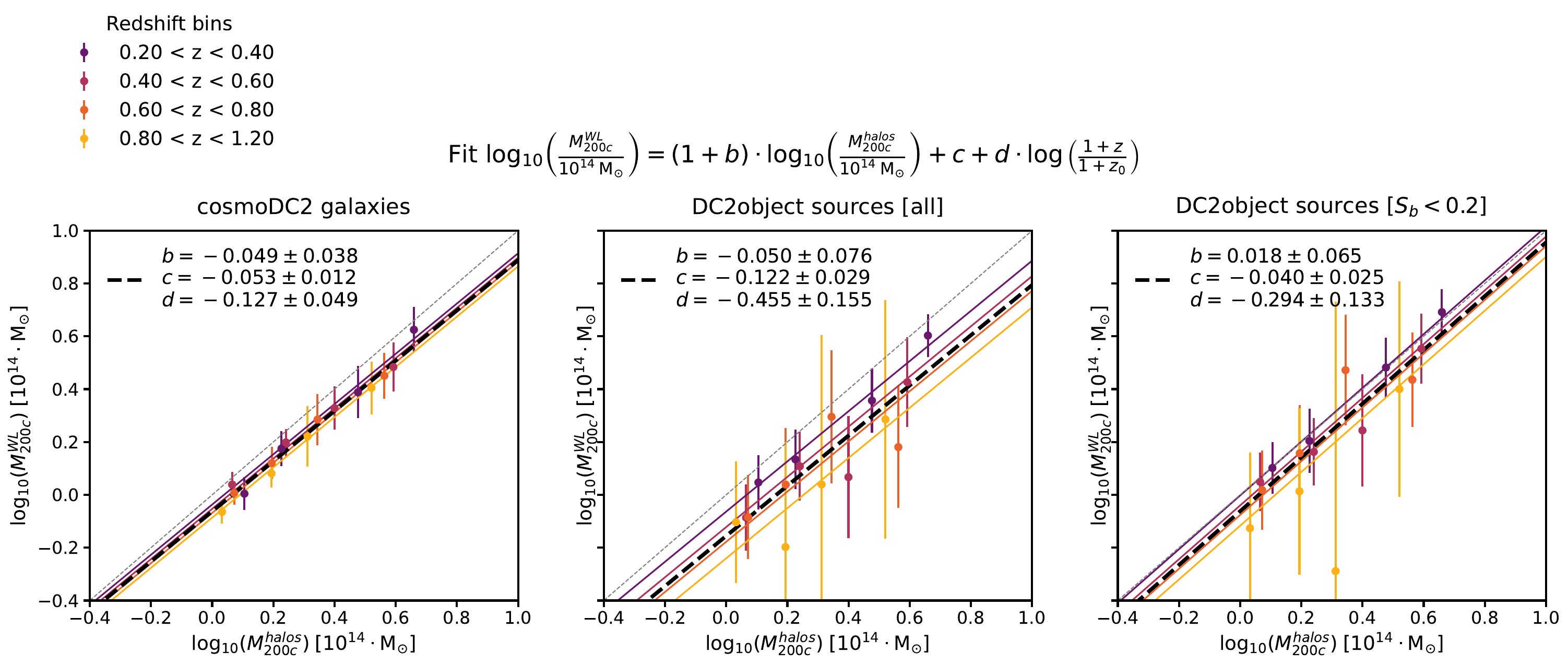}}
\caption{Estimated cosmoDC2 (first panel), DC2object (second panel) and DC2object with $S_b<0.2$ (third panel) stacked weak lensing masses as functions of the mean cosmoDC2 halo masses per bins of richness.
The colors indicate the different redshift bins. The black bold dashed lines correspond to the fits, plotted for the mean halo redshift $z = 0.62$ overall bins. The plain colored lines represent the fits per bins of redshift. The dashed gray thin lines are the $y=x$ lines. The pivot redshift $z_0$ is equal to $0.5$.}
\label{fig:fit_masses}
\end{figure*}

We fit the stacked lensing profiles within the radial range $\SI{1}{\mega\parsec}<R<\SI{3.5}{\mega\parsec}$, using a NFW halo model fit as explained in \cref{subsec:WL_theory}, and fixing the \cite{Duffy_mass_concentration} mass-concentration relation. It allows us to recover estimates of the stacked halos masses for every richness-redshift bin. In \cref{fig:fit_masses}, we show the inferred stacked weak lensing masses $M_{200c}^{\rm WL}$ as functions of the true mean halo masses $M_{200c}^{\rm halo}$ per bins of richness and redshift\footnote{Computed from the individual halo masses $M_{200c}$ within each richness-redshift bin.} and for (1)~cosmoDC2 galaxies, (2)~all DC2object sources and (3)~DC2object sources with $S_b < 0.2$ source selection cases. The errorbars correspond to the statistical errors obtained from the fitting procedure of the halo masses. These errorbars are larger when using DC2object catalog as background sources due to the propagation of errors from the stacked lensing profiles, impacted by the low number of background objects---especially at high redshifts---and the shape measurement. We fit the data using a redshift-dependent model, described by three free parameters $b, c$ and $d$ through the expression:
\begin{equation}
    \log_{10}\qty(\frac{M_{200c}^{\rm WL}}{10^{14}\:\rm{M_{\odot}}}) = (1+b) \cdot \log_{10}\qty(\frac{M_{200c}^{\rm halo}}{10^{14}\:\rm{M_{\odot}}}) + c + d \cdot \log{\qty(\frac{1+z}{1+z_0})},
    \label{eq:fit_masses}
\end{equation}
with the pivot redshift $z_0=0.5$.
When using all DC2object sources (i.e. including those affected by blending; see second panel), the weak lensing mass estimates are biased low compared to the cosmoDC2 halo masses (first panel). This effect is primarily driven by a redshift-dependent bias, with a best-fit value of $d=-0.46 \pm 0.16$, indicating an increasing discrepancy at higher redshift. The multiplicative factors $(1+b)$ remain consistent within uncertainties across the samples, leading to only slight mass-dependent biases.  
After removing high-entropy blended sources ($S_b > 0.2$, third panel), the bias between DC2object and cosmoDC2 mass estimates is partially mitigated. While some residual redshift-dependent bias remains ($d=-0.29 \pm 0.13$ compared to $d=-0.13 \pm 0.05$ in the cosmoDC2 case), the cut improves the agreement between DC2object and cosmoDC2. In addition, the multiplicative factor $(1+b)=1.02 \pm 0.07$ is consistent with unity, indicating that the blending entropy cut reduces the mass-dependent bias and significantly improves the agreement between the weak-lensing mass estimates and the true cosmoDC2 halo masses.
Contrary to our expectations, the weak lensing mass uncertainties decrease on average after applying the blending entropy cut (except for the highest redshift bin where statistics are limited), likely because the cut removes outlier sources while retaining galaxies with higher lensing weights. This trend is consistent with the behavior observed when comparing the \textit{top-hat} and \textit{gaussian} overlap definitions (see \cref{fig:overlap_cosmology}).

Furthermore, we notice that the cosmoDC2 fitted masses are underestimated relative to the true mean halo masses. This is expected because halos with a broad range of masses are stacked within the same richness–redshift bin due to the scatter in the mass–richness relation (see \cref{fig:richness}). The resulting halo mass distribution in each bin is non-Gaussian, so the single-mass NFW fit yields an \textit{effective} lensing mass rather than the mean mass, leading to a bias as shown in \cref{app:halo_mass_distrib} and discussed in \cite{Melchior_DeltaSigma}. However, throughout this paper we will refer to this effective lensing mass simply as the “mean” halo mass.

In addition, the cosmoDC2 dark matter halos are derived from the \textit{Outer Rim} simulation and subsequently modeled using halo mass functions and density profiles that are not native to the simulation, which can introduce further mass biases \citep[see, e.g.,][]{Bahe_mass_bias, Giocoli_mass_bias, Sommer_mass_bias}. Correcting these effects is beyond the scope of this work. Since our analysis focuses on the impact of blending, only the differential bias between cosmoDC2 and DC2object is relevant. \Cref{app:halo_mass_function} shows that our results remain robust to the considered halo mass function \citep[][]{Tinker_HMF, Despali_HMF, Bocquet_HMF}.

\subsubsection{Varying blending entropy cuts} \label{subsubsec:varying_Sb}

\begin{figure}
\centerline{\includegraphics[scale=0.35]{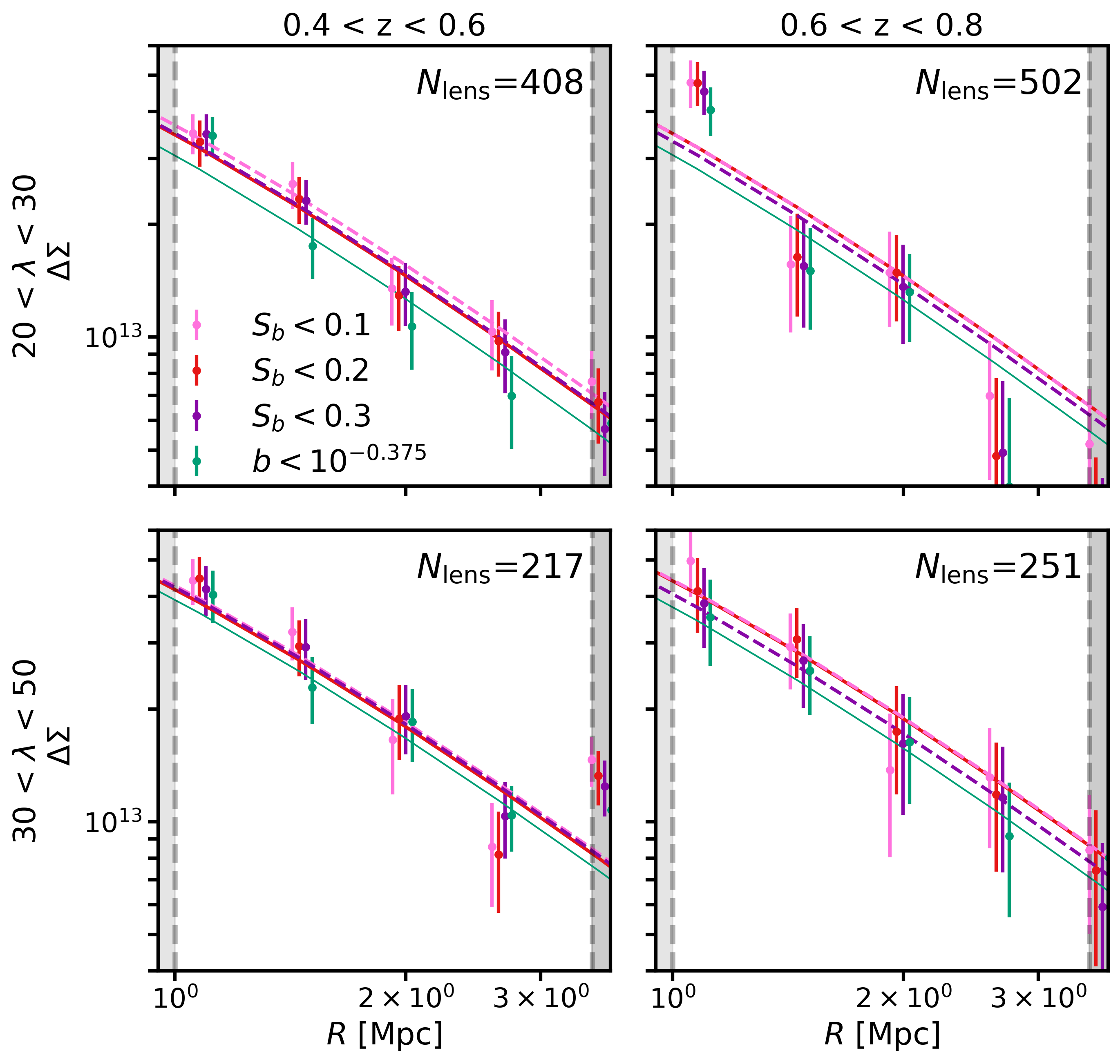}}
\caption{Measured DC2object stacked lensing profiles and associated NFW fits (solid lines) for background objects with blending entropy $S_b<0.1$ (dashed pink), $S_b<0.2$ (red), $S_b<0.3$ (dashed purple) and blendedness $b<10^{-0.375}$ (green) cuts. An offset to the bin points is applied for the readability of the plot. For display purposes we do not show the cosmoDC2 and DC2object profiles, and only show a subset of the richness and redshift bins of \cref{fig:binned_profiles} with high statistics.}
\label{fig:profiles_sb_b}
\end{figure}

In order to study the influence of the chosen blending entropy cut on the source selection and the stacked lensing profiles, we plot in \cref{fig:profiles_sb_b} the resulted DC2object profiles using $S_b<0.1$ (pink) and $S_b<0.3$ (purple) to compare with $S_b<0.2$ (red). We decide to show only four richness-redshift bins $\lambda_i = \{20, 30, 50\}$, $z_i = \{0.4, 0.6, 0.8\}$, with high statistics. Across all panels, we observe that the amplitudes of the lensing profiles for $S_b<0.1$ and $S_b<0.3$ cuts are similar to the $S_b<0.2$ one, therefore compatible with the cosmoDC2 case as shown in \cref{fig:binned_profiles}. Since the amplitudes of the DC2object profiles---when removing high-$S_b$ blended background objects---do not vary significantly in the $0.1-0.3$ range, the choice of the entropy cut is not unique and does not seem to be critical when studying  $\Delta \Sigma$ profiles.

In addition, and since the blendedness $b$ will be an available quantity of the future LSST galaxy catalogs, we also compared the obtained lensing profiles when using the $b < 10^{-0.375}$ cut on the blendedness, fixed in \cite{Mandelbaum_HSC} (see \cref{subsec:metrics}). We show in \cref{fig:profiles_sb_b} the resulted profiles in green. For display purposes we do not plot the DC2object (when considering blended background objects) lensing profiles, shown in \cref{fig:binned_profiles}. However, we observe that using a cut on the blendedness is not as effective at mitigating blends as the $S_b$ cuts since the resulted profiles systematically have lower amplitudes compared to $S_b<0.2$ (in red) case.

\subsection{Impact of blending on cluster cosmology} \label{subsec:blending_cosmo}

In this section, we study and mitigate the impact of blending on $(\Om, \sigma_8)$ cosmological parameters, combining halo number count and weak lensing masses \citep[][]{DES_Y3_cluster}. First, in \cref{subsubsec:cosmo_Om_s8}, we fix the $M-\lambda$ relation. Then, in addition to the cosmological parameters, we estimate the $M-\lambda$ relation parameters (see \cref{subsubsec:cosmo_Mlambda}). In \cref{subsubsec:cosmo_entropy_cuts}, we compare the influence of the chosen blending entropy cuts $(S_b<0.1, \: S_b<0.2$ and $S_b<0.3$) and of the blendedness cut $(b<10^{-0.375})$ on recovered posterior distributions. Finally, in \cref{subsubsec:cosmo_typeofoverlap}, we study the impact of the chosen overlap definition (i.e. \textit{top-hat} or \textit{gaussian}, see \cref{subsubsec:top_hat_gaussian_overlap}).

\subsubsection{Impact on cosmology at fixed $M-\lambda$ relation}
\label{subsubsec:cosmo_Om_s8}

\begin{figure}
\centerline{\includegraphics[scale=0.35]{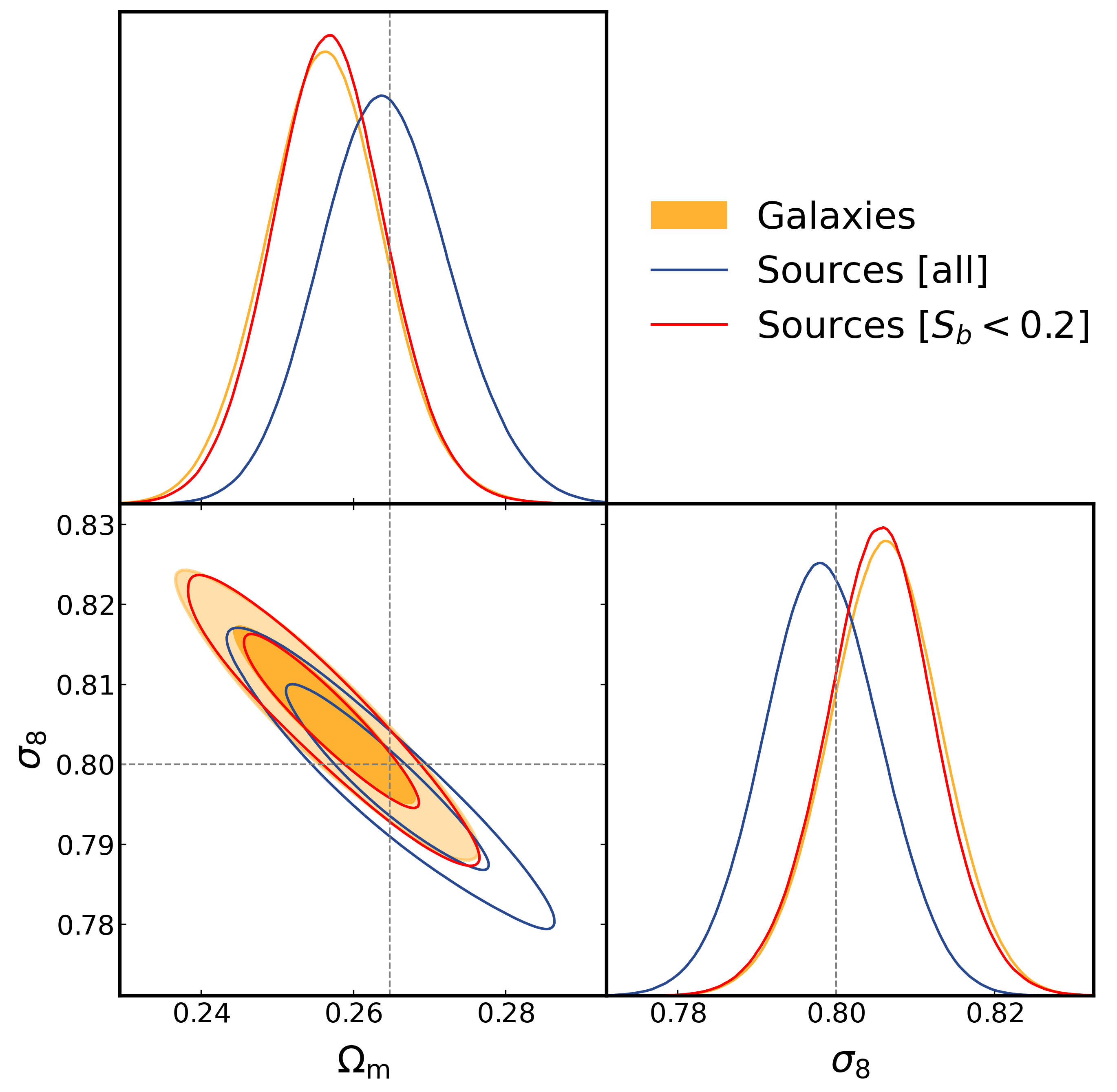}}
\caption{Posterior distributions of $\Om$, $\sigma_8$ with a fixed scaling relation \protect\citep[][]{Payerne_mass_richness} when combining halo abundance and weak lensing inferred masses. The weak lensing mass estimates have been calibrated using the \cref{eq:fit_masses} fit with $b,c,d$ parameters recovered from the comparison between cosmoDC2 weak lensing and halos masses (see \cref{subsubsec:mass_estimates}). The cosmoDC2 contours are plotted in orange. The DC2object contours with (respectively without) a 0.2 cut on the blending entropy are in red (respectively blue). The fiducial values from DC2 simulations are indicated with the dashed gray lines.}
\label{fig:blending_cosmology_Om_s8}
\end{figure}

We consider the inferred weak lensing masses in addition to the halo number count per richness-redshift bins to perform the cosmological analysis. For the theoretical predictions of the halo mass function we use the Core Cosmology Library (CCL, \citealt{Chisari_CCL}). We model the predicted halo number count $N_{ij}$ in the $i$-th richness bin and $j$-th redshift bin using the \cite{Tinker_HMF} halo mass function $\partial^2N(m,z)/\partial z \partial m$ through the expression:
\begin{equation}
    N_{ij} = \int_{\lambda_i}^{\lambda_{i+1}}\dd{\lambda}\int_{z_j}^{z_{j+1}}\dd{z}\int_{m_{\rm min}}^{+\infty}\dd{m} \pdv{N(m,z)}{z}{m} P(\lambda|m,z),
\end{equation}
with $P(\lambda|m,z)$ the fiducial log normal mass-richness scaling relation (see \cref{eq:mass_richness_relation}). We consider a Poisson likelihood, i.e., the number of halos in each richness and redshift bin is the result of a Poisson realization of the underlying cosmological abundance.\footnote{The Poisson likelihood is used as an approximation of the true abundance likelihood which is supposed to be Gauss-Poisson Compound, (GPC) and account for super-sample covariance (SSC), studied in \cite{Payerne_likelihood} and applied in \cite{Payerne_SSC}.}

Similarly, the predicted stacked mass $M_{ij}$ in the $(i,j)$ richness-redshift bin is given by:
\begin{equation}
    M_{ij} = \frac{1}{N_{ij}}\int_{\lambda_i}^{\lambda_{i+1}}\dd{\lambda}\int_{z_j}^{z_{j+1}}\dd{z}\int_{m_{\rm min}}^{+\infty}\dd{m} m \pdv{N(m,z)}{z}{m} P(\lambda|m,z).
\end{equation}
For each redshift-richness bin, we use a (measured) Gaussian likelihood with a diagonal covariance matrix, whose diagonal elements are given by the measured variances of the recovered stacked lensing masses. The halo abundance and weak lensing masses modeling have been performed using the \texttt{CLCosmo\_Sim}\footnote{\url{https://github.com/LSSTDESC/CLCosmo_Sim}} DESC public code.
In order to perform the cosmological analysis and estimate the posteriors, we use the implementation of the Markov Chain Monte Carlo (MCMC) in the \texttt{PyMultinest} \citep{Buchner_pymultinest} package, with \texttt{getdist} \citep{Lewis_getdist} for visualization. We first provide constraints on $\Om$ and $\sigma_8$ cosmological parameters only, fixing the $M-\lambda$ relation parameters to the best-fits, provided in \cref{tab:priors}. 

As discussed in \cref{subsubsec:mass_estimates}, the stacked weak lensing masses, estimated from the stacked lensing profiles in bins of richness and redshift, are biased relative to the mean halo mass within each bin. As a consistency check, and to compare the impact of blending on DC2object background sources, we calibrate the weak lensing masses, $M_{200c}^{\mathrm{WL}}$, by inverting \cref{eq:fit_masses} and applying the best-fit parameters found by comparing cosmoDC2 weak lensing and halos masses (see the left-hand panel of \cref{fig:fit_masses}). These calibrated masses are then used into the cosmoDC2 and DC2object (with and without blending) data vectors, in addition to the halo abundance, to provide unbiased cosmological constraints. In \cref{fig:blending_cosmology_Om_s8}, we present the posterior distributions of $\Om$ and $\sigma_8$. We show the contours for cosmoDC2 (orange), DC2object without a blending entropy cut (blue), and DC2object with a blending entropy cut of $S_b < 0.2$ on background sources (red). 

Using the halo abundance alongside the (calibrated) cosmoDC2 weak lensing masses, we obtain cosmological constraints on $\Omega_{\rm m}$ and $\sigma_8$ that are consistent with the fiducial values from the \textit{Outer Rim} simulation. When using all DC2object background sources, including blended ones, the resulting contours are still compatible with cosmoDC2 within the $1\sigma$ level, although a small shift is observed. However, when removing highly blended sources using a blending entropy cut ($S_b < 0.2$), the constraints from DC2object align more closely with those from cosmoDC2, with an offset of less than $0.1\sigma$. 
While the full DC2object sample appears slightly closer to the fiducial cosmology (likely driven by the $\Omega_m - \sigma_8$ degeneracy combined with residual modeling mismatches between the HMF used in the analysis and the underlying simulation), the sample selected with a blending entropy cut better reproduces the contours from the cosmoDC2 catalog, which is free from blending and benefits from perfect shape and redshift information. This demonstrates the effectiveness of the blending entropy metric in identifying and isolating blended sources, enabling mitigation of their impact on the lensing profiles, mass estimates, and cosmological parameter constraints.

\subsubsection{Impact on cosmology and $M-\lambda$ relation}
\label{subsubsec:cosmo_Mlambda}

\begin{table}
    \centering
    \begin{tabular}{c|c|c}
        Parameters & Fiducial values & Priors\\
        \hline \hline
        $\Om$ & 0.2648 &$\mathcal{U}(0.1, 0.5)$\\
        $\sigma_8$ & 0.8 &$\mathcal{U}(0.5, 1)$\\
        \hline
        $\ln{\lambda_0}$ & 3.35 &$\mathcal{U}(2, 4)$\\
         $\mu_z$ & 0.06 &$\mathcal{U}(-2, 2)$\\
         $\mu_m$ & 2.23 &$\mathcal{U}(0, 4)$ \\
         $\sigma_{\ln{\lambda_0}}$ & 0.56 &$\mathcal{U}(0.2, 1)$\\
        $\sigma_z$ & -0.05 &$\mathcal{U}(-2, 2)$\\
        $\sigma_m$ & 0.10 &$\mathcal{U}(-2, 2)$\\
        \hline
        $m_0$ & $10^{14.3}$ & -\\
        $z_0$ & $0.5$ & -\\
        
    \end{tabular}
    \caption{Fiducial cosmological and mass-richness relation parameter values (see \protect\citealt{Payerne_mass_richness}) and their associated priors.}
    \label{tab:priors}
\end{table}

\begin{figure*}
\centerline{\includegraphics[scale=0.45]{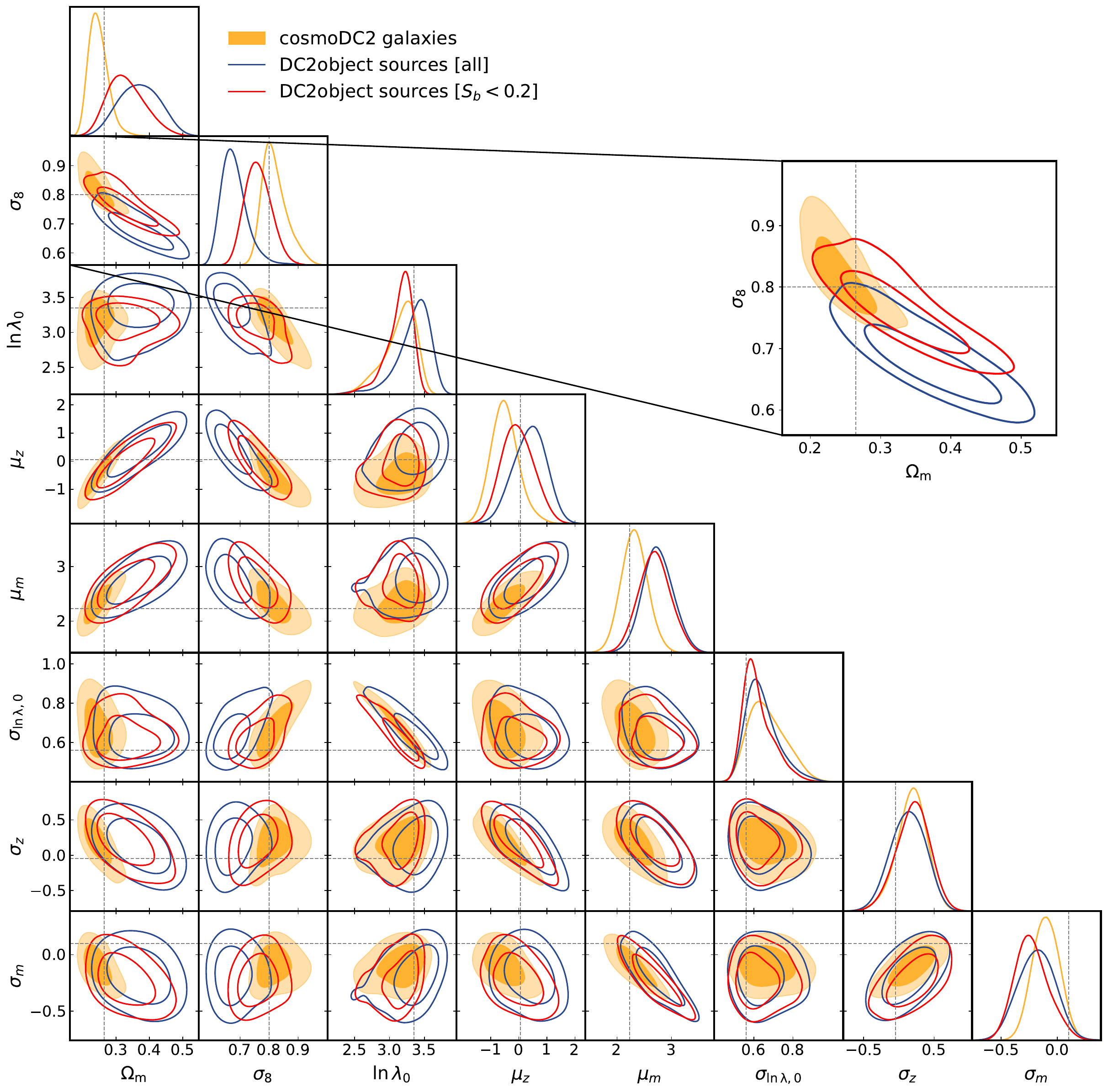}}
\caption{Posterior distribution of $\Om$, $\sigma_8$ and the scaling relation parameters when combining halo abundance and weak lensing inferred masses. The cosmoDC2 contours are plotted in yellow. The DC2object contours with (respectively without) a 0.2 cut on the blending entropy are in red (respectively blue). The fiducial values from DC2 simulations are indicated with the dashed gray lines.}
\label{fig:blending_cosmology}
\end{figure*}

Constraining the mass-richness relation using weak lensing measurements is essential, as the true relation between observable richness and halo mass is not known \textit{a priori}. Weak lensing provides a direct probe of halo mass, allowing us to calibrate this relation observationally and break degeneracies with cosmological parameters~\citep[see, e.g.,][]{Murata_mass_richness, Sunayama_mass_richness, Payerne_mass_richness}.
In this part, we provide constraints on the mass-richness parameters, in addition to $\Om$ and $\sigma_8$ cosmological parameters. We use the log-normal scaling relation of \cref{eq:mass_richness_relation} and the priors used for this analysis are indicated in \cref{tab:priors}. We do not use the mass calibration process explained in \cref{subsubsec:cosmo_Om_s8} as we assume that the bias between mean cosmoDC2 halos and weak lensing estimated masses is absorbed by the uncertainties on
the mass-richness relation. A more accurate correction would require explicitly modeling how the halo mass function and the scatter in the mass–richness relation populate each richness–redshift bin, which we leave for future work.

We show in \cref{fig:blending_cosmology} the 68\% and 95\% confident intervals on the mass-richness scaling relation parameters, as well as on $\Om$ and $\sigma_8$ cosmological parameters using the halo abundance and inferred weak lensing masses jointly. We plot the contours for cosmoDC2 (orange), DC2object with (red) and without (blue) a blending entropy cut $S_b<0.2$.

For the mass-richness relation parameters ($\ln{\lambda_0}$, $\mu_z$, $\mu_m$), we observe that the DC2object case deviates significantly from cosmoDC2, indicating that blending may affect the inferred mass-richness relation, essentially due to underestimates of the weak lensing masses (see \cref{subsec:blending_lensing_profiles}). The constraints on scatter parameters ($\sigma_{\ln{\lambda},0}$, $\sigma_z$, $\sigma_0$) also broaden when blending is included. Applying a $0.2$ cut on the blending entropy brings the mass-richness relation parameters closer to their cosmoDC2 estimates, particularly for $\ln{\lambda_0}$. It indicates that removing high-$S_b$ blends reduces the bias in cluster mass calibration. It also reduces the uncertainty on the scatter parameters, improving parameter constraints.

Concerning the $\Om$ and $\sigma_8$ cosmological parameters, we show that the cosmoDC2 contours recover the fiducial values. The DC2object contours are shifted toward higher $\Om$---indicating a $2 \sigma$ tension---and lower $\sigma_8$ values---with a $3\sigma$ tension---compared to cosmoDC2. This leads to a $\sim3\sigma$ tension between the cosmoDC2 and DC2object datasets. The entropy cut partially corrects this bias, especially on the $\sigma_8$ estimates and reduces the bias to $\sim1.3\sigma$, for a global tension of $1.6\sigma$. Removing the high-$S_b$ blends reduces but does not completely eliminate the bias.
We also notice that the DC2object contours are larger compared to the cosmoDC2 ones, indicating less precise mass measurements due to noisier lensing profiles (see \cref{subsec:blending_lensing_profiles}). The errors on $\Omega_m$ are multiplied by a factor of $\sim 2$ when considering DC2object background sources (with or without the entropy cut), essentially due to lower background source statistics. The larger contours can also be attributed to uncertainties in the DC2object photometric redshifts, in the HSM shape measurements and their calibration, and in the assumed halo mass function and mass–observable scaling relations. In particular, the BPZ photometric redshifts in DC2object are expected to be noisier than the photometric redshift performance anticipated for LSST, which amplifies the statistical uncertainties on the inferred cluster lensing signal and halo masses. However, this effect is mitigated in our analysis, since we do not explicitly model or propagate individual photo-$z$ error distributions.
It should be noticed here that we use the \cite{Tinker_HMF} halo mass function for this analysis. In \cref{app:halo_mass_function}, we compare the results on $(\Om, \sigma_8)$ contours for the \cite{Tinker_HMF}, \cite{Bocquet_HMF} and \cite{Despali_HMF} halo mass functions. We show that these three different halo mass functions give similar results.

\subsubsection{Varying blending entropy cuts} \label{subsubsec:cosmo_entropy_cuts}

\begin{figure}
\centerline{\includegraphics[scale=0.35]{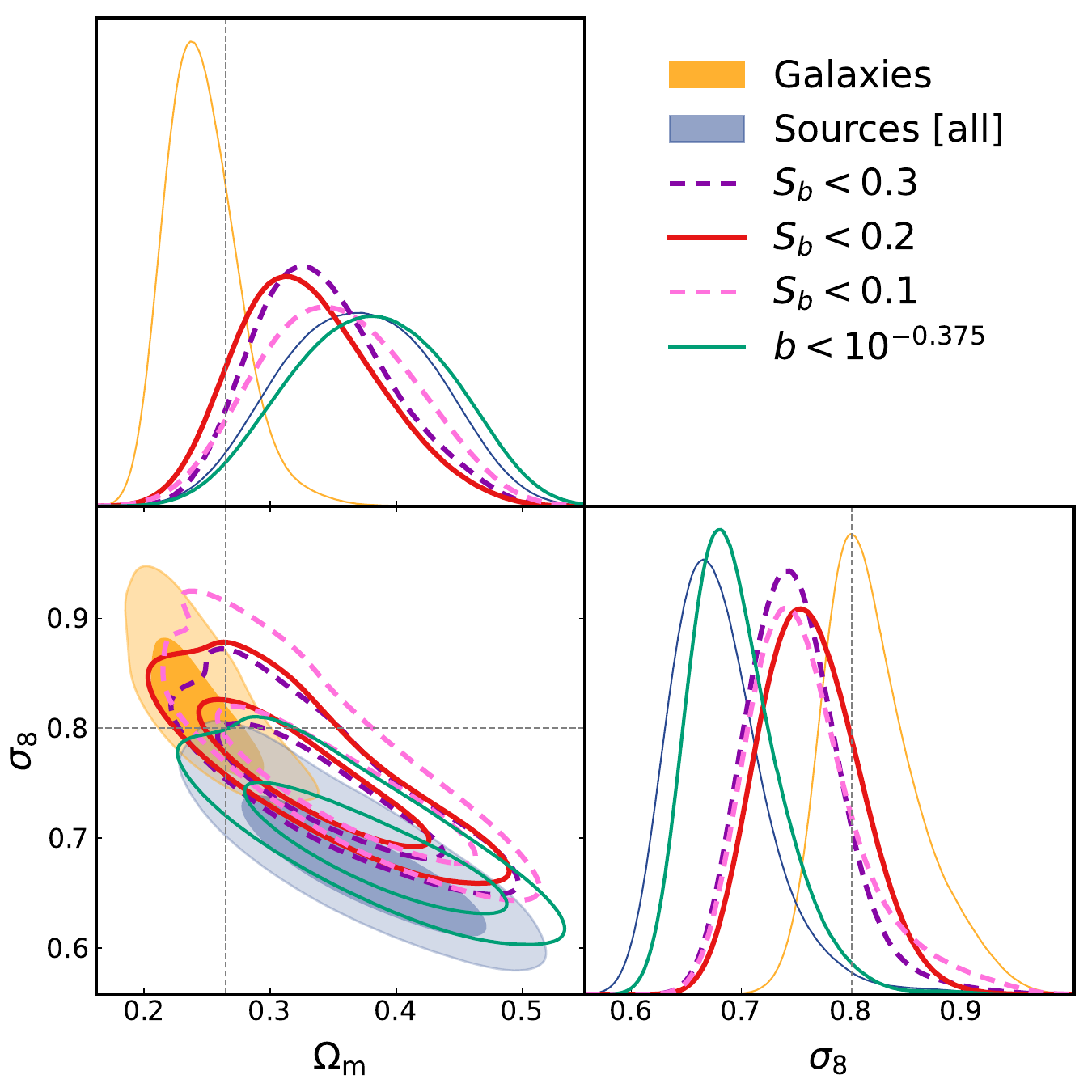}}
\caption{Posterior distributions of $\Om$ and $\sigma_8$ depending on different blending metric cuts and using the halo number count and weak lensing masses jointly. The cosmoDC2 contour is plotted in plain yellow. The DC2object contour without any cut on the blending effect is represented in plain blue. The DC2object contours using $S_b<0.1$, $S_b<0.2$ and $S_b<0.3$ are in pink, red and purple. The DC2object posterior distribution using a cut $b<10^{-0.375}$ on the blendedness is shown in green. The fiducial values from DC2 simulations are indicated with the dashed gray lines.}
\label{fig:sb_blendedness_cosmology}
\end{figure}

In addition to the use of $S_b<0.2$ cut to mitigate blended groups, we also compare the $(\Om, \sigma_8)$ estimates when using different cuts on the blending entropy, as well as on the blendedness. The resulted proportions of suppressed background sources using the different cuts are summarized in \cref{tab:percentage_suppressed_objects}. The results are shown in \cref{fig:sb_blendedness_cosmology}. We plot the DC2object $(\Om, \sigma_8)$ posterior distribution for $S_b<0.1$ (pink), $S_b<0.2$ (red), $S_b<0.3$ (purple) and $b<10^{-0.375}$ (green). As a comparison, we also plot the contours in the context of cosmoDC2 (plain yellow) and DC2object without blending metric cut (plain blue). First, we notice larger contours for $S_b<0.1$ due to a more severe cut leading to the suppression of a large number of background sources  $(\sim30\%)$ as illustrated in \cref{tab:percentage_suppressed_objects}, as well as larger errorbars for the $\Om$ parameter estimate. The use of a $0.1$ or $0.3$ cut on the blending entropy does not significantly improve the constraints compared to the $0.2$ case with a $\sim 1.6\sigma$ tension, showing the best cut value is not fixed in this range. However, when looking at the blendedness $b<10^{-0.375}$ contours, we observe that the tension between DC2object and cosmoDC2 data is still not corrected, with a bias of $\sim3\sigma$ between the two datasets. This demonstrates that the use of a blendedness cut may not be sufficient in identifying blended groups, especially unrecognized blends, and therefore fails to mitigate the impact of highly blended groups on cluster lensing cosmology and $\Om$, $\sigma_8$ parameter estimates. It emphasizes the need for a new blending metric, based on more information through the use of a reference dataset.

\subsubsection{Comparison between top-hat and gaussian overlaps} \label{subsubsec:cosmo_typeofoverlap}

\begin{figure}
\centerline{\includegraphics[scale=0.35]{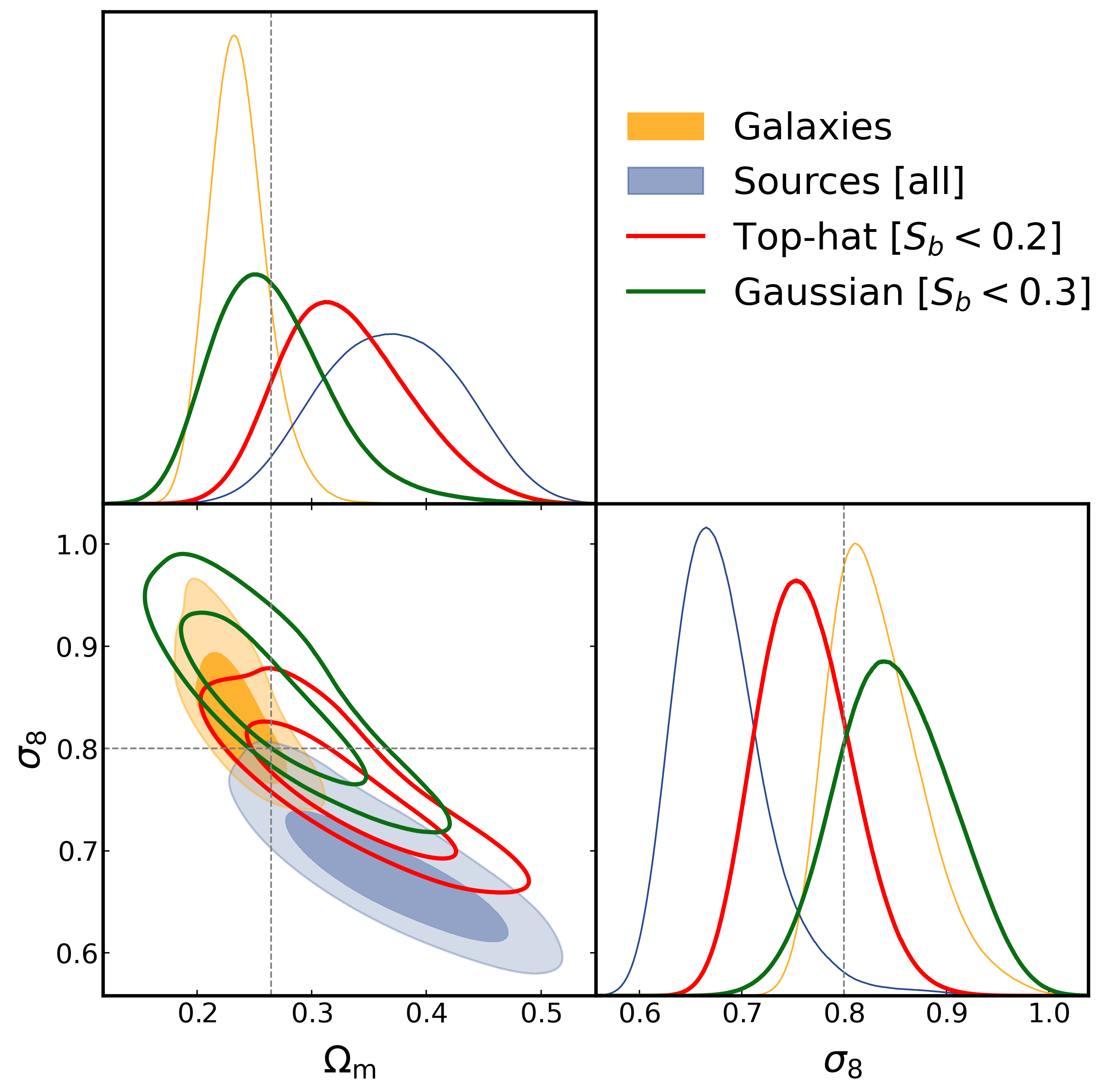}}
\caption{Posterior distributions of $\Om$ and $\sigma_8$ depending on the overlap definition and the associated $S_b$ cut, using the halo number count and weak lensing masses jointly. The cosmoDC2 contour is plotted in plain yellow. The DC2object contour without any cut on the blending effect is represented in plain blue. The DC2object contour using the \textit{top-hat} overlap with $S_b<0.2$ is in red. The one using the \textit{gaussian} overlap with $S_b<0.3$ is in green. The fiducial values from DC2 simulations are indicated with the dashed gray lines.}
\label{fig:overlap_cosmology}
\end{figure}

We can also compare the influence of the overlap used for the $S_b$ computation---\textit{top-hat} or \textit{gaussian}---on cosmological parameter estimates. In \cref{fig:overlap_cosmology} we compare the resulting posterior distributions of $\Om$ and $\sigma_8$ for the \textit{top-hat} (in red) and the \textit{gaussian} overlap (in green). We apply a cut of $S_b<0.2$ (respectively $S_b<0.3$) on the blending entropy defined with the \textit{top-hat} (respectively \textit{gaussian}) overlap (see \cref{subsec:blending_demography}). 
The tension between cosmoDC2 and the DC2object contours---using a \textit{top-hat} overlap definition---is reduced from $1.6\sigma$ to $0.54\sigma$ when considering the \textit{gaussian} definition. This suggests that both overlaps lead to mitigation of blending in the context of cluster lensing, with better consistency with the cosmoDC2 dataset for a blending entropy defined with the \textit{gaussian} overlap fraction. This definition also allows us to significantly reduce the bias on the $\Om$ parameter, (from $1.5\sigma$ for the \textit{top-hat} to $0.36\sigma$ for the \textit{gaussian}), as well as on $\sigma_8$. To obtain these results, a large fraction of DC2object background sources is suppressed ($\sim 62\%$, see \cref{tab:percentage_suppressed_objects}) compared to the \textit{top-hat} overlap definition, with only $\sim 23\%$ suppression. 

However, the size of the DC2object contours is only modestly affected: the \textit{gaussian} contours are visibly broader than the \textit{top‑hat} ones, but less than one might naively expect. From the effective source numbers (about 78\% retained with the \textit{top‑hat} cut and 39\% with the \textit{gaussian} cut) we would expect roughly 40\% larger statistical errors for the \textit{gaussian} selection, whereas we actually observe only about a $\sim 20\%$ increase in the width of the $\sigma_8$ posterior. This softer degradation is consistent with the fact that we simultaneously fit the mass-richness relation parameters, so the contour widths do not follow a simple scaling with the effective number of sources. In addition, the sources rejected only by the \textit{gaussian} cut have a lower mean weight than those kept by the \textit{top‑hat} cut, demonstrating that this selection preferentially removes low‑weight, lower‑quality sources.


\section{Conclusions} \label{sec:Conclusions}

Blending poses a significant challenge for future galaxy surveys, especially for ground-based observatories like the Rubin Observatory’s LSST, where atmospheric effects exacerbate source confusion on top of the higher galaxy density compared to previous surveys. The difficulty arises from blending’s dual impact: it not only biases measurements of galaxy properties---such as shapes and photometric redshifts---but also impairs basic source detection, leading to unrecognized blends. Consequently, studies that rely on per-object errors using one-to-one matching between detected objects and true galaxies fail to capture the full scope of the problem.

In this work, we address this limitation by introducing a novel probabilistic framework for characterizing blending when reference information is available. Our approach combines a flexible matching procedure---incorporating spatial positions, fluxes, and shapes---with a new metric, \textit{blending entropy}, which quantifies the ambiguity in associating detected objects with true galaxies. Unlike binary or heuristic indicators, blending entropy provides a continuous, probabilistically grounded measure of blending severity. This enables us to trace the effects of blending from catalog-level systematics through cosmological parameter inference.

We validate our framework using LSST DESC DC2 simulations by comparing the input cosmoDC2 catalog to the detected-object DC2object catalog. Our analysis explores the trade-off between accurately removing the most blended objects and the consequent loss of statistical power. We find that applying a blending entropy cut of 0.2 results in the removal of approximately 25\% of detected objects, at an $i$-band magnitude limit of 26.5. This $S_b$ cut effectively identifies and removes problematic blends, which are associated with higher shape errors and photometric redshift outliers, revealing that about 35\% of the remaining objects near the magnitude limit are unrecognized blends. Our metric complements the LSST pipeline’s blendedness parameter, which is computed solely from observed data and often fails to flag high-entropy blends near the survey’s depth limit. 

To demonstrate the cosmological relevance, we apply our methodology to a simplified cluster lensing analysis. Using cosmoDC2 halos and assuming a fiducial mass–richness relation, we show that blending systematically biases lensing profiles derived from the DC2object catalog---typically lowering them by injecting noise-like ellipticities misaligned with the lensing shear. This leads to underestimated halo masses and biased cosmological constraints, notably reducing the inferred value of $\sigma_8$ and introducing a >2$\sigma$ tension between cosmoDC2- and DC2object-based analyses. Removing high-entropy objects largely corrects these biases within statistical uncertainties. These improvements are robust across different overlap models (\textit{top-hat} and \textit{gaussian}) and entropy cuts. While this filtering reduces statistical power, it lays the foundation for more refined mitigation strategies enabled by blending entropy. 

We acknowledge that our results depend on simulation specifics (DC2 corresponds to LSST five-year depth) and pipeline versions---including detection, deblending, PSF modeling, shape and flux estimation, and photometric redshift calibration. Our cosmological analysis further relies on BPZ photometric redshift estimates without explicitly modeling individual photo-$z$ error distributions. Nevertheless, our conclusions are robust against magnitude cuts, variations in entropy computation, and modeling choices. Moreover, deblenders will remain limited by unrecognized blends, which typically have high blending entropy.

Importantly, our framework is general and readily extensible to other cosmological probes impacted by blending, such as cosmic shear, galaxy–galaxy lensing, and clustering. In addition, we focused our analysis on lensing measurements, but expect that blending will also bias cluster detection and therefore constraints derived from their abundance. Future work will explore the propagation of blending effects through full analysis pipelines and the design of optimal mitigation strategies for all static cosmic probes.

Finally, the probabilistic nature of our method makes it particularly well suited for integration with real observational data, especially when high-resolution space-based imaging is available for reference. By using LSST data alone in future work, we will identify which combinations of observables best correlate with blending entropy, enabling the design of optimized, data-driven selection cuts or weighting schemes. Alternatively, LSST will overlap significantly with datasets from the Hubble Space Telescope \citep[][]{Piotto_HST}, James Webb Space Telescope \citep[][]{McElwain_JWST}, \textit{Euclid}, and the Nancy Grace Roman Space Telescope, offering valuable opportunities for cross-matching. To effectively operate across diverse optical and infrared photometric bands---and to account for real-world complexities and data quality variations---our method will require targeted adaptation. However, preliminary tests with Rubin Data Preview 1 \footnote{\citet[][]{LSST_DP1_data}}, public \textit{Euclid} \citep[][]{Aussel_Euclid_Q1} and Hubble \citep[][]{Guo_CANDELS_HST} data show promising results in identifying unrecognized blends, mostly limited by the depth of the reference data set. We plan to develop and calibrate mitigation strategies in deep fields, where space-based observations can be treated as effectively noise-free, allowing us to refine and extend these techniques across the full LSST footprint. This approach will not only improve the accuracy of blending characterization but also enhance the overall scientific return of the next generation of surveys.

Our implementation is publicly available at: \url{https://github.com/LSSTDESC/friendly}.




\section*{Authors' contributions}

MR: led the software development, investigation on DC2 data, methodology, validation, visualization, and writing and editing.
CD: led the conceptualization and methodology, co-supervision, and writing and editing.
MK: led the methodology, co-supervision, and writing and editing.
MA: contributed to the development of CLMM and provided a review of the manuscript, helped with editing and analysis.
CC: contributed to CLMM, provided feedback on the cluster analysis, temporary supervision, and editing.
SL: contributed to the conceptualization, initial software development, and editing.
CP: contributed to CLMM and \texttt{CLCosmo\_Sim}, provided feedback on the cluster analysis, and assisted with editing.
CA: made significant contributions to the CLMM software.
AM: contributed to the conceptualization and initial software development.
MR: contributed to the development of CLMM, provided  feedback on the cluster modeling and analysis.
NS: contributed to the development, maintenance, testing, and validation of CCL.

\section*{Acknowledgements}

The DESC acknowledges ongoing support from the Institut National de Physique Nucléaire et de Physique des Particules in France; the Science \& Technology Facilities Council in the United Kingdom; and the Department of Energy and the LSST Discovery Alliance in the United States. DESC uses resources of the IN2P3 Computing Center (CC-IN2P3 Lyon/Villeurbanne - France) funded by the Centre National de la Recherche Scientifique; the National Energy Research Scientific Computing Center, a DOE Office of Science User Facility supported by the Office of Science of the U.S. Department of Energy under Contract No. DE-AC02-05CH11231; STFC DiRAC HPC Facilities, funded by UK BEIS National E-infrastructure capital grants; and the UK particle physic grid, supported by the GridPP Collaboration. This work was performed in part under DOE Contract DE-AC02-76SF00515.

This paper has undergone internal review in the LSST Dark Energy Science Collaboration. The authors thank the internal reviewers Michel Aguena and Shuang Liang for their valuable comments.
The authors thank Ismael Mendoza, and Prakruth Adari for providing feedback and helping the project inception.

\bibliographystyle{apsrev4-2}
\bibliography{example}

\appendix

\section{Ellipse formalism}
\label{app:ellipse}

Ellipses can be described by their general quadric ellipse equation in the $(x,y)$ Cartesian plane:
\begin{equation}
    {\displaystyle Ax^2+Bxy+Cy^2+Dx+Ey+F=0},
\end{equation}
where
\begin{align}
    A &= a^2\sin^2{\theta} + b^2\cos^2{\theta}, \\
    B &= a^2\cos^2{\theta} + b^2\sin^2{\theta}, \\
    C &= 2(b^2 - a^2)\sin{\theta}\cos{\theta}, \\
    D &= -2Ax_0 - 2By_0, \\
    E &= -Cx_0 - 2By_0, \\
    F &= Ax_0^2 + Cx_0y_0 + By_0^2 - a^2b^2.
\end{align}
$x_0, y_0$ are the coordinates of the ellipse center, $a, b$ the semi-major/minor axis and $\theta$ the orientation angle. From the $A, B, C, D, E, F$ factors, ellipses describing the shapes of galaxies and objects can be plotted and analyzed in order to properly study blending.

For objects, the $a, b, \theta$ parameters are computed from the second moments $I_{xx}, I_{yy}$ and $I_{xy}$ of the luminous flux in \textit{i}-band using a 2D Gaussian flux distribution defined as follows:
\begin{equation}
     G(\textbf{r}) = \frac{F}{2\pi\sqrt{\abs{\Sigma}}}\exp(-\frac{\textbf{r}\cdot\Sigma^{-1}\cdot \textbf{r}}{2}),
\end{equation}
where $F$ is the amplitude scaling parameter and:
\begin{equation}
     \Sigma = \begin{bmatrix}I_{xx} & I_{xy}\\I_{xy} & I_{yy}\end{bmatrix} \quad\mathrm{and}\quad
     \Sigma^{-1} = \frac{1}{I_{xx}I_{yy}-I_{xy}^2}\begin{bmatrix}
         I_{yy} & -I_{xy} \\ -I_{xy} & I_{xx}
     \end{bmatrix}
\end{equation}
The relations between $a, b, \theta$ and $I_{xx}, I_{yy}, I_{xy}$ are given by:
\begin{align}
    &a^2 = \frac{2\ln{2}(I_{xx}\cos{\theta}^2-I_{yy}\sin{\theta}^2)}{\cos{2\theta}}, \\
    &b^2 = \frac{2\ln{2}(I_{yy}\cos{\theta}^2-I_{xx}\sin{\theta}^2)}{\cos{2\theta}}, \\
    &\tan{2\theta} = \frac{2I_{xy}}{I_{xx}-I_{yy}}.
\end{align}

\section{Blending entropy from \textit{gaussian} overlap definition}
\label{app:Sb_gaussian_distrib}

\begin{figure*}
\centerline{\includegraphics[scale=0.35]{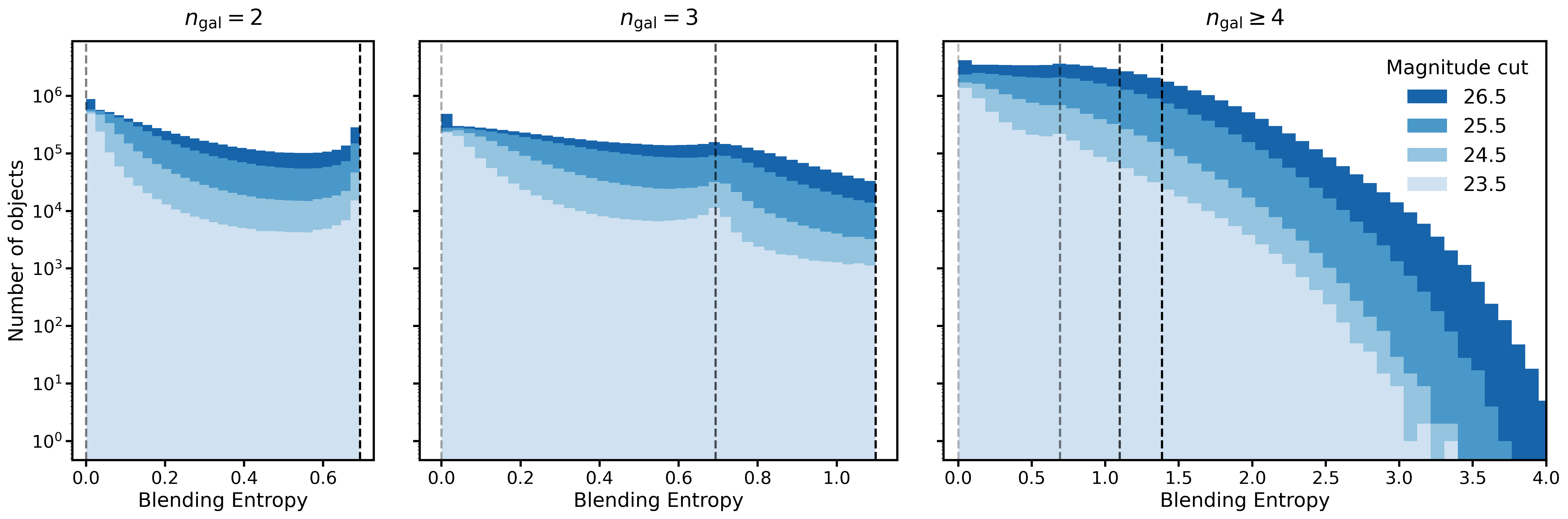}}
\caption{Distributions of the blending entropy (computed from the \textit{gaussian} overlap) for groups composed of $\n$ galaxies where $\n=2$ (first plot), $\n=3$ (second plot) and $\n \geq 4$ (third plot). The different colors represent the cut on magnitude (in $i$-band) applied to the DC2object catalog. The dashed lines indicate the maximum values of the entropy $S_{b,\rm{max}}=\ln{n}$ for groups composed of $n=1,\dots,\n$ galaxies.} 
\label{fig:Sb_mag_cut_gaussian}
\end{figure*}

\begin{figure}
\centerline{\includegraphics[scale=0.35]{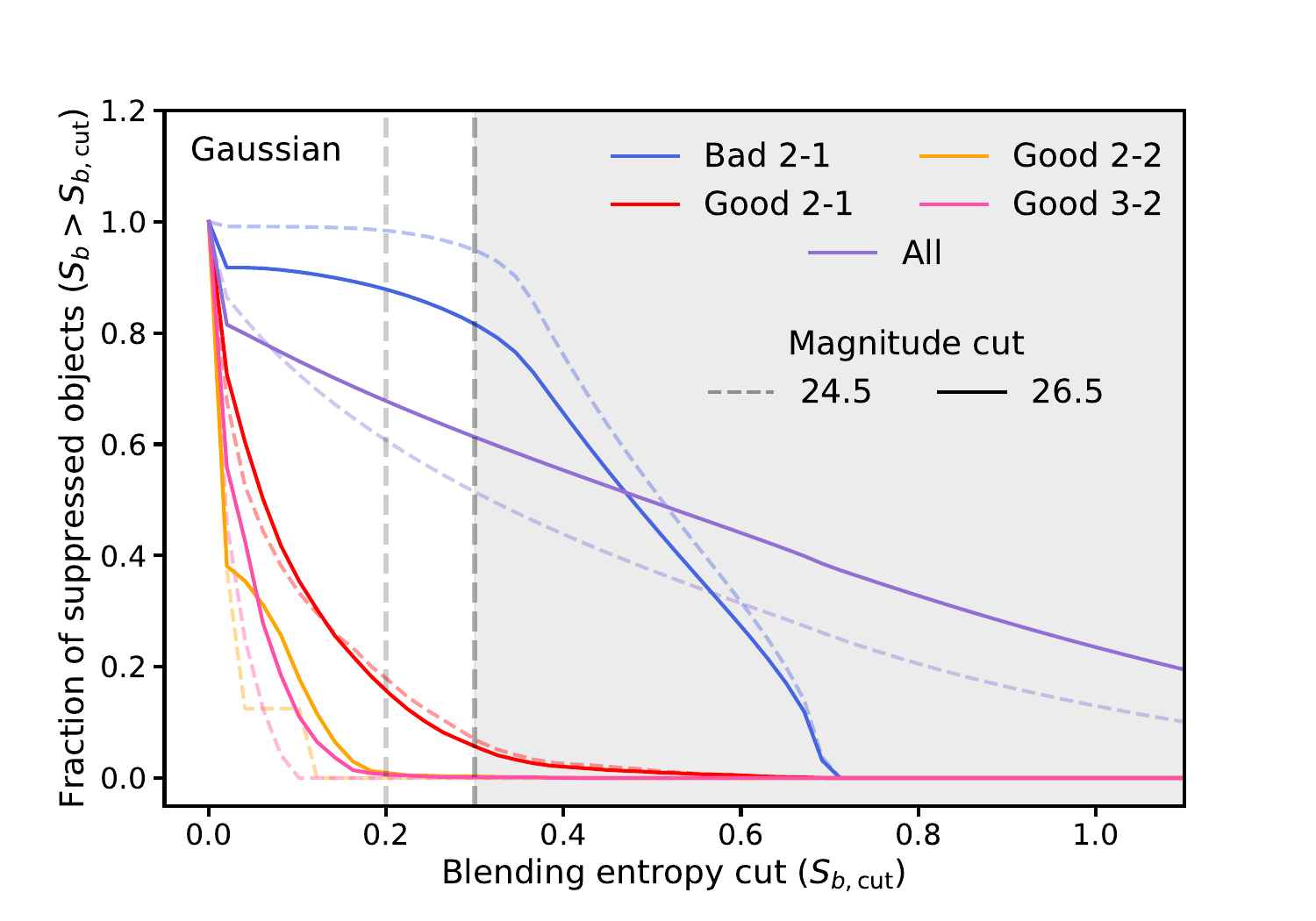}}
\caption{Fractions of suppressed objects belonging to \textit{bad} $2-1$ (blue) and \textit{good} $2-1$ (red), $2-2$ (orange) $3-2$ (pink), and all (purple) groups according to the $S_{b,\ \rm{cut}}$ cut on their blending entropies ($S_b$). The curves have been plotted for two DC2 limiting magnitudes in \textit{i}-band (solid and dashed curves). The dashed vertical lines represents respectively the cut of $S_b = 0.2$ (light line, adopted in \cref{subsubsec:Sb_top_hat}) and $S_b = 0.3$ (dark line) for the separation of \textit{bad} and \textit{good} blended objects. The blending entropy has been computed using the \textit{gaussian} overlap definition.}
\label{fig:Sb_good_bad_gaussian}
\end{figure}

Using 2D Gaussians representation of galaxies instead of ellipses, we can recover the \textit{gaussian} overlap fraction (see \cref{subsubsec:top_hat_gaussian_overlap}), defining the blending entropy metric ($S_b$). In \cref{fig:Sb_mag_cut_gaussian}, we plot the distributions of the blending entropy (computed from the \textit{gaussian} overlap) for groups composed of $\n$ galaxies where $\n=2$ (first plot), $\n=3$ (second plot) and $\n \geq 4$ (third plot). We represent through different colors the obtained distributions using a \textit{i}-band magnitude cut on DC2object catalog. The dashed lines indicate the maximum values of the entropy $S_{b,\rm{max}}=\ln{n}$ for groups composed of $n=1,\dots,\n$ galaxies. We observed a smaller range of values for the entropy, compared to the \textit{top-hat} overlap case (see \cref{fig:Sb_mag_cut}). Moreover, the distributions are less skewed around low values of the entropy, and the observed peaks representing maximum values of the entropy for the different groups (dashed lines) are less pronounced. This indicates that a possible $S_b$ cut to isolate highly blended objects may be less obvious and defined compared to the \textit{top-hat} definition of the overlap fraction.

In \cref{fig:Sb_good_bad_gaussian} we show the fractions of suppressed objects belonging to \textit{bad} 2–1 (blue), and \textit{good} 2–1 (red), 2–2 (orange), and 3–2 (pink) groups, as well as all groups combined (purple), as a function of the cut $S_{b,\ \rm{cut}}$ applied to their blending entropies $S_b$, using the \textit{gaussian} overlap. Compared to \cref{fig:Sb_good_bad}, which uses the \textit{top-hat} overlap definition, \cref{fig:Sb_good_bad_gaussian} displays similar trends in the suppression of \textit{bad} and \textit{good} blended objects, but with notable differences in sharpness and separation power. With the \textit{gaussian} overlap definition, the drop in the suppression fraction of \textit{bad} $2-1$ objects (blue curve) is smoother, and the separation between \textit{bad} and \textit{good} objects is less pronounced. For instance, at $S_b = 0.2$ (light dashed line), the suppression of \textit{bad} objects is no longer near unity, and $\sim 20\%$ of \textit{good} $2-1$ groups are suppressed.
Furthermore, the \textit{good} group curves (red, orange, pink) decline less steeply with the \textit{gaussian} definition, leading to more overlap between \textit{bad} and \textit{good} populations in the $S_b$ distribution. This makes the selection of an optimal cut more ambiguous. The overall suppression (purple curve) is also less sensitive to the cut, indicating that the \textit{gaussian} overlap definition results in blending entropy distributions that are less discriminative. 
Moreover, with this definition, the difference between the two magnitude cuts ($24.5$---dashed curves and $26.5$---solid curves) is more gradual and less steep than in the \textit{top-hat} case. For example, at $S_b = 0.2$, the difference in the suppression of all objects (purple curves) between the two magnitude cuts is smaller than in \cref{fig:Sb_good_bad}, indicating that the \textit{gaussian} definition is less sensitive to changes in source depth. A larger fraction of \textit{bad} $2-1$ groups is suppressed when considering brighter objects.
With this overlap definition, we fix $S_b>0.3$ in order to identify and select problematic blends, resulting in $\sim 61\%$ of suppressed objects (see \cref{tab:percentage_suppressed_objects}).

\section{HSM shapes calibration}
\label{app:HSM_calib}

\begin{figure*}
\centerline{\includegraphics[scale=0.42]{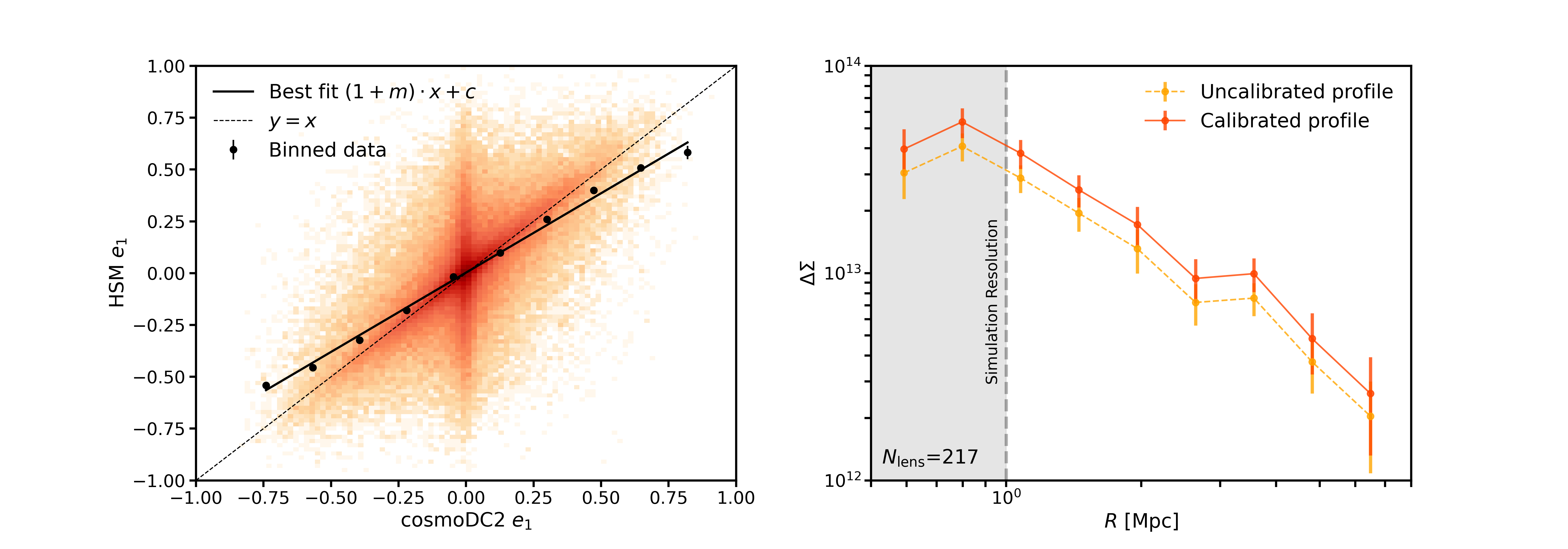}}
\caption{\textit{Left:} observed HSM $e_1$ ellipticities vs. cosmoDC2 true ellipticities. The 1-to-1 matching between objects and galaxies have been realized by selecting pairs with relative probabilities of matching greater than 0.8. The black dots represent the binned data. The applied linear fit is represented by the plain black line. The final fit has been done jointly on $e_1$ and $e_2$ ellipticities. \textit{Right:} stacked $\Delta\Sigma$ profiles using either background DC2 objects with HSM calibrated (dark orange) or uncalibrated (light orange) ellipticities.}
\label{fig:HSM_ell}
\end{figure*}

To calibrate the HSM ellipticities \citep[][]{Mandelbaum_HSM_2}, we perform a simplified calibration based on matching between cosmoDC2 galaxies and DC2object detection. We select pairs of galaxies and objects based on their relative probability of matching (see \cref{subsec:blending_entropy}). We impose this probability to be greater than 0.8, leading to at least 80\% chance for a specific object to match its true galaxy. This allows to select the $1-1$ groups but also the \textit{good} blended objects (see \cref{subsec:blending_demography}). On the left plot of \cref{fig:HSM_ell}, we show the object $e_1$ ellipticity components recovered by HSM in comparison with their matched true (lensed) shapes from cosmoDC2. The procedure is the same for the $e_2$ component. By performing a joint linear fit $e_i^\mathrm{HSM} = (1+m)e_i^\mathrm{true}+c$ on $e_1, e_2$, we find $(1+m) \approx 0.76$ and $c \approx \num{-0.27e-2}$. The right plot of \cref{fig:HSM_ell} compares the calibrated (dark orange) and non-calibrated (light orange) stacked $\Delta\Sigma$ lensing profiles for DC2object background catalog. The gray area indicates the region ($R < \SI{1}{\mega\parsec}$) where the simulation has not enough resolution to be reliable (see \cref{subsec:stacked_lensing_profiles}). The applied calibration to HSM ellipticities of DC2 objects divides the lensing profile amplitude by a factor of $\sim 0.76$, leading to a boost of the calibrated lensing signal.

\section{WL cuts for weak lensing sources}
\label{app:clean_cuts}

\begin{figure}
\centerline{\includegraphics[scale=0.35]{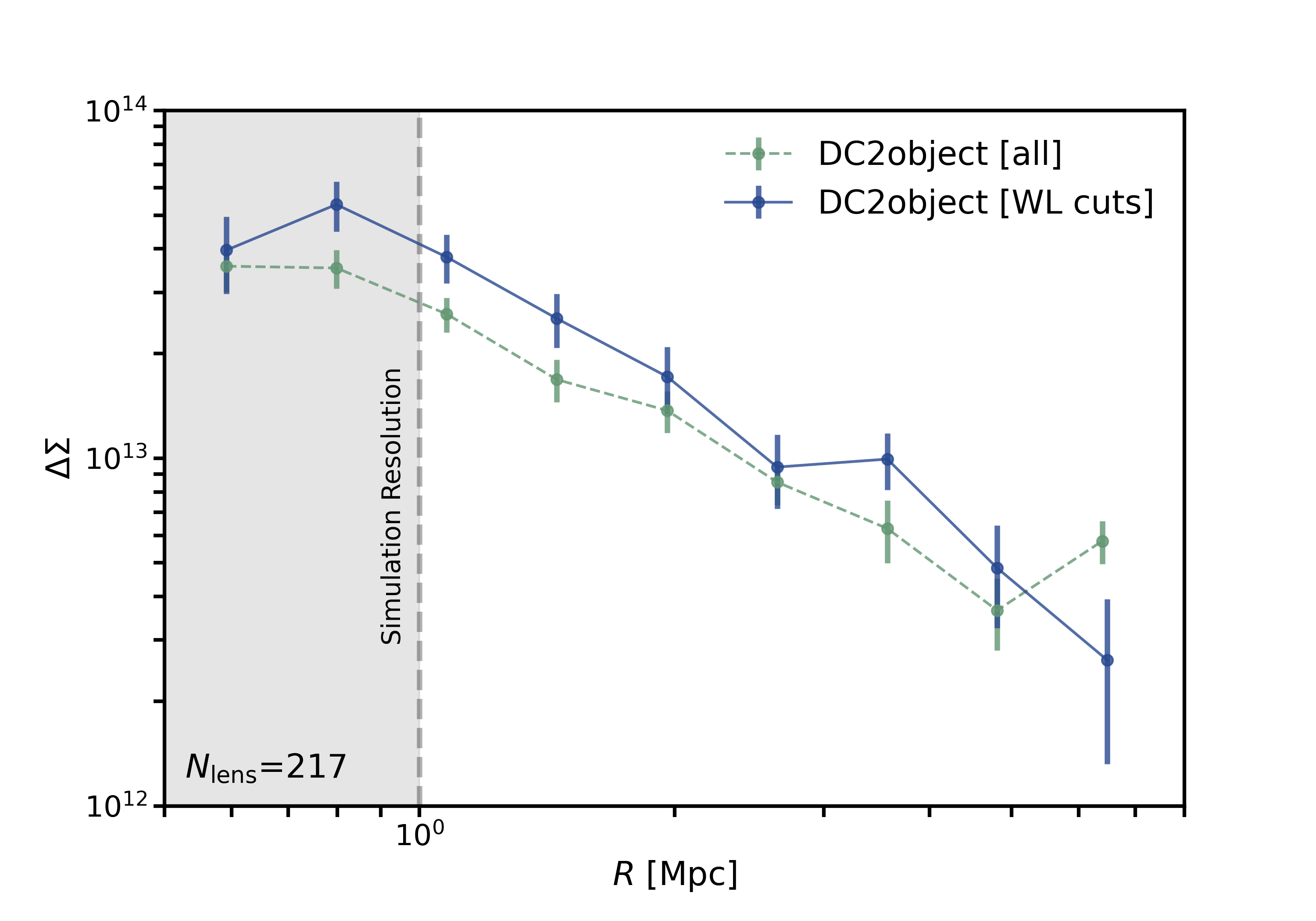}}
\caption{Stacked $\Delta\Sigma$ profiles with and without WL cuts (i.e. $\mathrm{SNR} > 10$, $\mathrm{resolution^{HSM}} \geq 0.3$, and $\mathrm{errors^{HSM}} \leq 0.4$) applied to the DC2object background sources.}
\label{fig:clean_cuts}
\end{figure}

To consider higher quality data in the weak lensing source selection, we adopt some cuts, labeled in this work as WL cuts. Following \cite{Mandelbaum_HSC}, we adopt a cut on the DC2 object Signal-to-Noise Ratio (SNR), such that $\rm{SNR} > 10$. In order to consider sufficiently resolved galaxies compared to the PSF, we apply a cut $\geq 0.3$ on the resolution of HSM shapes. A cut $\leq 0.4$ on the shape measurement errors has also been adopted. In \cref{fig:clean_cuts}, we compare the stacked $\Delta\Sigma$ profiles of the DC2object catalog with (blue) and without (green) the previously discussed cuts. These less noisy object shape selections allow to boost the lensing signal, leading to an increase in the lensing profile amplitudes.

\section{DC2 halo mass distributions and WL estimated masses}
\label{app:halo_mass_distrib}

\begin{figure*}
\centerline{\includegraphics[scale=0.35]{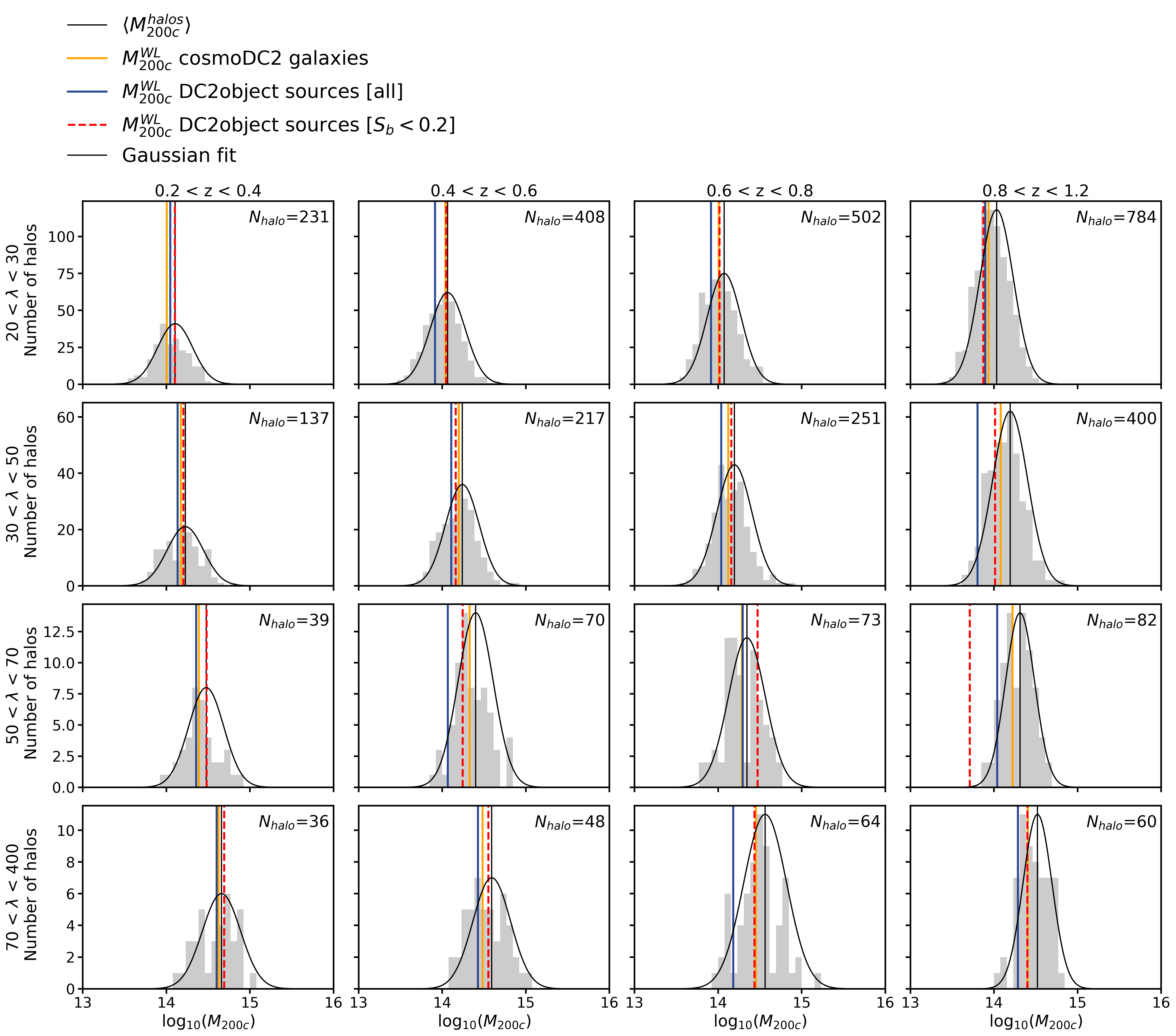}}
\caption{Halo mass distribution histograms (gray) per bins of richness and redshift. The halo mean mass per bin is indicated with the black solid line. The estimated weak lensing masses using cosmoDC2 background sources are shown with orange lines. The estimated weak lensing masses using DC2object with (respectively without) a $0.2$ cut on $S_b$ is represented with red (respectively blue) lines.
Gaussian distributions with $\mu$ (respectively $\sigma^2$) equals to the mean (respectively variance of) halo masses are over-plotted in solid black lines.}
\label{fig:mass_distrib}
\end{figure*}

In \cref{fig:mass_distrib}, we plot the cosmoDC2 halo mass distributions (gray histograms) within each bin of richness and redshift. On top of it we show Gaussian distributions with means $\mu$ (respectively variances $\sigma^2$) equal to the mean (respectively dispersion of) halo masses. We indicate with orange lines the cosmoDC2 weak lensing masses estimated from the $\Delta\Sigma$ profiles (see \cref{subsubsec:lensing_profiles}) and with dashed black lines the mean halo masses. As noticed in \cref{subsubsec:mass_estimates}, we show that the stacked masses are often spread from $10^{14}$ to $10^{15} \mathrm{M_{\odot}}$ with a larger number of low masses that can explain the bias between cosmoDC2 weak lensing masses and halo mean masses. As indicated in \cref{subsubsec:mass_estimates}, this bias may also arise from the spread of the mass-richness relation and the modeling of the halo mass function and density profiles.
In addition, we observe in \cref{fig:mass_distrib} that the DC2object weak lensing masses (indicated with blue lines) are biased low compared to cosmoDC2 masses (in orange) which can be recovered using a $S_b<0.2$ cut (red dashed lines).

\section{Impact of the halo mass function on cosmological parameter estimates}
\label{app:halo_mass_function}

\begin{figure}
\centerline{\includegraphics[scale=0.35]{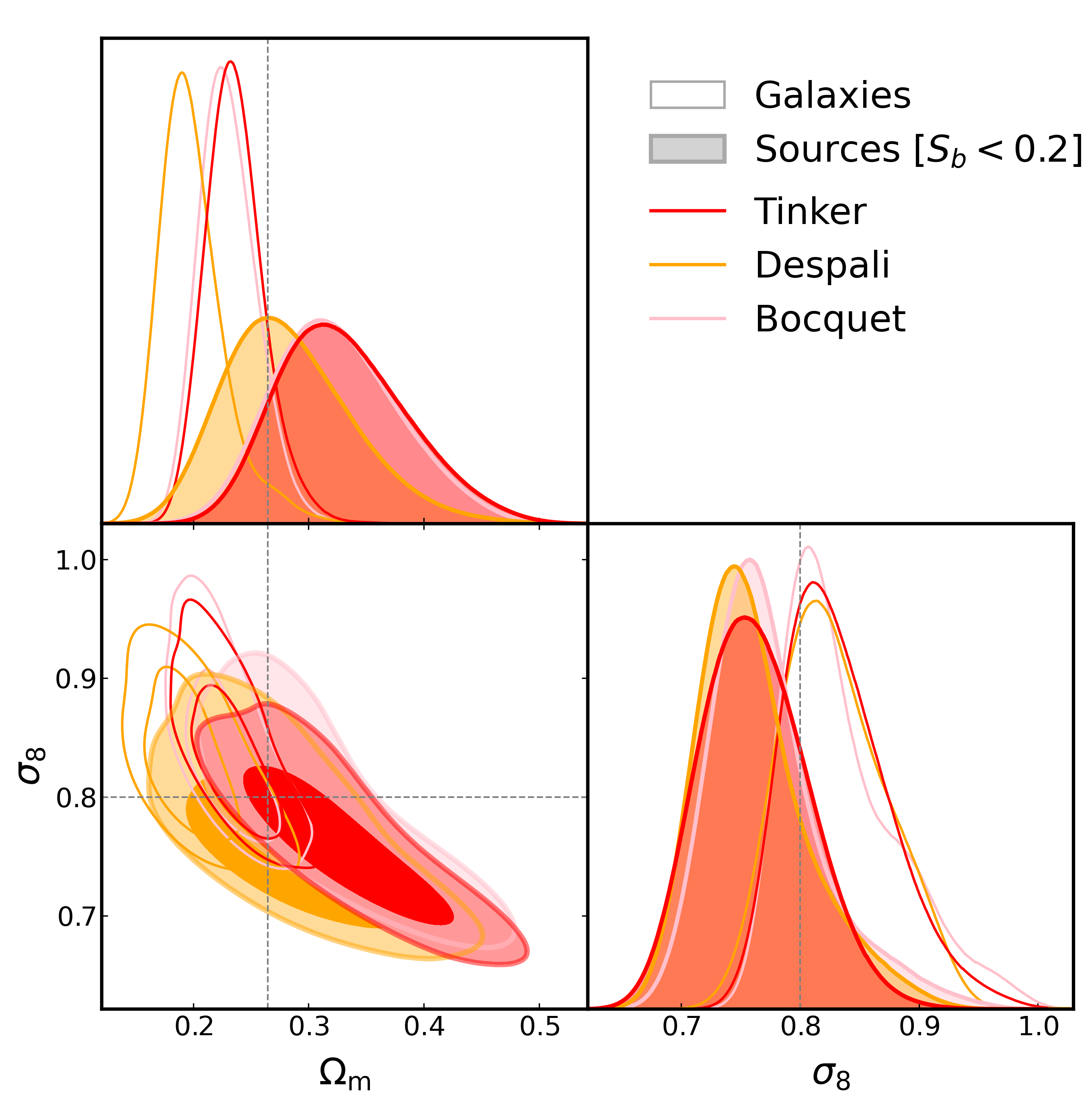}}
\caption{Posterior distributions of $\Om$ and $\sigma_8$ depending on the halo mass function definition, using the halo number count and weak lensing masses jointly. The contours are plotted for the \protect\cite{Tinker_HMF} (in red), \protect\cite{Despali_HMF} (in orange), and \protect\cite{Bocquet_HMF} (in pink) halo mass functions. The filled (respectively unfilled) contours are plotted using the DC2object sources with a blending entropy cut $S_b < 0.2$ (respectively cosmoDC2 galaxies). The fiducial values from DC2 simulations are indicated with the dashed gray lines. The DC2object contours (when considering blended sources) are not shown here for the clarity of the plot.}
\label{fig:HMF_blending_cosmo}
\end{figure}

The main analysis of the impact of blending on posterior distributions, combining halo number counts and weak lensing masses, is conducted using the halo mass function (HMF) from \citet{Tinker_HMF}, with the main results presented in \cref{fig:blending_cosmology}. However, the cosmological constraints may depend on the specific HMF model adopted, since the HMF is not an input to the \textit{Outer Rim} simulation. To test the robustness of our results against the choice of HMF, we show in \cref{fig:HMF_blending_cosmo} the posterior distributions of $\Om$ and $\sigma_8$ obtained using both cosmoDC2 (unfilled contours) and DC2object background sources after applying a blending entropy cut of $S_b < 0.2$ (filled contours). 
We compare results from three different HMFs: \citet{Tinker_HMF} (red), \citet{Despali_HMF} (orange), and \citet{Bocquet_HMF} (pink). Overall, we find consistent results across the three HMFs, particularly for the $\sigma_8$ parameter, for both cosmoDC2 and DC2object datasets. For $\Om$, the \citet{Despali_HMF} model results in slightly lower values (around $\sim 0.20$) compared to the \citet{Tinker_HMF} and \citet{Bocquet_HMF} models (around $\sim 0.24$) in the cosmoDC2 case. However, a similar offset is observed in the DC2object constraints, indicating that the mitigation of blending is not significantly affected by the choice of HMF.
The overall agreement between the different HMF models remains satisfactory, with all constraints overlapping at the $1\sigma$ level when high-$S_b$ blended sources are excluded. This indicates that the cosmological inference is relatively insensitive to the choice of HMF when blending is properly accounted for. These results further support the use of a blending entropy cut at $S_b < 0.2$ as an effective approach to mitigate blending-induced biases in weak lensing measurements, allowing for reliable estimates of cluster masses in cosmological studies.
For the clarity of the plot, we do not show the DC2object contours, when considering blended sources. However, they are consistent with the trend observed in \cref{fig:blending_cosmology}, using the \cite{Tinker_HMF} halo mass function, and are biased towards higher $\Om$ and lower $\sigma_8$ values, compared to the cosmoDC2 contours.

\end{document}